\documentclass{aa}  
\usepackage{natbib}
\usepackage{lscape}
\usepackage{longtable}
\usepackage{graphicx}
\usepackage{txfonts}
\usepackage[pdftitle="The VEGAS survey"]{hyperref}
\def\Re{\mbox{$R_{\rm e}$}}

\begin{document}

   \title{VEGAS: A VST Early-type GAlaxy Survey.}
   \subtitle{I. Presentation, wide-field surface photometry, and substructures in
 NGC 4472}

   \author{Massimo Capaccioli
          \inst{1,2}
          \and
          Marilena Spavone\inst{1}
\and 
Aniello Grado\inst{1}
\and 
Enrichetta Iodice\inst{1}
\and 
Luca Limatola\inst{1}
\and 
Nicola R. Napolitano\inst{1}
\and 
Michele Cantiello\inst{3}
\and   
Maurizio Paolillo\inst{2}
       \and
          Aaron J. Romanowsky\inst{4,5}
       \and
          Duncan A. Forbes\inst{6}
       \and
          Thomas H. Puzia\inst{7,8}
       \and
          Gabriella Raimondo\inst{3}
       \and
          Pietro Schipani\inst{1}
          }
          
 \institute{INAF-Astronomical Observatory of Capodimonte, Salita Moiariello 16, I80131, Naples, Italy\\
              \email{capaccioli@na.infn.it}
         \and
             University of Naples Federico II, C.U. Monte Sant'Angelo,
             Via Cinthia, 80126, Naples, Italy\\
         \and
             INAF-Astronomical Observatory of Teramo, Via Maggini,
             64100, Teramo, Italy\\
         \and
Department of Physics and Astronomy, San Jos{\'e} State
            University, One Washington Square, San Jos{\'e}, CA 95192,
            USA\\
\and
             University of California Observatories, 1156 High Street,
             Santa Cruz, CA 95064, USA\\
         \and
            Centre for Astrophysics \& Supercomputing, Swinburne
            University, Hawthorn, VIC 3122, Australia\\
\and
Institute of Astrophysics, Santiago, Chile,\\
\and
National Research Council Canada, Victoria, Canada\\
             }

   \date{Received ....; accepted ...}


  \abstract
   {We present the VST Early-type GAlaxy Survey (VEGAS), which
is designed to obtain deep multiband photometry in $g, r, i$, of about one hundred nearby galaxies down to 27.3, 26.8, and 26 mag/arcsec$^2$ respectively, using the ESO facility VST/OmegaCAM.}
   {The goals of the survey are 1) to map 
the light distribution up to ten effective radii, $r_{e}$, 2) to trace color gradients and surface brightness 
fluctuation gradients out to a few $r_{e}$ for stellar population characterization, and 3) to obtain a
full census of the satellite systems (globular clusters and dwarf galaxies) out to 20\% of the galaxy virial radius.
The external regions of galaxies retain signatures of the formation and evolution mechanisms that shaped them, and the study of nearby objects enables a detailed analysis of their morphology and interaction features. 
To clarify the complex variety of formation mechanisms of early-type
galaxies (ETGs), wide and deep photometry is the primary observational step, which at the moment has been pursued with only a few dedicated programs. The VEGAS survey has been designated to provide these data for a volume-limited sample with exceptional image quality.}
   {In this commissioning photometric paper we illustrate the capabilities of the survey using \emph{g-} and \emph{i}-band VST/OmegaCAM images of the nearby galaxy NGC 4472 and of smaller ETGs in the surrounding field.}
   {Our surface brightness profiles reach rather faint levels and agree excellently well with previous literature. Genuine new results concern the detection of an intracluster light tail in NGC 4472 and of various substructures at increasing scales. We have also produced extended (\emph{g-i}) color profiles.}
   {The VST/OmegaCAM data that we acquire in the context of the VEGAS survey provide a detailed view of substructures in the optical emission from extended galaxies, which can be as faint as a hundred times below the sky level.}

\authorrunning{Capaccioli et al.}
\titlerunning{The VEGAS survey}
   \maketitle 
%

\section{Introduction}\label{intro}  

Recent years have witnessed renewed interest in
bright early-type galaxies (ETGs). Observations at high redshift
revealed that ETGs have undergone remarkable amounts of size evolution
over time (e.g., \citealt{Daddi05}; \citealt{vanDokkum10}).
Theory suggests this growth to be a basic aspect of hierarchical structure formation, 
with mergers building up extended bulges and stellar halos (\citealt{Oser10}; \citealt{Hopkins10}).

The new paradigm of ``two-phase'' or ``inside-out'' galaxy assembly, pictured by cosmological simulations, outlines two regimes in the formation of the baryonic structure of a galaxy. In a first early ($z \gtrsim 2$) phase there is rapid in situ star formation from infalling cold gas, followed by a longer accretion phase where the system considerably grows in size and mass by accreting smaller satellites.
This new paradigm motivates
a return to classical studies of nearby ETGs, searching for the
expected signatures of formational processes, particularly at large radii.

Pilot studies have indeed revealed extensive
evidence of outer galaxy assembly: from pervasive photometric substructures
(\citealt{Tal09}; \citealt{Janowiecki10}) to metallicity gradients 
(\citealt{Coccato10}; \citealt{Forbes11}), rotational changes 
(\citealt{Proctor09}; \citealt{Coccato09}; \citealt{Arnold11}), 
and accretion signatures in chemo-dynamical phase space \citep{Romanowsky12}.

A full understanding of any galaxy begins with photometry. The situation for nearby ETGs is the following:
the central regions are studied in much detail (e.g., \citealt{Ferrarese06}, F+06 hereafter; \citealt{Cote07}), while the faint outskirts are still poorly investigated, even if they are becoming a hot topic with multiple surveys being carried out
(e.g., \citealt{Kormendy09}, K+09 hereafter, \citealt{Duc15}).
(e.g., \citealt{Kormendy09}, K+09 hereafter, \citealt{Duc15}).
There is a critical need for modern, wide-field (WF), multiband CCD photometry of a large
sample of galaxies in a broad range of environments, replacing the photographic
and narrow-field CCD work of decades past (e.g., \citealt{Peletier90}; \citealt{Caon94}). The aim is to systematically gauge the basic global properties of ETGs over a wide baseline of sizes:  
luminosity profiles, isophote shapes, substructure characteristics,
color gradients, surface brightness fluctuations, inventories of satellite galaxies
and globular clusters (GCs), etc.
The wide range of science results and applications available
from such a dataset,
beyond the general goal of testing two-phase assembly models,
cannot be covered here, therefore we briefly highlight a few topics.

Multiband surface brightness (SB) mapping of ETGs allows us to measure key physical parameters  through the fit of generalized $R^{1/n}$ profiles \citep{Caon93}: total luminosity, Sersic index $n$, effective surface brightness and radius, $\mu_e$ and $R_e$, boxy- or diskyness,  etc.  (\citealt{Caon93}; \citealt{Balcells07}). The correlations between them, such as $\mu_e$ vs $R_e$, mass vs size or photometric plane (\citealt{Kormendy85}; \citealt{Capaccioli92}; \citealt{Shen03}) help shedding light into formation processes. Along the same line, outer breaks in the SB profiles might correlate in a non-trivial way with the inner core or cusp transition (e.g., \citealt{Cote07}); this has not been studied so far. Moreover, photometry is a way to identify and gauge substructure and/or light excesses as expected from the diffuse stellar components (e.g., \citealt{Zibetti05}), especially in the intracluster environment (\citealt{Mihos05}; \citealt{Mihos13}), through deviations from the regular $R^{1/n}$ behavior. Radial color gradients are
critically related to the formation mechanisms \citep{Carlberg84} because they give a hint of the different distributions in stellar ages and metallicities (\citealt{Saglia02}; \citealt{Pipino08}; \citealt{Tortora11}).
The combination of color distribution with surface brightness fluctuations (SBF, \citealt{Tonry88}) supplies further information on the chemical properties of the stellar populations and helps lift the age-metallicity degeneracy out to a few effective radii \citep{Cantiello13}.

\begin{figure*}
\centering
\hspace{-0.cm}
 \includegraphics[width=15cm, angle=-0]{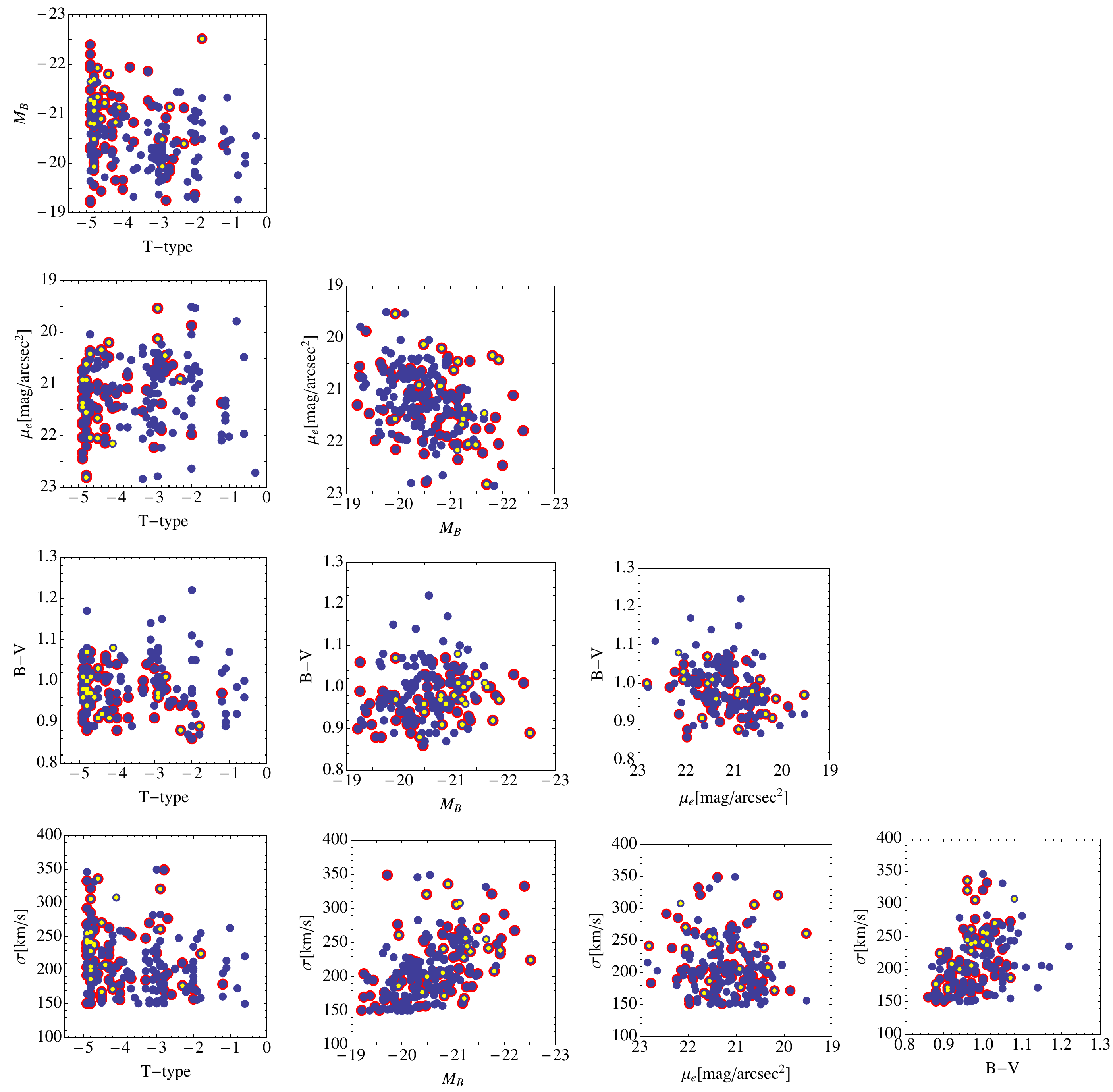}
\caption{VEGAS potential target distribution in the parameter space (blue points). Higher priority is given to galaxies with
HST data (red circles) and Chandra/XMM data (yellow dots). We
note that close systems are observed in a 
single VST/OmegaCAM pointing in many cases.}\label{fig:fig1}
\end{figure*}

Furthermore, accurate photometry up to 10 $R_e$ is  mandatory (and still lacking) for dark matter studies because of the advent of efficient kinematical tracers such as the planetary nebulae (PNe, e.g., \citealt{Romanowsky03}; \citealt{Napolitano09}) and globular clusters (GCs, e.g., \citealt{Romanowsky09}, \citealt{Napolitano14}).  In particular, extended deep photometric mapping will naturally provide a fairly complete census of galaxy satellites, from globular clusters (GCs) to satellite galaxies: a multipurpose database
that is also useful for testing the formation scenarios.

Among the key questions that still remain open there is the well-known bimodality of the color distribution of GCs in galaxies (e.g., \citealt{Peng06}). This has different possible explanations: i) either high-redshift, two-phase formation of elliptical systems (e.g., \citealt{Forbes97}), ii) the dissipative merging of late-type spirals (e.g., \citealt{Ashman92}); iii) the hierarchical feeding of a bright (metal-rich) elliptical by (metal-poorer) dwarfs (e.g., \citealt{ Cote98}), iv) the recent proposal of a unimodal metallicity distribution that is transformed into a bimodal color distribution because of the nonlinearity of color-metallicity relations (e.g., \citealt{Yoon06}, \citealt{Cantiello15}, and references therein). Although a unique consensus on the interpretation of this phenomenon 
is still lacking, we note that as in the Milky Way GC system, the systems 
of several other early-type galaxies are clearly bimodal in metallicity (\citealt{Brodie12, Brodie14}; \citealt{Usher12, Usher15}).

Finally, satellite galaxies are important because they can be tidally disrupted in their journey around larger systems. These events are possibly the mechanisms producing the diffuse halos around galaxies (\citealt{Ibata94}; \citealt{Zibetti04}; \citealt{Arnaboldi12}) or even the intragroup or cluster light (\citealt{Mihos13}; \citealt{Zibetti05}).

In view of all this and considering the special characteristics of the VLT Survey Telescope (VST; \citealt{Capaccioli11}), a project for a photometric survey of nearby ETGs, dubbed VEGAS, has been undertaken on the Italian Guaranteed Time Observation (GTO). This is the first paper of a series where we present the survey project and its strategy, the data reduction and analysis techniques, and report on a test case conducted to assess and certify the quality of our products.

The paper is organized as follows. In Sect. \ref{sec:vegas} we briefly describe the VEGAS survey aims and objectives. The observations of a test galaxy are described in Sect. \ref{sec:data}. In Sect.
\ref{phot} we illustrate the strategies adopted for the data analysis, with a particular emphasis on the determination of the sky background and the measurement of accurate surface brightness profiles. In Sect. \ref{results} we discuss the surface brightness profiles of the objects in this study and compare results with previous literature, while in Sect. \ref{scatter} we discuss the effects of the scattered light on the surface brightness profiles. Finally, in Sect. \ref{conc} we discuss that the VST/OmegaCAM data are of the highest quality for wide-field imaging and why we believe that this machinery is a powerful tool for an ``industrial'' analysis of optical photometry of nearby galaxies. In Appendix
\ref{app} and \ref{PSF1} we describe the details of the data reduction and of the point spread function.

We adopt a distance modulus for the Virgo cluster  of $31.14 \pm\ 0.05$ mag as in \citet{Mei07}. This correspond to a distance of 16.9 Mpc, so 1 arcsec is 81.9 pc. The magnitudes throughout the paper are in the AB system. Our surface brightness data are not corrected for Galactic extinction, but the total magnitude values listed in Table \ref{basic} have been corrected assuming the recipe of \citet{Battaia12}.

\section{VEGAS survey}\label{sec:vegas}

The VST Elliptical GAlaxies Survey (VEGAS) is a deep multiband ($g, r, i$) imaging survey of early-type galaxies in the southern hemisphere carried out with VST at the ESO Cerro Paranal Observatory (Chile). The large field of view (FOV) of the OmegaCAM mounted on VST (one square degree matched by pixels 0.21 arcsec wide), together with its high efficiency and spatial resolution (typically better than 1 arcsec; \citealt{Kuijken11}) allows us to map with a reasonable integration time the surface brightness of a galaxy out to isophotes encircling about 95\% of the total light. Observations started in October 2011 (ESO Period 88), and since then, the survey has acquired exposures for about 20 bright galaxies (and for a wealth of companion objects in the field), for a totality of $\sim$80 hr (up to Period 93).

Since the OmegaCAM detector is a mosaic of 32 CCDs, a dithering strategy has to be devised to fill the blind gaps among the 2000 $\times$ 4000 pixels of individual CCDs. The actual implementation of the dithering strategy has consequences for setting the  weight
map of the various pixels of the final combined image, as well as for mixing and averaging the residual errors in the engineering of the individual CCDs because of the overlapping of adjacent CCDs. 

The survey project is designed to map the surface brightness of galaxies with $T_{type} < 0$, $\sigma > 150$, $Dec < +5$, $V_{rad}<4000$ km/s, and $B_{tot} < -19.2$, sampling all environmental conditions and the whole parameter space. To this end, we selected from the catalog of nearby galaxies by \citet{Prugniel96} a large sample of about 240 potential E/S0 targets (Fig. \ref{fig:fig1}) with the aim of optimizing the observing strategy throughout
the year so as to observe half of this sample in five years and to uniformly cover the galaxy parameter space.

The distribution of parameters in Fig. \ref{fig:fig1} refers to the central targets of the VEGAS pointings, while we expect to simultaneously observe many lower luminosity systems.
Higher priority is given to galaxies with ancillary data (e.g., {\it HST} or {\it Chandra}/{\it XMM}, see Fig. \ref{fig:fig1}).

The expected depths at a signal-to-noise ratio $\mbox{(S/N) of}>3$ in the $g$, $r,$ and $i$ bands are 27.3, 26.8, and 26 mag arcsec$^{-2}$ , respectively. They are the result of a compromise between a reasonable exposure time and the need to detect signatures of a diffuse stellar component around galaxies (see, e.g., \citealt{Zibetti05}) and the dynamical interaction of ETGs with the intergalactic medium. 

The main products of the VEGAS survey are 1) a 2D light distribution out to 8-10 $R_e$: galaxy structural parameters and diffuse light component, inner substructures as a signature of recent cannibalism events, inner disks and bars fueling active nuclei present in almost all the objects of our sample; 2) radially averaged surface brightness profiles and isophote shapes out to 10 $R_e$; 3) color gradients and the connection with galaxy formation theories; 4) detection of external low-surface brightness  structures of the galaxies and the connection with the environment; 5) census of small stellar systems (SSS: GCs, ultra-compact dwarfs and galaxy satellites) out to $\sim$20 $R_{e}$ from the main galaxy center, and their photometric properties (e.g., GC luminosity function and colors, and their radial changes out to several \Re), allowing us to study the properties of GCs in the outermost ``fossil'' regions of the host galaxy. 
This latter subproject is also called VEGAS-SSS \citep{Cantiello15}. We note that the majority of studies on the photometric properties of the GC system in ETGs cover the central (few arcmin) region of the host galaxy (e.g., ACS Virgo \& Fornax cluster surveys, \citealt{Cote04} and \citealt{Jordan07}). An exception to the inner imaging studies is the SLUGGS survey that uses the Subaru/Suprime camera (e.g., \citealt{Blom12}).

As a natural byproduct of the survey (for the depth and high S/N in the central galaxy regions), a galaxy SBF, and a SBF-gradient analysis is planned to chemically characterize the stellar population within $\sim 2R_e$ (or more, for the nearest  ETGs in the sample). 

A fundamental aspect of  the survey resides in the legacy value of the data-set for ETGs, to be used for a wide range of research lines. The survey area extends from $-70$ to +5 degrees in Dec and 0--24h in RA (see Fig. \ref{fig:fig1}), which ensures observability throughout the year and an advantageous overlap with the KiDS survey area \citep{deJong13}.

VEGAS will provide a volume-limited survey in the South complementary  to the Next Generation Virgo Cluster Survey (NGVS, \citealt{Ferrarese12}), with similar depth but no environmental restrictions, and will be the southern equivalent to MATLAS \citep{Duc15}. The updated status of VEGAS observations is posted at the link \url{http://www.m2teamsoftware.it/vst/index.php/science/gto-surveys/vegas}.


\section{NGC 4472 field: observations and data reduction}\label{sec:data}
\begin{figure*}
   \centering
   \includegraphics[width=13cm]{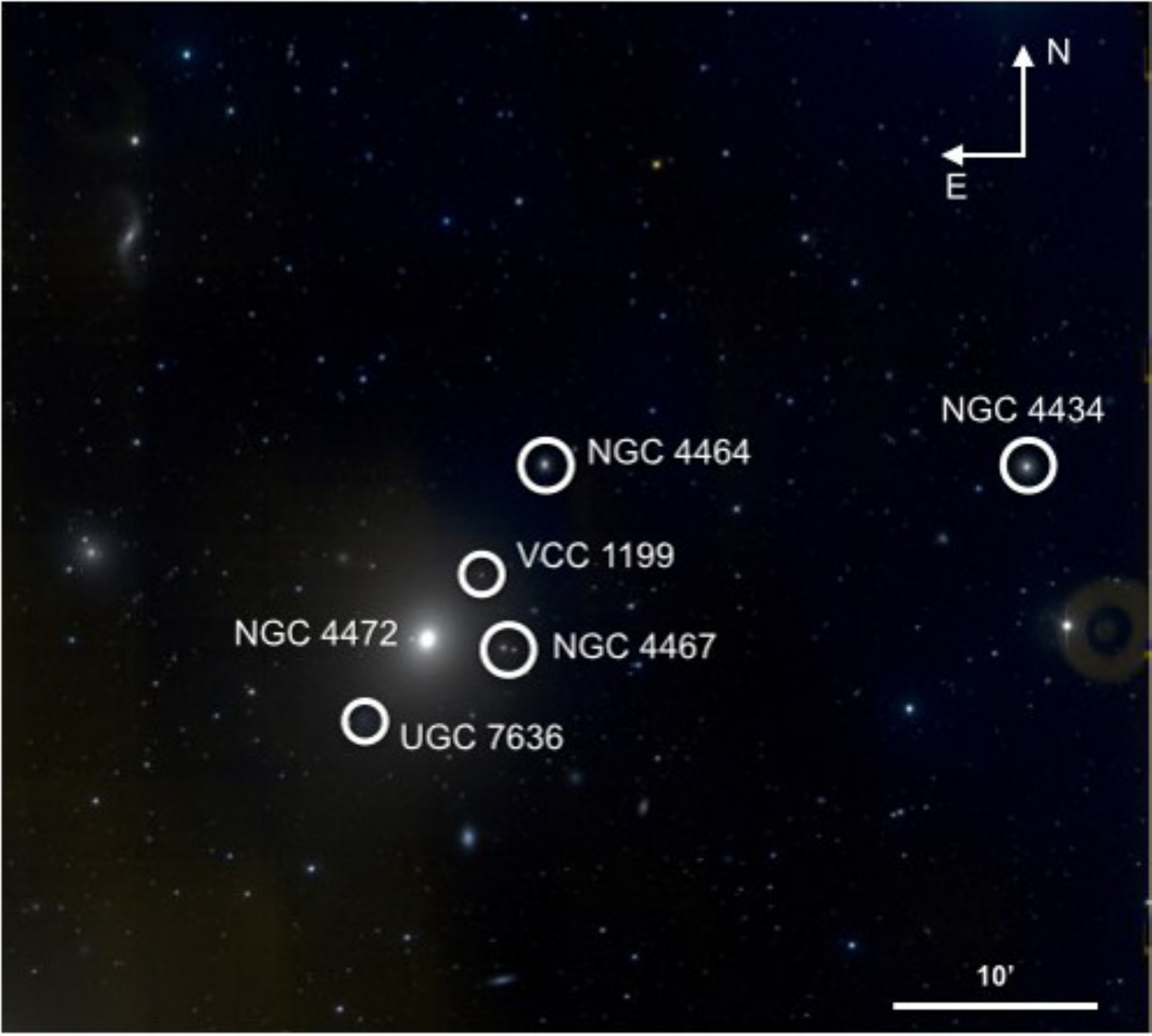}
   \caption{VST color composite image of the $0.9^{\circ} \times\ 0.8^{\circ}$ field around the giant galaxy NGC 4472 from {\it g} and {\it i} band VEGAS images. Circles mark the other four ETGs studied in this paper and the interacting system UGC 7636.}
              \label{field2}
    \end{figure*}

This first VEGAS paper presents a deep photometric analysis of the ETGs in the VST field of the galaxy NGC 4472 (M 49),  
the brightest member of the Virgo cluster (Table \ref{basic}). 
We have chosen this field for the following reasons: 
\begin{itemize}
\item it is well-studied with an ample scientific photometric literature (\citealt{Kim00, Ferrarese06, Kormendy09, Janowiecki10, Mihos13});
\item it offers a wide range of cases for investigating the ability of VEGAS to map the faint galaxy outskirts. 
Together with this supergiant nearby object that fills almost the entire OmegaCAM field, there are smaller ETGs either embedded in the light of NGC 4472 or close to the edges of the frame (see Fig. \ref{field2}). Each one of these cases requires a different data reduction strategy and calls for an independent verification.
\end{itemize}

\begin{table}
\caption{\label{basic}Parameters of  NGC 4472} \centering
\begin{tabular}{lccc}
\hline\hline
Parameter & Value & Ref.\\
\hline \vspace{-7pt}\\
Morphological type & E2 & RC3 \\
R.A. (J2000)& 12h29m46.7s& NED \\
Dec. (J2000) & +08d00m02s & NED\\
Helio. radial velocity & 981 km/s&NED\\
Distance& 16.9 Mpc& \citet{Mei07} \\
Mean axis ratio&0.81& NED\\
Absolute magnitude $M_{g}$& -22.85\tablefootmark{a}& This work\\
Absolute magnitude $M_{i}$& -24.22\tablefootmark{a}& This work\\
\hline
\end{tabular}
\tablefoot{ \tablefoottext{a}{Corrected for interstellar extinction as in \citet{Battaia12}.}}
\end{table}
The data used in this paper consist of exposures in {\it g} and {\it i} SDSS bands (Table \ref{data}) obtained with VST + OmegaCAM in service mode under photometric sky conditions and with the following constraints: 
\begin{itemize}
\item S/N $\geq$ 3 per arcsec$^{2}$;
\item dark time;
\item seeing $\leq$ 1'';
\item airmass $\leq$ 1.2.
\end{itemize}

For the sake of clarity, we repeat that the FOV of each frame covers one square degree, with a scale of 0.21 arcsec pixel$^{-1}$. 
The total integration time is 5695\,seconds in {\it g} and 4590\,seconds in {\it i}.  
The different exposures have the same center, which has been chosen not to coincide with that of NGC 4472, principally in order to move the galaxy core out of the central crossing of the gaps. 
More details about the dithering strategy can be found in the VST manual at the following link: \url{https://www.eso.org/sci/facilities/paranal/instruments/omegacam/doc/}.

\begin{table}
\caption{\label{data}VST exposures used in the photometry of the NGC 4472 field.} \centering
\begin{tabular}{lccccc}
\hline\hline
Band & Date & Nr. frames & Total exp. time & FWHM\tablefootmark{a} \\
     &      &            & [sec] & [arcsec] \\
\hline \vspace{-7pt}\\
{\it g} & 2013-03-19 & 5 & 1225 & 0.83 \\
        & 2013-03-20 & 5 & 1225 & 1.40\\
        & 2013-04-15 & 10 & 2120 & 0.85 \\
        & 2013-04-16 & 5 & 1125 & 0.92 \\
{\it i} & 2013-03-19 & 5 & 1250 & 0.66 \\
        & 2013-04-16 & 10 & 1670 & 0.73 \\
        & 2013-05-14 & 10 & 1670 & 0.77 \\
\hline
\end{tabular}\label{tab:tab_data}
\tablefoot{ \tablefoottext{a}{Median value of the FWHM.}}
\end{table}

The data were processed with a pipeline specialized for the VST-OmegaCAM observations (dubbed VST-tube; \citealt{Grado12}), which performs the 
following main steps:
\begin{itemize}
\item prereduction;
\item astrometric and photometric calibration;
\item mosaic production.
\end{itemize}

Science images are first treated to remove the instrumental signatures, applying overscan, bias, and flat-field corrections, as well as gain harmonization of the 32 CCDs, illumination correction and, for the {\it i} band, defringing.
Relative and absolute astrometric and photometric calibrations are applied before creating the final coadded image mosaics. 
In Appendix \ref{app} we describe the various steps of the procedure in detail.

\section{NGC 4472 field: photometric processing}\label{phot}

 \subsection{Sky background subtraction}\label{back}
The background estimate and subtraction is the most critical operation in deep photometric analysis because it affects the ability of
detecting and measuring the faint outskirts of galaxies. 

There are at least two ways to model the sky background. The first one, extensively tested in classical photographic  surface photometry \citep{Capaccioli88}, consists of fitting a surface, typically a 2D polynomial, to the pixel values of the mosaic that is unaffected by celestial sources or defects. The advantage comes from the simultaneity of the exposure of the galaxy and the background, which is particularly relevant in wide-field images owing to the differential effects of refraction and to the moon light, if any. Minor glitches in the CCDs' sensitivity are averaged as well. The second method mimics the ON-OFF procedure devised in IR astronomy that is made possible by the use of digital detectors. The background is estimated from exposures taken as close as possible, in space and time, to the scientific ones. The main advantage is that the risks in guessing which pixels belong to celestial sources and which to the background are largely reduced, particularly in the target galaxy outskirts. A shortcoming of this strategy, in addition to the already mentioned lack of simultaneity in the galaxy and background exposures, is that it consumes more telescope-time. 

In this first paper we have adopted the direct polynomial interpolation procedure described below. The reason is that it this is capable of exploring the background for galaxies embedded in the light of more extended sources, as is the case for all the objects of this study except NGC 4472. At the same time, we have tested the procedure on the giant galaxy whose size competes with that of the VST frame.

\begin{figure*}
   \centering
  \hspace{-1cm}    
  \includegraphics[width=8.5cm]{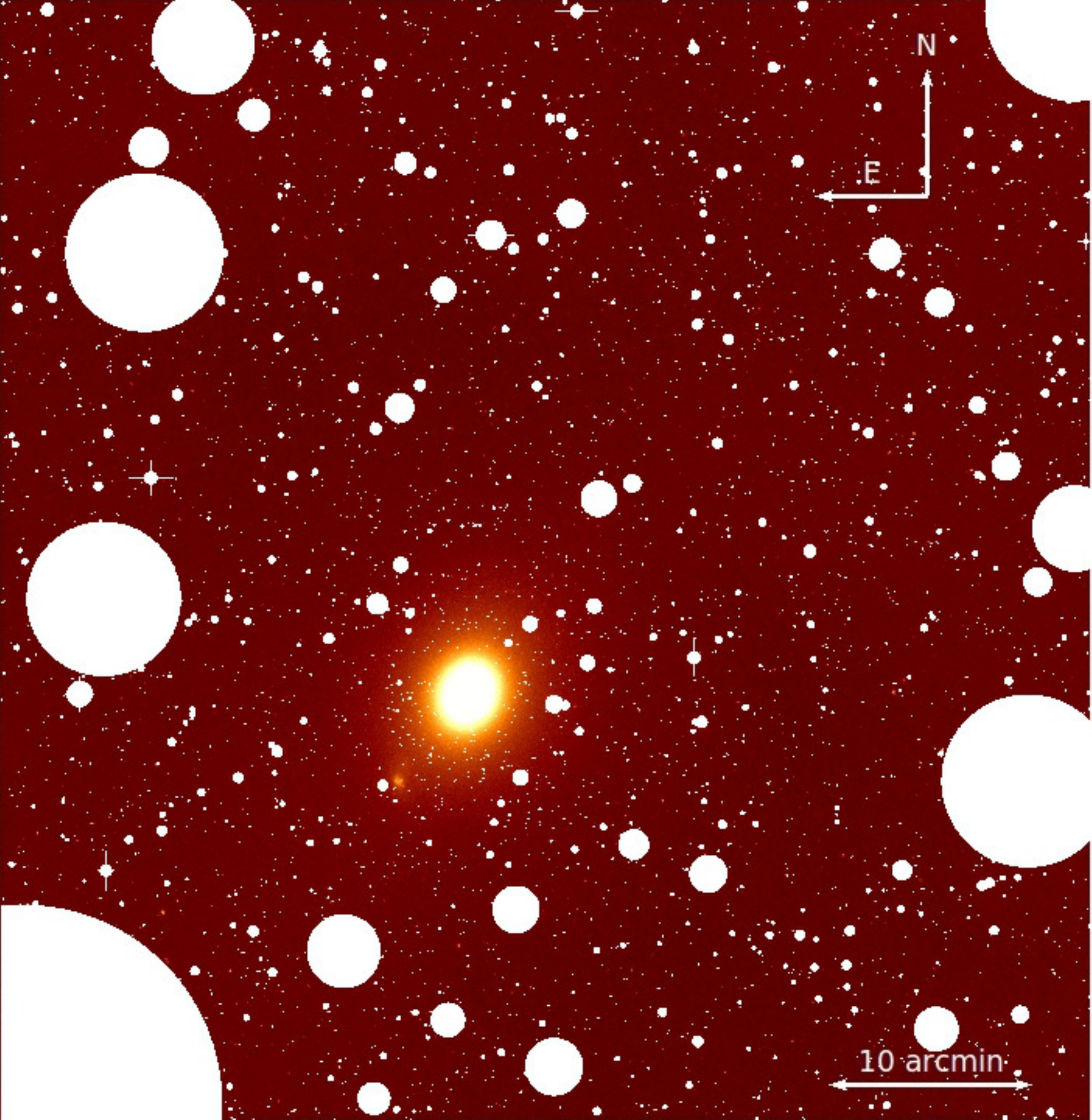}\\
   \caption{VST {\it g} -band mosaic of NGC 4472 showing the masking of bright sources in the field, done either automatically or manually depending on the brightness of the sources.  }
              \label{mask}
    \end{figure*}

VST images contain a very large number of sources (stars, galaxies,
and image
defects). They have to be masked out to define the subset of bona fide 
background pixels to perform the interpolation. To this end, we used ExAM\footnote{ExAM is a code developed by Z. Huang during his PhD. A detailed description of the code can be found in his PhD thesis, available at the following link: \url{http://www.fedoa.unina.it/id/eprint/8368}} \citep{Huang11}, a program based on
SExtractor \citep{Bertin96}, which was developed to accurately mask background and 
foreground sources, reflection haloes, and spikes from saturated stars. Very bright stars and galaxies were masked manually. Figure \ref{mask} shows a 0.89 $\times$ 0.91 square degrees OmegaCAM {\it g} -band image\footnote{The reduced size with respect to the nominal VST FOV of one square degree results from a trimming of the low-weight pixels at the rim of the mosaic.} 
of the NGC 4472 field to which the masking procedure has been applied. Masked areas are marked as blank circles.

The most critical step is to optimize the size of
the galaxy mask. In principle the problem is very simple. The pixels to be removed from the image are all and only those belonging to the galaxy: {\it a)} ``all'' because we wish the residual galaxy halo to avoid causing an overestimate of the background that induces spurious cutoff in the outer light profiles, {\it b)} ``only'' because we wish to avoid unnecessarily widening  the blank area where the computed surface interpolates the background, which might again induce unreal trends in the faint end of the light distribution.
The problem is particularly difficult for ETGs compared to spirals and irregulars because the outermost light distribution smoothly fades.

We solved the problem by creating a set of elliptical masks of increasing sizes centered on the galaxy, with fixed flattening and orientation mimicking the mean behavior of the outer galaxy halo. For each mask we then computed the fifth-order Chebyshev polynomial that best fit the residual source-free image. We then analyzed the median values of the differences between the image and the fitted surface in elliptical annuli around each
mask as a function of the mask size to find the smallest mask with vanishing
residuals.
Clearly this procedure hardly converges when the targeted galaxy fills a significant portion of the OmegaCAM FOV.

\begin{figure*}
   \centering
   \includegraphics[width=13cm]{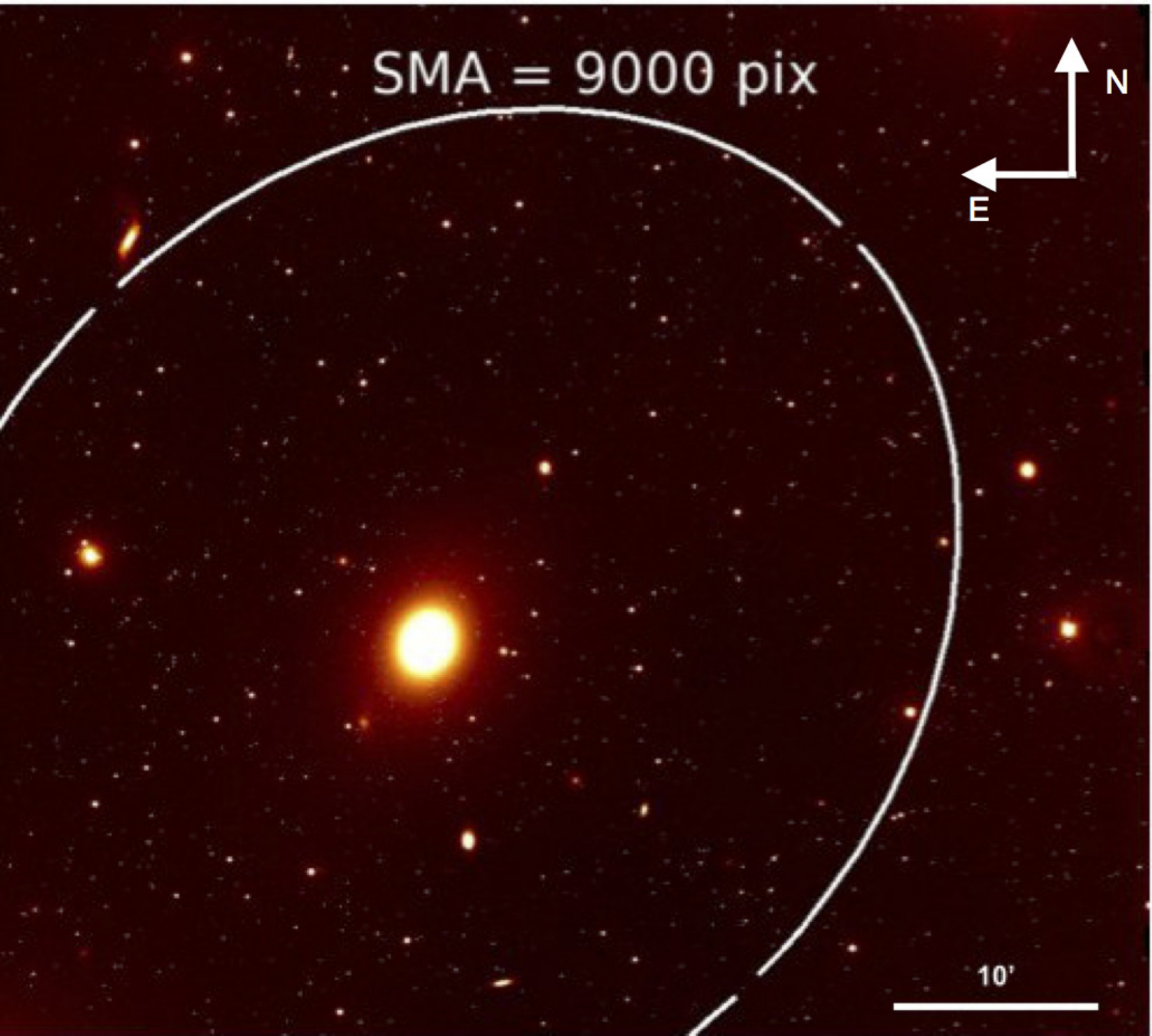}
   \caption{VST {\it g} -band mosaic of NGC 4472. The isophotal contour (in white)
   represents the last isophote fitted to obtain the surface
   brightness profile. The image size is $0.9^{\circ} \times\ 0.8^{\circ}$.}
              \label{out}
    \end{figure*}

This is the case for NGC 4472 (see Fig. \ref{out}). 
Our compromise strategy here assumes that the background level is the median value over the outer annuli of the mosaic. 
This rough constant estimate is first subtracted from the image and then further improved by randomly picking $5\times5$ pixel$^2$ 
boxes at the edge of the image and averaging the median counts.
By this approach we have estimated a further correction of 
$\Delta c=-0.3$ ADU over $\sim$ 100 ADU for the {\it g} band and $\Delta c=-1.3$ ADU over $\sim$ 600 ADU for the {\it i} band.

As a test we assumed that the surface brightness profile of the galaxy (see Sect. \ref{light}) 
can be well approximated by an $r^{1/4}$ law \citep{deV48}, and fitted\footnote{We used MINUIT
\citep{James75}, which is a program written by staff of CERN (European Organization for Nuclear Research). It searches for minima
in a user-defined function with respect to one or more parameters
using several different methods as specified by the user.} the function $I(a) = I_{0}
\times\ 10^{\left(-3.3307 \times\ \left(a/a_{e}\right)^{1/4}\right)}+\Delta c'$, where $a$ is the galaxy semi-major axis, to the azimuthal light profiles derived in Sect. \ref{phot}. The free parameters are $I_{0}$, $a_{e}$
and $\Delta c'$. It turns out that $\Delta c'$, meaning that the second-order correction of $\Delta c$ is about zero with an uncertainty of 0.1$\%$ in the less favorable case ({\it g} band). 

\begin{figure*}
   \centering
   \includegraphics[width=13cm]{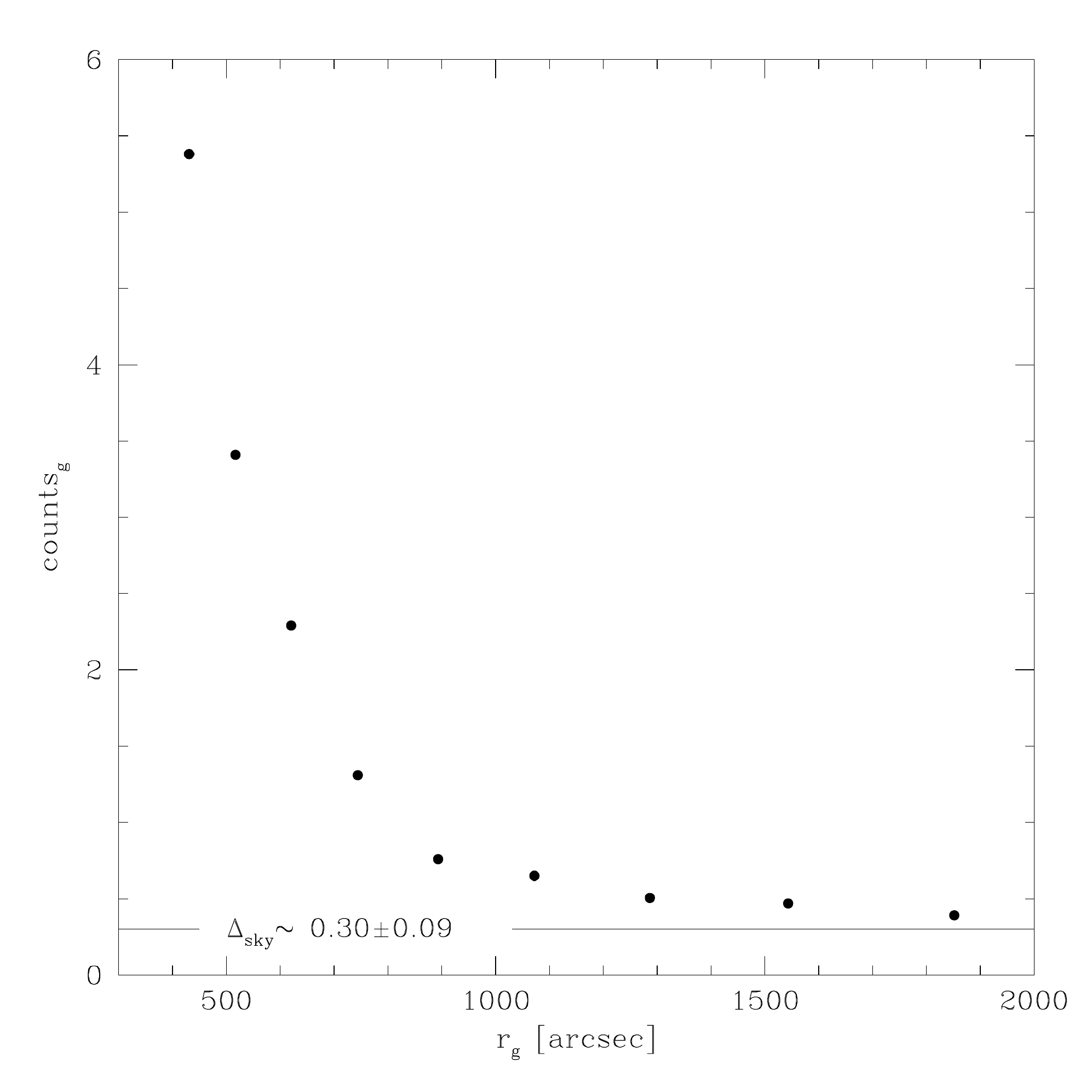}
   \caption{Azimuthally averaged intensity profile of NGC 4472 as a function of the semi-major axis. The horizontal line indicates the residual background counts of $\Delta_{sky}\sim 0.3$.}
              \label{sky}
    \end{figure*}

\begin{figure*}
   \centering
   \includegraphics[width=9cm]{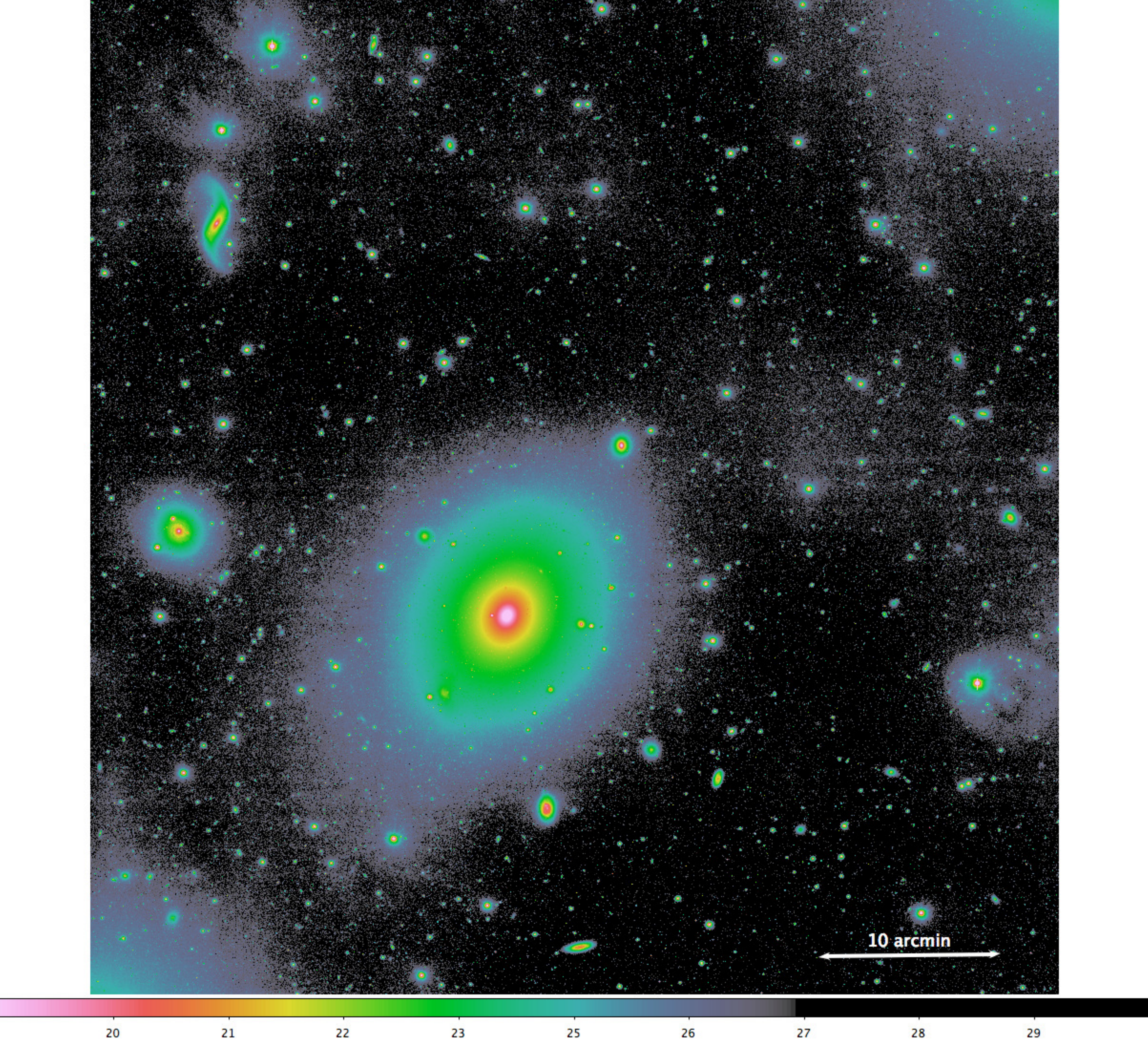}   
\includegraphics[width=8.6cm]{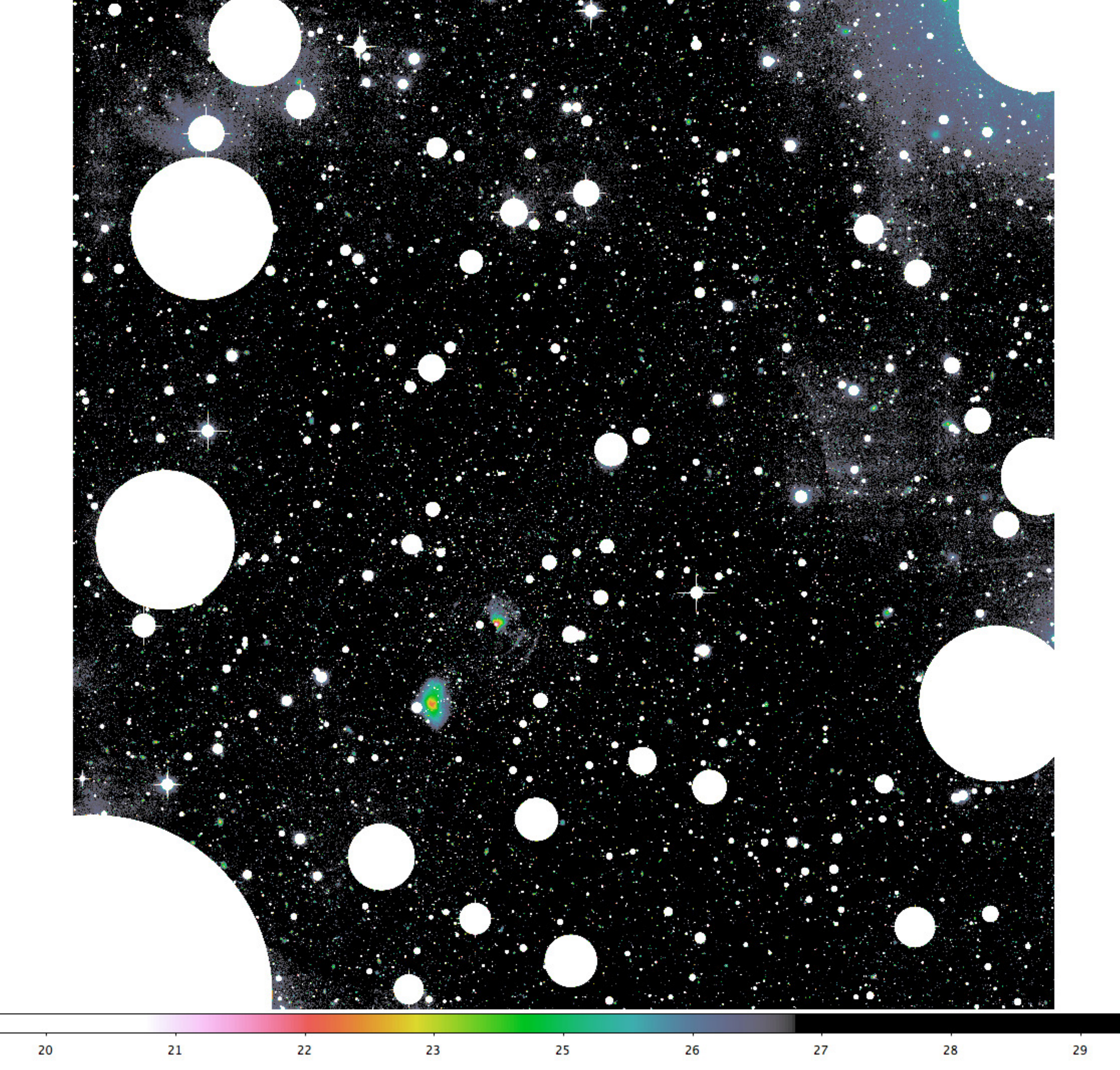}
   \caption{{\it Left}: False-color image of the VST pointing of NGC 4472, trimmed to the same size as that in Fig. \ref{out}. The magnitude scale adopted to produce the picture is shown at the bottom. The last clearly visible isophote is at $\mu_{g} \sim 27$ mag/arcsec$^{2}$. {\it Right}: Residual image obtained by subtracting from the left picture the galaxy model described in Sect. \ref{morph}. Masks adopted to exclude features from the fitting are shown as circles.}
              \label{lowSB}
    \end{figure*}

Moreover, we applied the methodology described by \citet{Pohlen06} to quantify the sky variations. As described in the following
subsection, we extracted from the sky-subtracted image of NGC 4472 the azimuthally averaged intensity profile out to the edges of the frame by fixing both the position angle and the ellipticity of the galaxy. From this profile (Fig. \ref{sky}) we estimated a residual background of $\sim 0.3\pm 0.09$ counts by extrapolating the outer trend. The uncertainty in the extrapolated value is lower than 0.1\% of the sky background, which means that it becomes relevant at a level of 29 mag/arcsec$^{2}$. This limit is not intrinsic to VST, but arises from the fact that NGC 4472 practically fills the field of view of the camera. In Fig. \ref{lowSB} we show a false-color image of the NGC 4472 field (left) together with its 2D residuals (right) obtained by subtracting the galaxy model described in Sect. \ref{morph}. The white circles mask the areas ignored in the isophotal fitting. The bluish foggy patch in the middle of the right side in both images is due to the malfunctioning of  CCD  82 of OmegaCam (a problem now solved by the replacement of the board).

In conclusion, we stress that the background-subtraction procedure for the OmegaCAM images is sometimes made quite difficult by the residual unevennesses left in the mosaic by the combination of the 32 independent CCDs. For this reason, we will evaluate the ON-OFF background-subtraction procedure
in another paper.

\subsection{Isophotal analysis}\label{iso}

The isophotal analysis of the VEGAS galaxies is performed on the final mosaic in each band with the {\small IRAF}\footnote{IRAF ({\it
Image Reduction and Analysis Facility}) is distributed by the National
    Optical Astronomy Observatories, which is operated by the
    Associated Universities for Research in Astronomy, Inc. under
    cooperative agreement with the National Science Foundation.} task {\small ELLIPSE}.
Briefly, {\small ELLIPSE} computes the intensity, $I(a,\theta)$, azimuthally
sampled along an elliptical path described by an initial
guess for the isophote center, $(X, Y)$, ellipticity, $\epsilon$, and semi-major
axis position angle, $\theta$, at different semi-major axis lengths, $a$. 
At a given $a_0$, $I(a_0, \theta)$ is expanded into a Fourier series as

\begin{equation}
I(a_{0}, \theta)=I_0+\sum _k (a_k \sin(k\theta)+b_k \cos (k\theta))
\end{equation}
according to Jedrzejewski (1987). The best-fit parameters are those minimizing the 
residuals between the actual and the model isophotes; $a_k$ and $b_k$
are the coefficients measuring the deviations from a pure ellipse, including the signature of boxiness and/or diskiness \citep{Bender89}.

\subsection{Light and color distribution}\label{light}

Together with the geometrical parameters, the task {\small ELLIPSE} provides the light distribution azimuthally averaged either over each isophote or within isophotal annuli of specified thickness. 

The error associated with the surface brightness measurements was computed with the formula 
\begin{equation}
\sigma_{\mu}=\sqrt{\left(\frac{2.5}{I \ln 10}\right)^2 \left( \sigma_I+\sigma_{sky}\right)^2+\sigma_{ZP}^2},
\end{equation}
where the flux $I$ and the errors $\sigma$ for the flux $I$, the sky, and the photometric ZP, and the resulting $\sigma_{\mu}$ are in counts. We assumed simple Poissonian behavior, therefore  $\sigma_I = \sqrt{I/n}$, where $n$ is the number of pixels producing the median value $I$. The errors on the background are those discussed in Sect. \ref{back}, while those on the ZP are listed
in Table \ref{pho}. 

The resulting light profiles are presented and discussed in the next section, and the tables with the corresponding data for each galaxy are published in the online version of this journal. Here we comment on the resolution of the innermost and the reliability of the outermost measurements. Although our data have a good overall resolution, as shown by the FWHM values of the PSF (see Tab.\ref{data}),   
we did not attempt any deconvolution to improve the resolution since our galaxies have previously been observed by {\it HST}. The direct comparison with {\it HST} profiles (\citealt{Kormendy09}; see next section) shows our profiles to be unaffected by seeing for $r > 2$ arcsec in the {\it g} band; this limit is also valid for the {\it i} band. When we present the light profiles below, we also show and quote the seeing-blurred innermost measurements, but they will not be used for fitting the data with empirical photometric laws.

The faintest end of the luminosity profiles has large errors. They do not reach the same threshold value in all cases because of the different nature of the background to be subtracted combined with the size of the object (the smaller the better).  

\section{Individual galaxies: results and comparisons}\label{results}

In this section we present and discuss the results for the objects of this study and compare them with the available literature. Tables with the photometric and geometric profiles are available online; for the sake of clarity, we repeat that these data are not corrected for interstellar extinction. The effective parameters and the total magnitudes are listed in Table \ref{Re}, while Table \ref{tab:devauc} provides the effective parameters of the $r§^{1/4}$ models that best fit our profiles outside of the seeing-convolved cores. 

The effects of the scattered light are illustrated for NGC 4472 in Appendix \ref{PSF1}. At the end of this section, we list the effects for the smaller companions.
\begin{table*}
\caption{Magnitudes and effective parameters.} \label{Re}
\begin{tabular}{lcccccccccccc}
\hline\hline
Name   &            Band  &     $a_{L}^{1/4}$  &        $m_{L}$  &      $\Delta m$ &    $m_{T}$ &       $a_{e}^{1/4}$ & $\mu_{e} $&$     \langle\epsilon\rangle$ &       $r_{e}^{1/4}$&$ (\mu_{e})_{V}$ &$      (r_{e}^{1/4})_{V} $\\
            &                &[arcsec $^{1/4}$] & [mag]& [mag]&[mag]&[arcsec $^{1/4}$] &[mag] & & [arcsec $^{1/4}$] & [mag] &[arcsec$ ^{1/4}$] \\
      (1)         &       (2)           &  (3)                  & (4) & (5) & (6) & (7) & (8) & (9) & (10) &  (11) & (12) \\
\hline \vspace{-7pt} \\
NGC 4472&       $g$ &   5.47    &       8.55    &       0.05\tablefootmark{a}         &       8.50    &                        3.49   &        22.59&  0.16    &       3.42    &       22.73 &         3.73\\
NGC 4472&       $g$&    6.56    &       8.48    &       0.11\tablefootmark{b}   &       8.37    &                  3.71   &       23.03&  0.19    &       3.61    &                   &    \\
NGC 4472&       $g$&    6.56    &       8.37    &       0.10\tablefootmark{c}   &       8.27    &                  3.86   &       23.31&  0.16    &       3.78  &              &       \\
NGC 4472&       $i$&    6.47    &       7.19    &       0.23\tablefootmark{d}   &       6.96    &                  3.99   &       22.27&  0.18    &       3.89  &                  &     \\
NGC 4472&       $i$&    6.56    &       7.10    &       0.10\tablefootmark{e}   &       7.00    &                   3.90  &       22.09&  0.16    &       3.82  &                  &   \\
\vspace{-7pt} \\
NGC 4434&       $g$&    2.89    &       12.52&          0.60    &       12.46   &        1.83     &       21.08&  0.05    &       1.82    &       20.08   &       1.83\\
NGC 4434&       $i$&    2.89    &       11.45  &         0.50   &       11.40   &        1.83     &       19.81&  0.05    &       1.82   &                 &             \\
\vspace{-7pt} \\
NGC 4464&       $g$&    2.76    &       13.04&          0.40    &       13.00    &       1.67    &       20.43&  0.27    &       1.61    &       19.92   &       1.66\\
NGC 4464&       $i$&    2.76    &       11.84&          0.20    &       11.82    &       1.67    &       19.05&  0.27    &       1.61  &                  &             \\
\vspace{-7pt} \\
NGC 4467&       $g$&    2.33    &       14.63&          0.40    &       14.59    &       1.57    &       21.31&  0.22    &       1.52    &       20.91   &       1.56\\
NGC 4467&       $i$&    2.33    &       13.50&          0.20    &       13.48    &       1.55    &       20.03&  0.23    &       1.50 &                   &             \\
\vspace{-7pt} \\
VCC 1199&       $g$&    2.02    &       15.92  &         0.10   &       15.94    &       1.25    &       20.81&  0.11    &       1.23    &       20.28   &       1.22\\
VCC 1199&       $i$&    2.02    &       14.71  &         0.10   &       14.70    &       1.25    &       19.57&  0.12    &       1.23 &                   &           \\      
\vspace{-7pt} \\
UGC 7636& $g$ &     3.41 & 14.22           & 0.07             &       14.15        &       2.62       &      24.91&      0.39 &     1.41 &                 &              \\
UGC 7636& $i$ &     3.41 & 13.38           & 0.53             &       12.85        &       3.13       &      21.26&      0.39 &     2.95 &                    &              \\ 
\hline
\hline
\end{tabular}
\tablefoot{ Column 3: Major axis of the faintest isophote for which SB is measured. Column 4: Magnitude within $a_{L}$, computed assuming a fixed mean ellipticity $\langle\epsilon\rangle$ (Col. 7). Column
5: Extrapolation of the growth curve to infinity. Column 6: Total magnitude $m_{T}=m_{L}+\Delta\ m$. This value is not corrected for interstellar extinction. According to \citet{Battaia12}, the correction would be $A_{g}=0.074$ and $A_{i}=0.044$ mag. Column 7: Major axis of the effective isophote of flattening $\langle\epsilon\rangle$ that
encircles half of the total light. Column 8: SB at the effective semi-major axis $a_{e}$. Column 9: Adopted mean ellipticity. Column 10: Mean effective radius $r_{e}=a_{e} \sqrt {1-\langle \epsilon\rangle}$. Columns11 and 12: Effective parameters for the V band \citep{Kormendy09}.}
\tablefoot{ \tablefoottext{a}{Excluding the ICL tail.} \tablefoottext{b}{Including the ICL tail.} \tablefoottext{c}{Including the ICL tail, but flattening the ellipticity profile from 1.75 arcmin on.}\tablefoottext{d}{With measured ellipticity.} \tablefoottext{d}{With ellipticity modified as for the {\it g} band (note c).}}
\end{table*}

\begin{table*}
\caption{Effective parameters of the $r§^{1/4}$ models that
best fit our light profiles outside the seeing-blurred cores.} \label{tab:devauc}
\centering
\begin{tabular}{lcccc}
\hline\hline
\scriptsize

 & {\it g} band & {\it g} band& {\it i} band & {\it i} band \\
 \hline
   Galaxy  &  $r_{e}^{1/4}$    &  $\mu_{\rm e}$  & $ r_{e}^{1/4} $ &  $\mu_{\rm e}$ \\
       &  [arcsec$^{1/4}$]   &   [mag/arcsec$^2$] &   [arcsec$^{1/4}$]   &  [mag/arcsec$^2$] \\
\hline
   NGC 4472& 3.51 & 22.52  & 3.51  & 21.25 \\
   NGC 4434&  1.85 & 21.13 &   1.75 & 19.64 \\
   NGC 4464& 1.59 & 20.16 &   1.50 & 18.48 \\
   NGC 4467&  1.33 & 20.18 &   1.30 & 18.86 \\
   VCC 1199&  1.00 & 18.73 &   0.97 & 17.11 \\
\hline
\hline
\end{tabular}
\end{table*}

\subsection{NGC 4472}

\begin{figure}[!h]
   \centering
   \includegraphics[width=8.2cm]{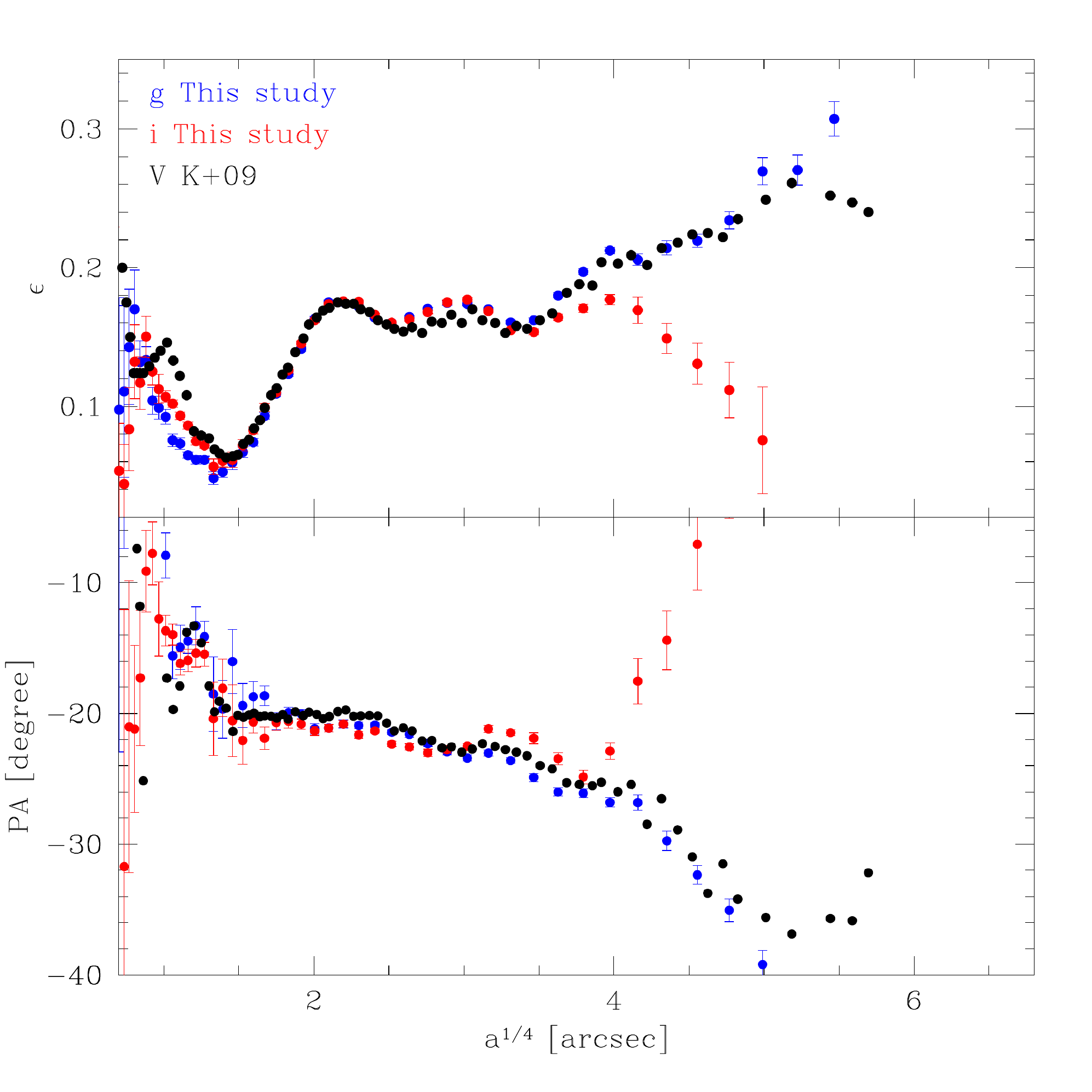}\\
 \includegraphics[width=8.2cm]{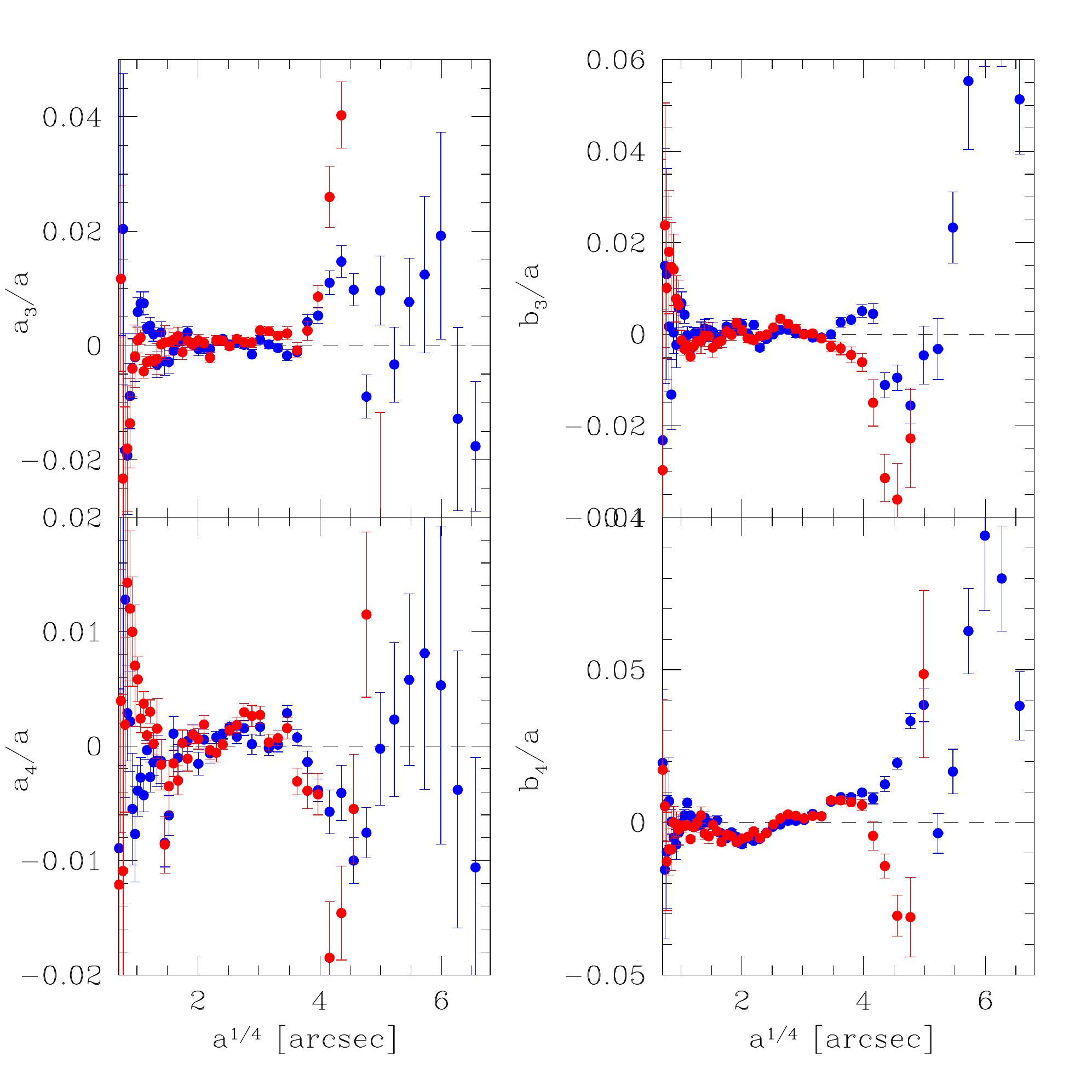}
    \caption{NGC 4472. Top: position angle (P.A.) and ellipticity
     ($\epsilon$) profiles in the {\it g} and {\it i} bands compared with those in K+09. 
     Bottom: isophotal shape parameters in the {\it g} and {\it i}
     bands.}
              \label{ell_4472}
    \end{figure}

Figure \ref{ell_4472} shows the results of the isophotal analysis performed by {\small ELLIPSE}. Some comments are in order.

The profiles in the two bands are substantially similar out to $a \sim 15'$ or $(a/a_{e})_{g}\sim 4.83$, where $a_{e}$ is the effective semi-major axis. The rapid change in the inner region is due to the well-known peculiarity of the nucleus of NGC 4472. \citet{Ferrarese06} in fact
detected a ``boomerang-shaped'' dust lane crossing the central regions
of the galaxy. Beyond $a^{1/4} \simeq 4$ arcsec both the ellipticity and the position angle profiles diverge in the two bands: the {\it g} isophotes flatten outward, while in the {\it i} band they have a rounded shape. The deviations are far larger than the formal errors provided by {\small ELLIPSE}. Nonetheless, we doubt that this behavior is spurious; it may be due to the excessively large size of the supergiant elliptical that almost fills the OMEGACam FOV. A comparison with \citet{Kormendy09} suggests that the diverging {\it g} -band flattening profile might not be real. We return to this point below.

The shape parameters in both the {\it g} and {\it i} bands (Fig. \ref{ell_4472}) show a
moderate boxiness of the isophotes, which confirms the presence of dust in
the central regions of the galaxy. Since the dust optical depth decreases toward longer
wavelengths, the {\it i} -band profiles are less affected by dust.

\begin{figure}[!h]
   \centering
\hspace{-0.6cm}   \includegraphics[width=9.5cm]{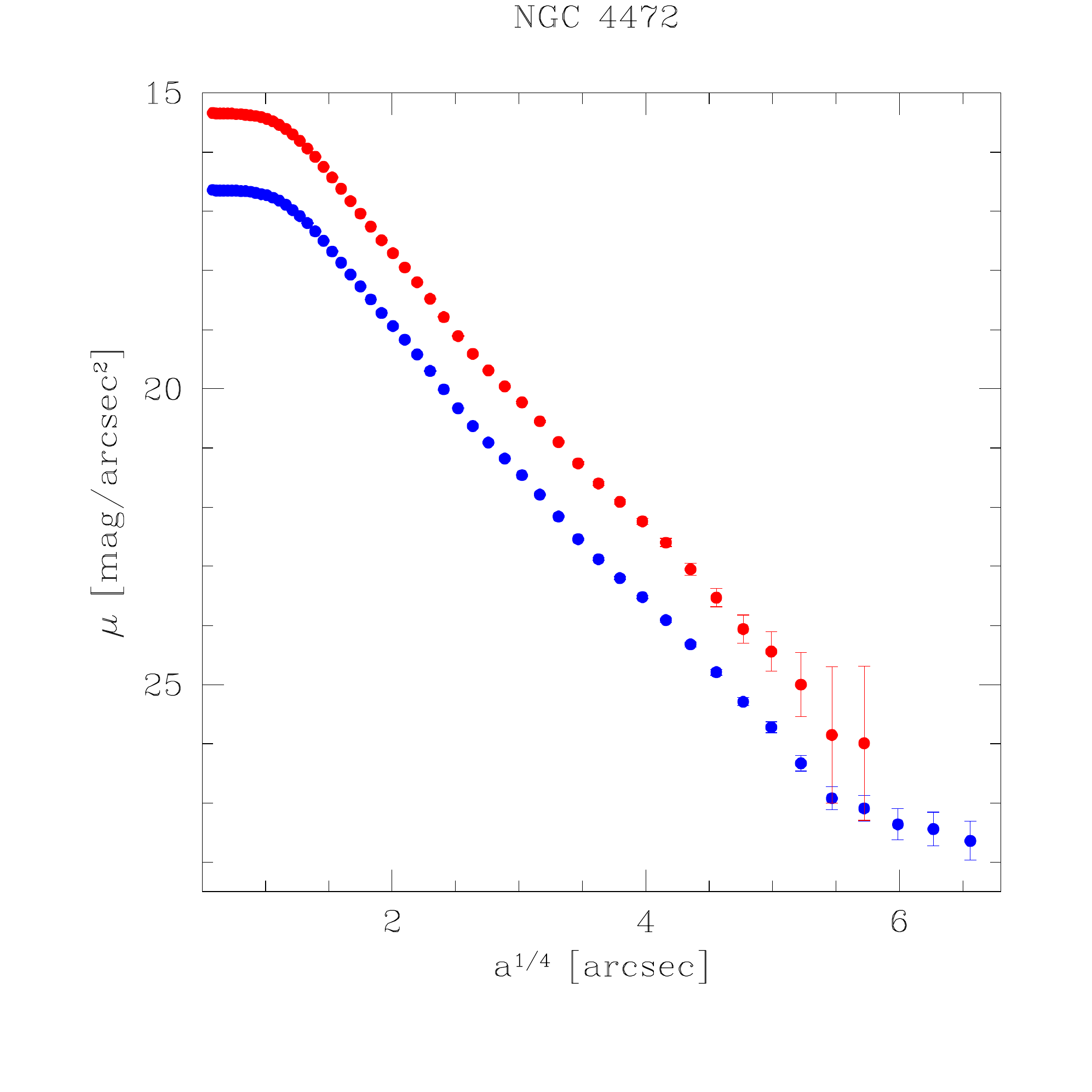}\\
 \includegraphics[width=9cm]{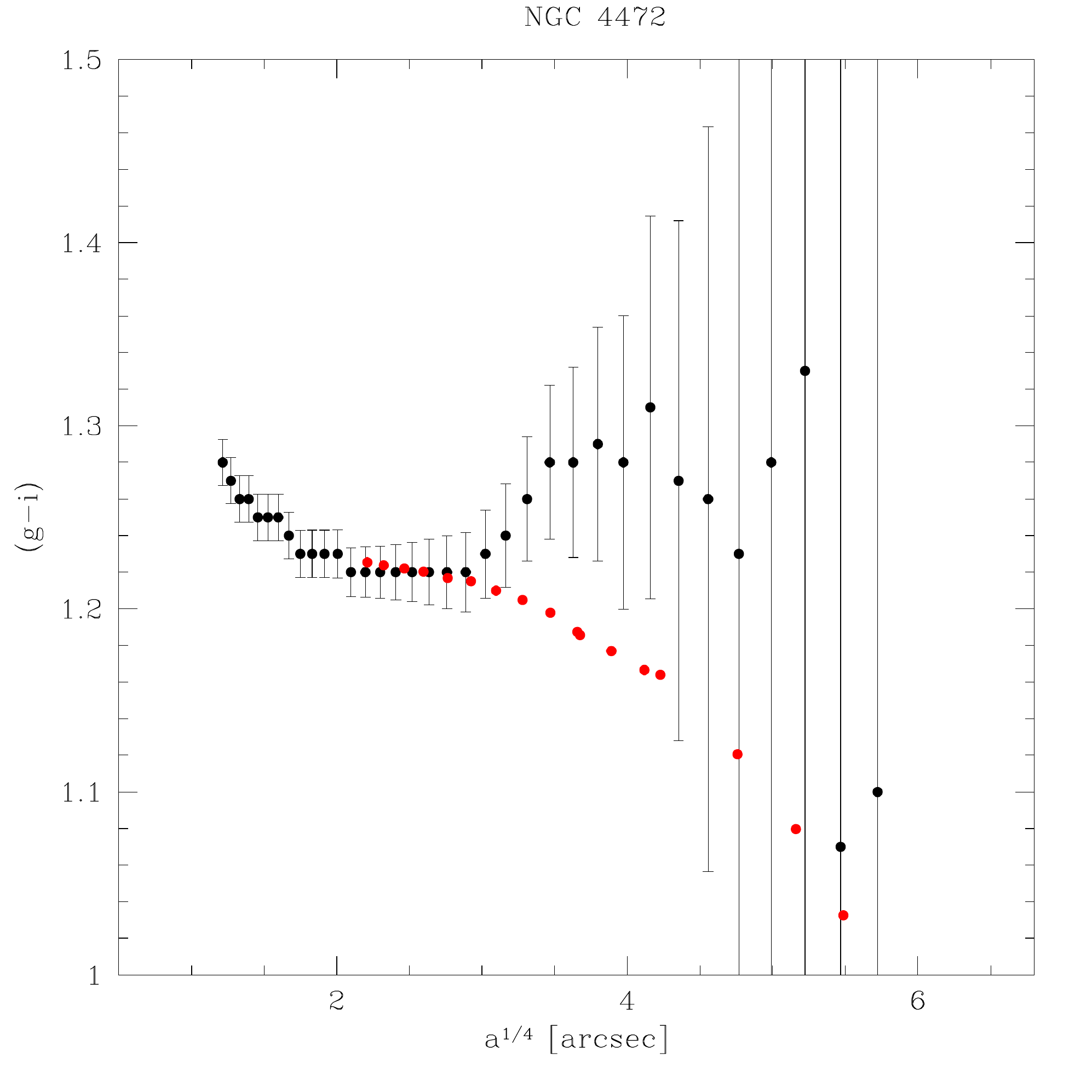}\\
   \caption{NGC 4472. Top: azimuthally averaged light profiles in the {\it g} (blue) and {\it i} (red) bands. 
   Bottom: {\it (g-i)} color profile in the region ($a > 2$ arcsec) unaffected by differential seeing. The red dots trace the {\it (B-V)} profile published by \citet{Mihos13} shifted by +0.25 mag, measured in the regions of high S/N for both datasets. The comparison supports the blueward gradient that we found for $a/a_{e} > 4$, although with very large errors.}
              \label{prof}
    \end{figure}

The azimuthally averaged light profiles in the {\it g} and {\it i} bands are shown in Fig. \ref{prof} as a function of the isophote semi-major axis $a$.
The average surface brightness extends out to $a \simeq 30.6$ arcmin from the galaxy 
center for the {\it g} band, with the largest formal errors of about 0.3 mag, while in the {\it i} band we reach 
$\sim 17.6$ arcmin with errors four times larger. 

The effect produced by the extended PSF onto the image of NGC 4472, and therefore onto its azimuthally averaged light profile, was estimated by the methods outlined in Appendix \ref{PSF1}. The result is that no significant contribution is present in the light distribution out to the faintest measured point. This conclusion is particularly important because it verifies that the observed bending in the light profile occurs at $\mu_{g} \sim 27$ mag/arcsec$^2$. This cannot be due to scattered light.

The surface brightness profiles in both the {\it g} and {\it i} band are fairly linear in $r^{1/4}$ units except at the center. When forcing a de Vaucouleurs (1948) law \citep{deV48} over the range $1''$ to $625''$, the best-fit parameters are $r_{e}=152''\pm 6''$ and $\mu_{e}=(22.52\pm0.05)$
mag/arcsec$^2$ in {\it g} band and $r_{e}=152''\pm7''$ and $\mu_{e}=(21.25\pm0.05)$
mag/arcsec$^2$ in {\it i} band (see also Table \ref{tab:devauc}). Interestingly enough, the effective radii are exactly the same, with the same error in both bands. The color at $r_{e}$ is $(g-i) = 1.27 \pm 0.07$.

The $r^{1/4}$ fit highlights a neat change in the slope of the {\it g} -band light profile at $a_{e}^{1/4} \simeq\ 5.5$, where $\mu_{g} \sim 27$ mag arcsec$^{-2}$. Is this bending, just outlined by the less extended {\it i}-profile and by the B-band major axis profile of \citet{Caon94}, a signature of intracluster light (ICL)? To determine whether it might be an artifact of the turn-up of the flattening of the outer isophotes (see Sect. \ref{iso}), we simulated an $r^{1/4}$ galaxy using the {\it g} -band interpolation parameters of NGC 4472 for two isophotal geometries: a fixed ellipticity $\epsilon = 0.25,$ which in the second case increases linearly from $a_{e}=750 ''$ and mimicks the {\it g} -band ellipticity profile of Fig. \ref{ell_4472}. The outer light profile of the second case remains brighter where the ellipticity increases, but the effect is quantitatively negligible compared to what we observe. Moreover, we note that as suggested by \citet{Gonzalez05}, the presence of an outer and more elliptical component with a significant gradient in the P.A. is most likely due to a population of some ICL.

There is another possibility of how a spurious change of slope in the SB profile might be produced: an incorrect setting of a background level. However, this is not the case here because a too faint value for the background would produce a smooth change in the slope instead of a sharp break. Finally, we note that the level at which the break occurs is compatible with the typical SB values at which \citet{Zibetti05}
have observed changes of slope induced by the ICL in a series of stacked 
galaxy clusters. 

\begin{figure}
   \centering
   \includegraphics[width=9cm]{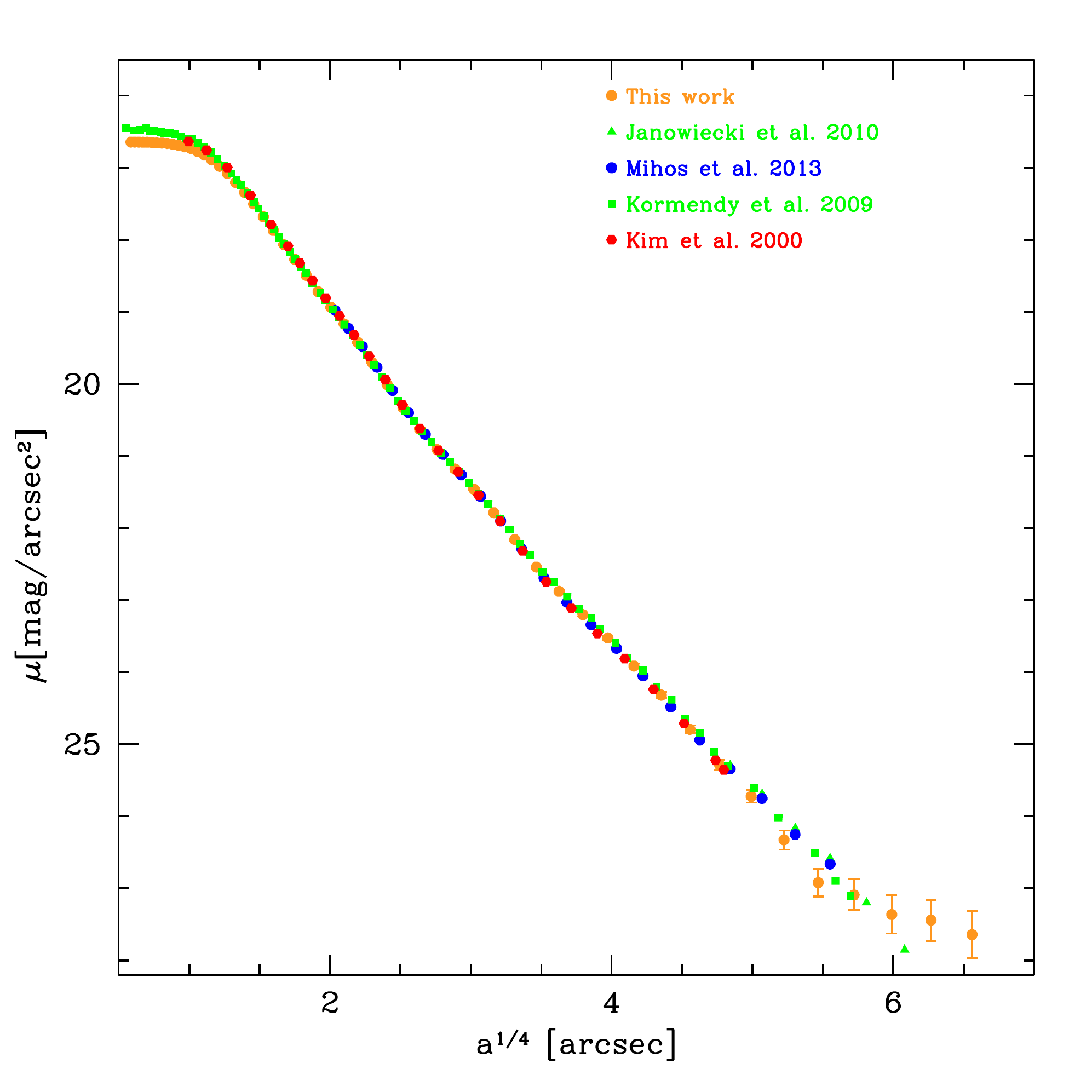}
   \includegraphics[width=9cm]{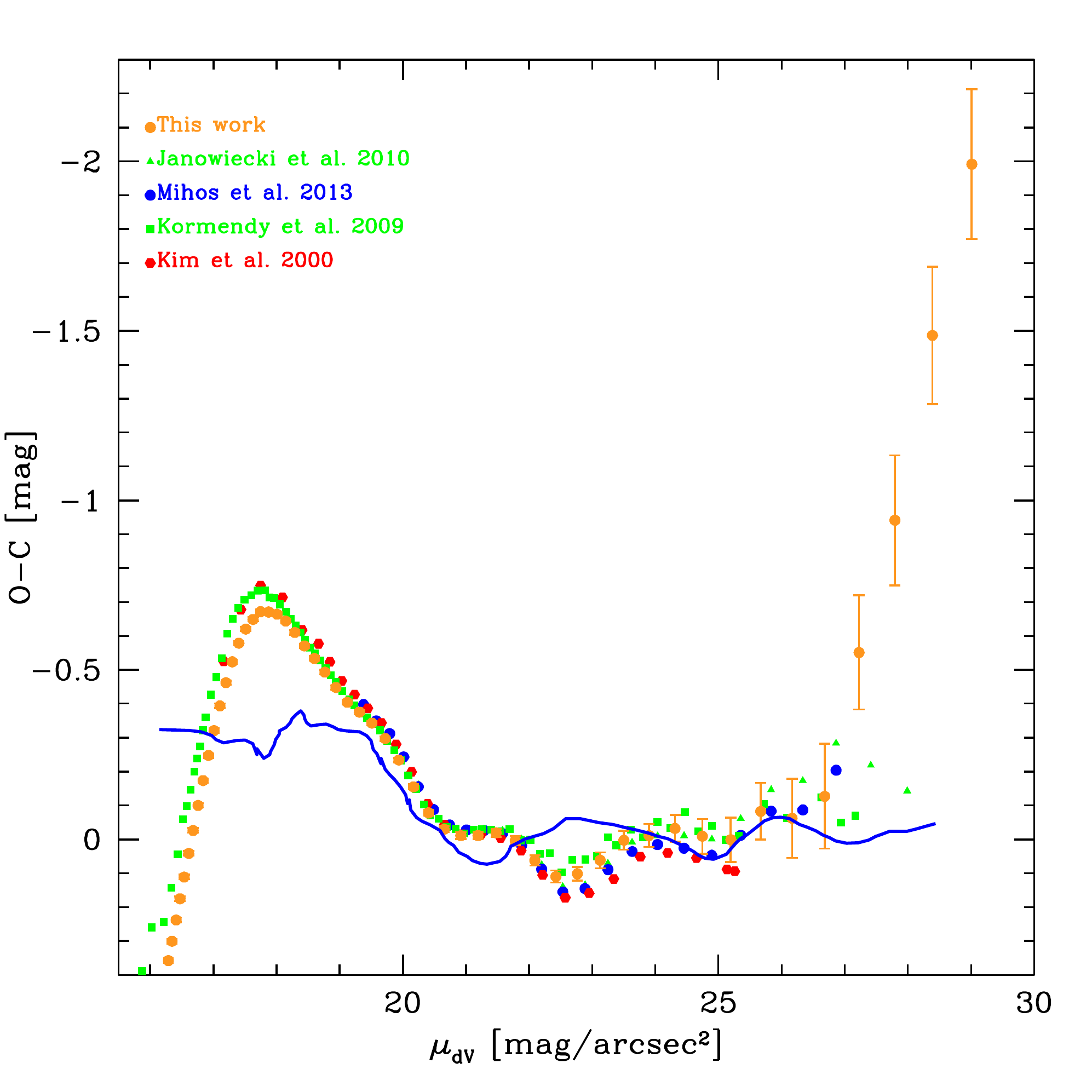}
   \caption{NGC 4472.  {\it Top panel:}  Azimuthally averaged g-band profile from VEGAS compared to
     literature. The color code of the symbols mimicks the corresponding
     photometric bands. Arbitrary shifts have been used to match with the VEGAS profile. In particular, the B-band profile by\citet{Mihos13} has been
     shifted by -0.35 mag, the V band from \citet{Kormendy09} and
     \citet{Janowiecki10} by +0.35 mag, and the profile by \citet{Kim00} by
     +0.92 mag. {\it Bottom panel:}  (O-C) residuals of mean profiles from
     a best-fitting $r^{1/4}$ model used only to remove the main
     gradient and facilitate comparison. The blue solid
     line plots the (O-C) residuals for the east-west photometric cross-section of
     the standard elliptical galaxy NGC3379 from \citet{deV79}. There
     are clear similarities: the bright core and a
     wavy trend overimposed on the smooth $r^{1/4}$ trend. }
   \label{conf_prof4472}
\end{figure}

Our azimuthally averaged {\it g} -band profile is compared with results from the available literature in Fig. \ref{conf_prof4472}. The offsets providing the best match to our photometry are -0.35 mag for the $B$-band profile of \citet{Mihos13}, +0.35 for the $V$
photometry of \citet{Kormendy09} and \citet{Janowiecki10}, and +0.92 for the $R$-like band of \citet{Kim00}.
\citet{Caon94} have not been considered here because these authors provided main axes and no azimuthal profiles. In spite of the different color bands, the agreement among the various profiles is good from outside the seeing-blurred core to $\mu_{g} \sim\ 27$ mag $arcsec^{2}$. \citet{Janowiecki10}, whose data extend far enough out, did not confirm the ICL tail exhibited by our profile.

There is instead a problem in the zero points of the various photometric analyses of NGC 4472. In particular, by adding the offsets to the B band \citep{Mihos13} and the V band \citep{Kormendy09}, we obtain a $\langle$(B-V)$\rangle=0.70$, which is largely inconsistent with the known average color of NGC 4472 (e.g.,  $\langle$(B-V)$\rangle=0.96$ from RC3 \citep{deV91}). Comparison with the stellar population synthesis models by \citet{BC03} with standard assumptions\footnote{We adopted a star formation history with an exponentially decreasing rate, as is typically used for ETGs in the local Universe, with a Salpeter IMF in a metallicity range between $Z_{\odot}$ and  $2.5 Z_{\odot}$} provide $\langle (B-g)_{BC}\rangle=0.49$ and $\langle (g-V)_{BC}\rangle=0.48$, which turn into zero-point residuals of $\Delta\langle (B-g)_{BC}\rangle=0.14$ and $\Delta\langle (g-V)_{BC}\rangle=-0.13,$ which might be the zero-point shifts in both \citet{Mihos13} and \citet{Kormendy09}. The very small error estimated for our photometry by the comparison with 2MASS (see Appendix \ref{calib}) is confirmed by the comparison of our photometry of NGC 4472 with that of Ferrarese et al. (private communication) which in the range from 18 to 26 mag/arcsec$^{2}$ provides an average value of $\Delta \mu_{g} = 0.002\pm 0.016$.

A clearer way to compare these different data is to plot their residuals with respect to $r^{1/4}$ fits all with the same slope (Fig. \ref{conf_prof4472}). The agreement is spectacular: the scatter is better than the formal error computed for our photometry for all $\mu_{g}$ brighter than $\sim\ 27$ mag arcsec$^{2}$. Thereafter, the scatter increases significantly with no apparent dependence on the color band. In the same figure we have plotted as a solid line the residuals for the East-West photometric cross-section of the standard elliptical galaxy NGC3379 from \citet{deV79}, scaled in such a way that the effective surface brightness of the two galaxies coincides. We note in NGC 4472 the same inner core as was discovered in NGC 3379 by \citet{deV79} and the occurrence of a wavy pattern of the residuals of similar amplitude, which calls for an explanation. A recent study of the M96
galaxy group \citep{Watkins14} has revealed faint shells around NGC 3379 and a dusty disk in the
inner regions. The observed trend in the observed minus calculated (O-C) residuals seems to be
typical for galaxies with such substructures. We intend to verify with VEGAS whether this behavior is a common feature for ETGs.

The bottom  panel of Fig. \ref{prof} plots the mean {\it (g-i)} color profile for NGC 4472,
obtained from the two azimuthally averaged luminosity profiles above. 
Data points affected by differential seeing ($a<2$ arcsec) were removed.
On average, the center of the
galaxy has a redder color, with a maximum value up to {\it (g-i)} $\sim (1.3 \pm
0.18)$ mag. Our {\it (g-i)} color profile is fully consistent with that published by \citet{Chen10}, which only  extends up to $a^{1/4}\sim 2.8$ or $a/a_{e} \sim 1$, however.
The color stays bluer in the range $5'' \leq\ a \leq\ 150''$ ($1.5 \leq\ a^{1/4} \leq\ 3.5$), then it turns redder again, and the gradient is almost flat, although the errors here are too large to robustly assess whether there are color gradients outside this radial range. However, a comparison with the {\it (B-V)} color profile published by \citet{Mihos13} (red dots in Fig. \ref{prof}, plotted with a shift $\Delta\ (B-V) = 0.25$, measured in the regions of high S/N for both datasets) seems to confirm the steep blueward gradient in the galaxy outskirts, from approximately $a\sim 10'$ or $a/a_{e} > 4$.

\subsubsection{Total magnitudes}\label{mag}
Total magnitudes require a careful examination of the trends of the light profiles as well as a critical analysis of the geometry of the isophotes.  Direct integration over all pixels encircled by a given outermost isophote is out of consideration because
it is difficult to interpolate the light profile around contaminating sources (satellite galaxies, GCs, background galaxies, foreground start, etc.). The procedure we adopted consists of summing the areas encircled between successive isophotes multiplied by an average flux value. These growth curves, built using the azimuthally averaged light profiles and the flattening profiles under the assumption of elliptical isophotes, are then plotted against the reciprocal of the outer semi-major axis $1/a$ of the various elliptical annuli to estimate the extrapolation to $1/a \rightarrow 0$. There is no need to correct for resolution since the convolution with the PSF preserves the energy. In contrast, much care must be placed 1) in judging the meaning of the ellipticity measurements at faint levels because they may significantly affect the result, and 2) in the method of extrapolating a signal there where the trend of the light profile is totally unknown. Errors in the total magnitude reflect onto the estimates of the effective radius, which is thus a rather poorly defined parameter. It can be shown that for an $r^{1/4}$ galaxy, an error $\Delta\ m$ in the extrapolation turns into a relative error $\Delta\ r_{e}/r_{e} =1.84 \Delta\ m$.

The case of NGC 4472 is particularly complex for two reasons: 1) it shows a stretched tail in the outermost {\it g} -band profile, which is interpreted as intracluster light that may be cut off in computing the total luminosity of the galaxy, and 2) the trend of the flattening with radius for $a > 150$ arcsec, which is just opposite in the two bands (see Fig. \ref{ell_4472}) and poses the question of whether this is real or if the truth is in between these two curves. The difference is non-negligible. Table \ref{Re} reports the total magnitude in the {\it g} and {\it i} bands, computed using the nominal ellipticity curves shown in Fig. \ref{ell_4472}. The integration is performed out to the last observed point at $a_{L}$. The extrapolation term $\Delta\ m$ was estimated assuming an $r^{1/4}$ extension mimicking the behavior of the main body of the galaxy, that is, cutting out the ICL tail. The exercise was repeated including ICL, but in this case, the extrapolation is large and indeed uncertain. It is very difficult to set a reliable figure for the error on $m_T$. The overall uncertainty in the light profile combines with those on the isophotal shape and on the extrapolation to give an uncertainty of at least 0.1 mag. In any case, it seems that ICL contributes some 15\% of the total {\it g} -band light of NGC 4472.

The effective semi-major axes were derived by the growth curves at 50\% of the total luminosity given by $m_T$, while the corresponding surface brightness was interpolated at $a_e$ in the light profiles. 

\subsubsection{Substructures of NGC 4472}\label{morph}

\begin{figure*}
   \centering
   \includegraphics[width=15cm]{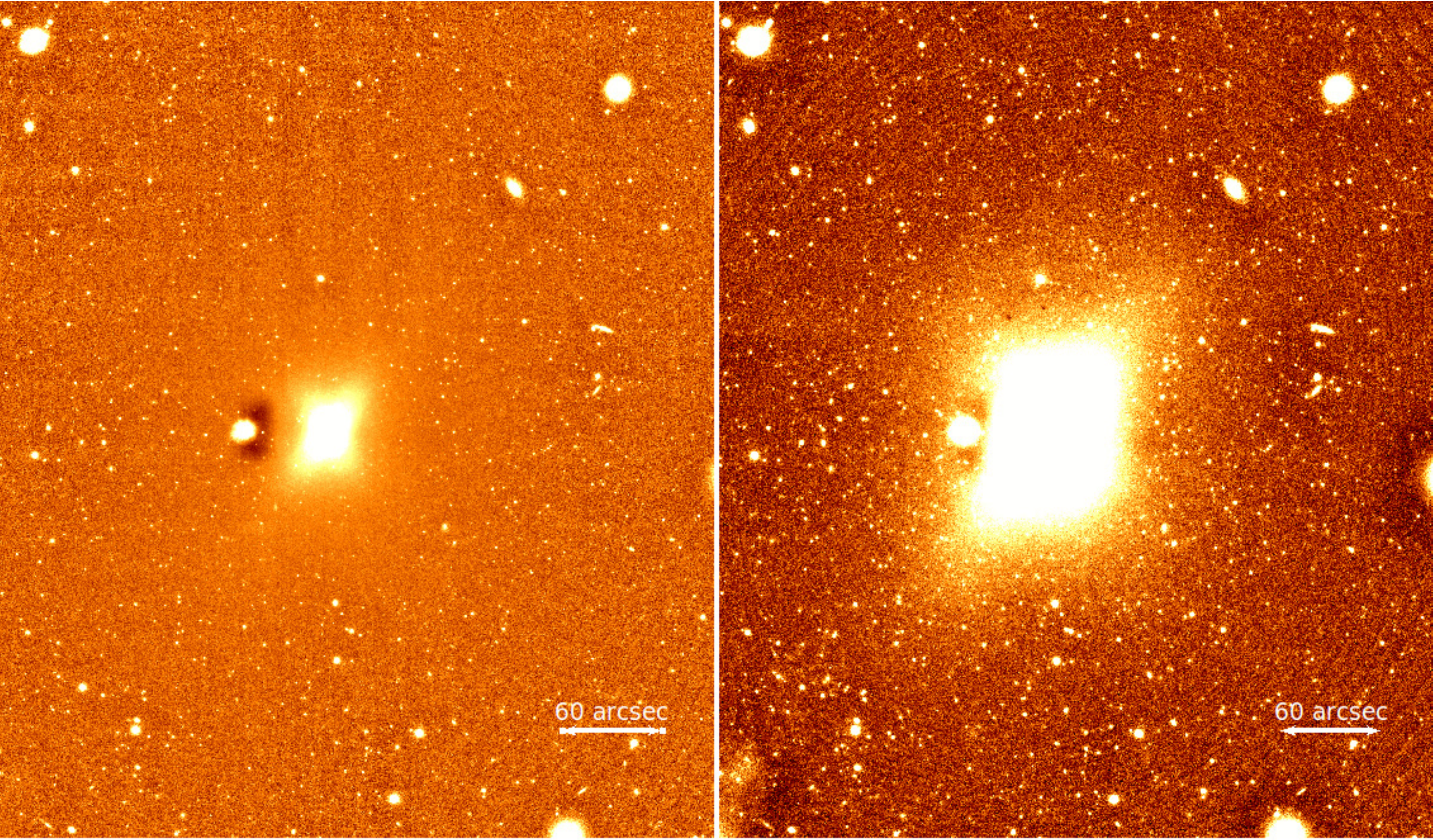}
   \caption{NGC 4472. Unsharp masked image, extracted from the whole VST mosaic ($500 \times\ 600$ arcsec) in the {\it i} (left
     panel) and {\it g} band (right panel). Lighter colors correspond
     to brighter features.}
              \label{unsharp}
    \end{figure*}

To examine the inner structure of NGC 4472 and detect the high-frequency structures, we first smoothed the images in the two bands with the IRAF task
{\small FMEDIAN}, which takes a median in a 2D window of $150\times 150$ pixels in {\it i} band
and of $300\times 300$ pixels in {\it g} band. These sizes were chosen by trial and error to best emphasize the inner structure of the galaxy.
Each image was then divided by its smoothed version to remove the low-frequency components.
The final unsharp masked images are shown in Fig. \ref{unsharp}. They both show an X-shaped pattern in the inner regions that
most likely is the signature of boxy isophotes, as pointed out in
Sect. \ref{iso}. Boxy isophotes are indicative of an interaction or
a mass transfer from a passive satellite (\citealt{Binney85, Whitmore88}) and of the presence of dust.

\begin{figure*}
   \centering
  \includegraphics[width=17cm]{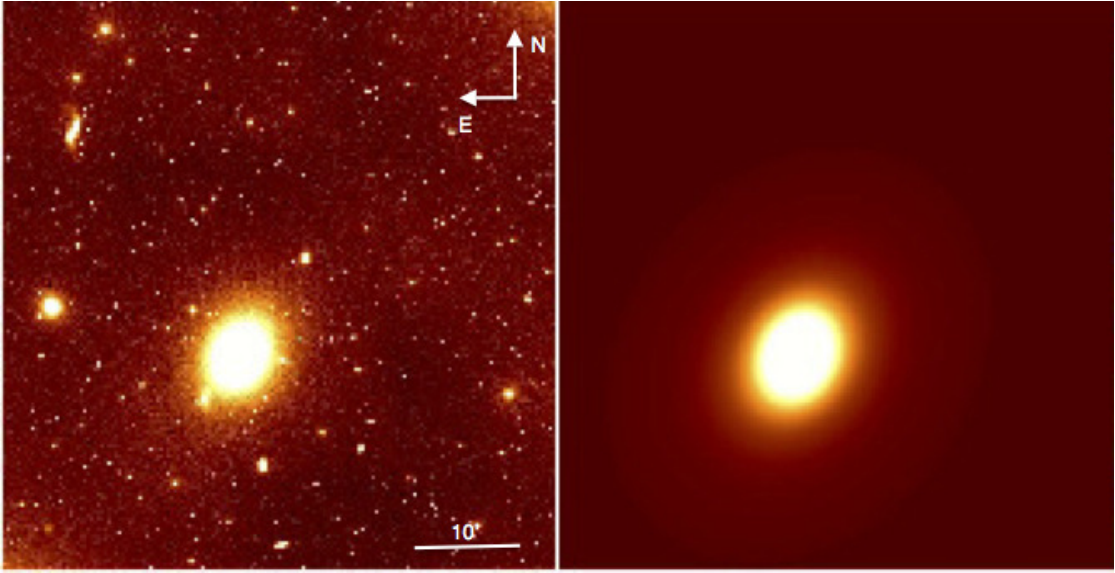}
  \caption{NGC 4472. Left panel: A region of $55 \times\ 56$ arcmin of the VST {\it g} band mosaic. 
      Right panel: 2D model (see text).}
              \label{mod}
    \end{figure*}

\begin{figure*}
   \centering
  \includegraphics[width=13cm]{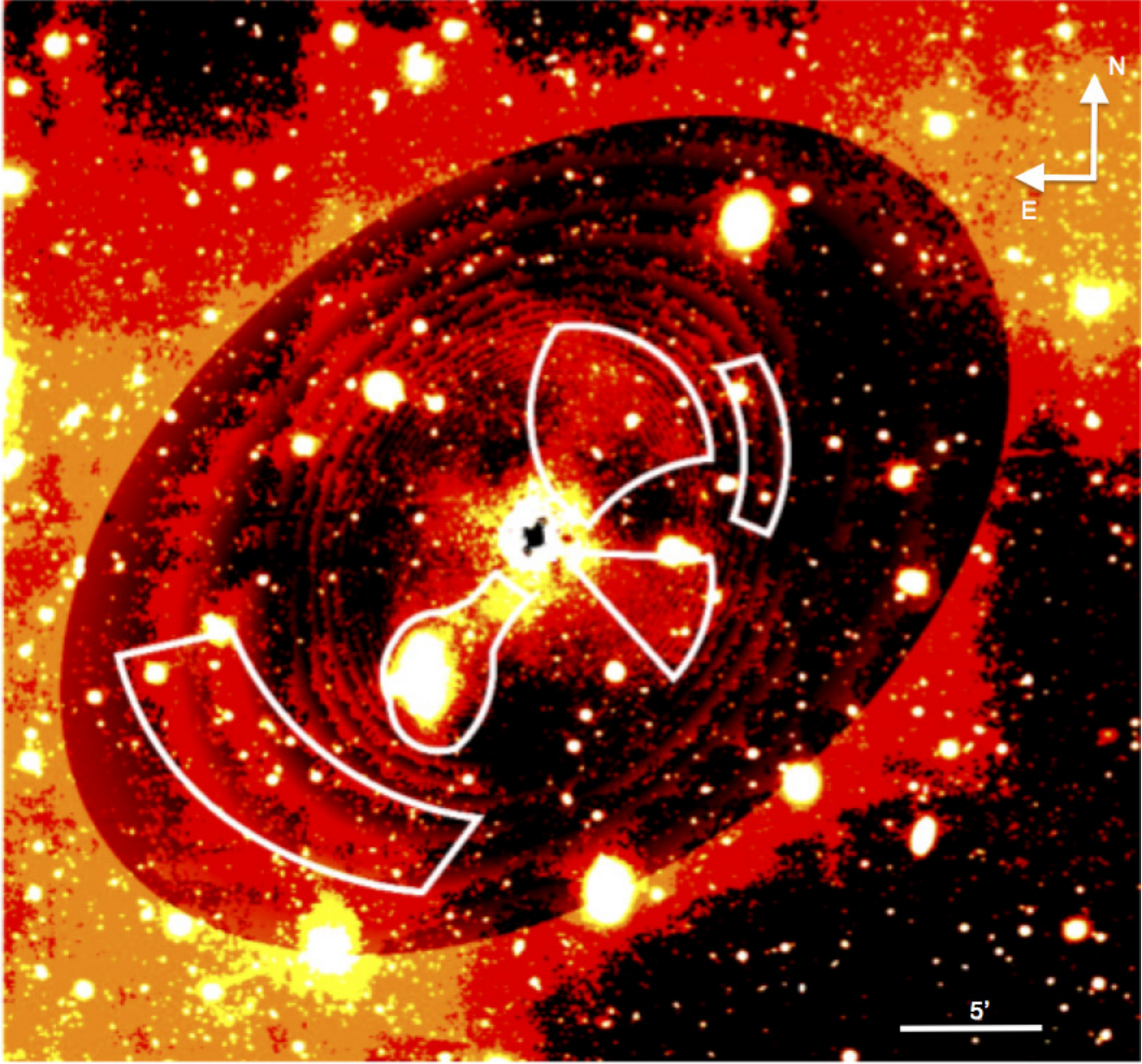}
\caption{NGC 4472. Zoom ($37 \times\ 35$ arcmin) of the median-smoothed residual image. The outermost elliptical contour marks the region where the b-model subtraction was not applied.
The superimposed white contours are 1) the tail
    connecting UGC 7636 to the giant ETG, and 2) shells and fan of
    material identified by \citet{Janowiecki10} and \citet{Battaia12} that is visible in our
    residual image. The wave-like residuals, concentric with the NGC 4472 nucleus, are spuriously introduced by the numerical procedure.}
              \label{shell}
    \end{figure*}

To highlight possible larger substructures, we produced a 2D model of NGC 4472 that best fit the azimuthally averaged isophotes with the {\small IRAF} task {\small BMODEL}. Only the {\it g} -band image was considered here because of its higher S/N ratio.
The image and its model are shown in Fig. \ref{mod}, while Fig. \ref{shell} shows the difference between them. 
This residual map shows a clear asymmetry in the nuclear region and some diffuse features, such as a tail
associated with the dwarf irregular galaxy UGC 7636 interacting with NGC 4472 and concentric shells
and fans of material  (white contours) that were also identified photometrically by
\citet{Janowiecki10} and \citet{Battaia12} and by \citet{Dabrusco15} using globular clusters. The outer boundaries of these shells and substructures mimic the pattern of the minima in the O-C residuals of the azimuthal light profile with respect to a smooth $r^{1/4}$ interpolation, shown in the bottom panel of Fig. \ref{conf_prof4472}.

\begin{figure*}
   \centering
   \includegraphics[width=13cm]{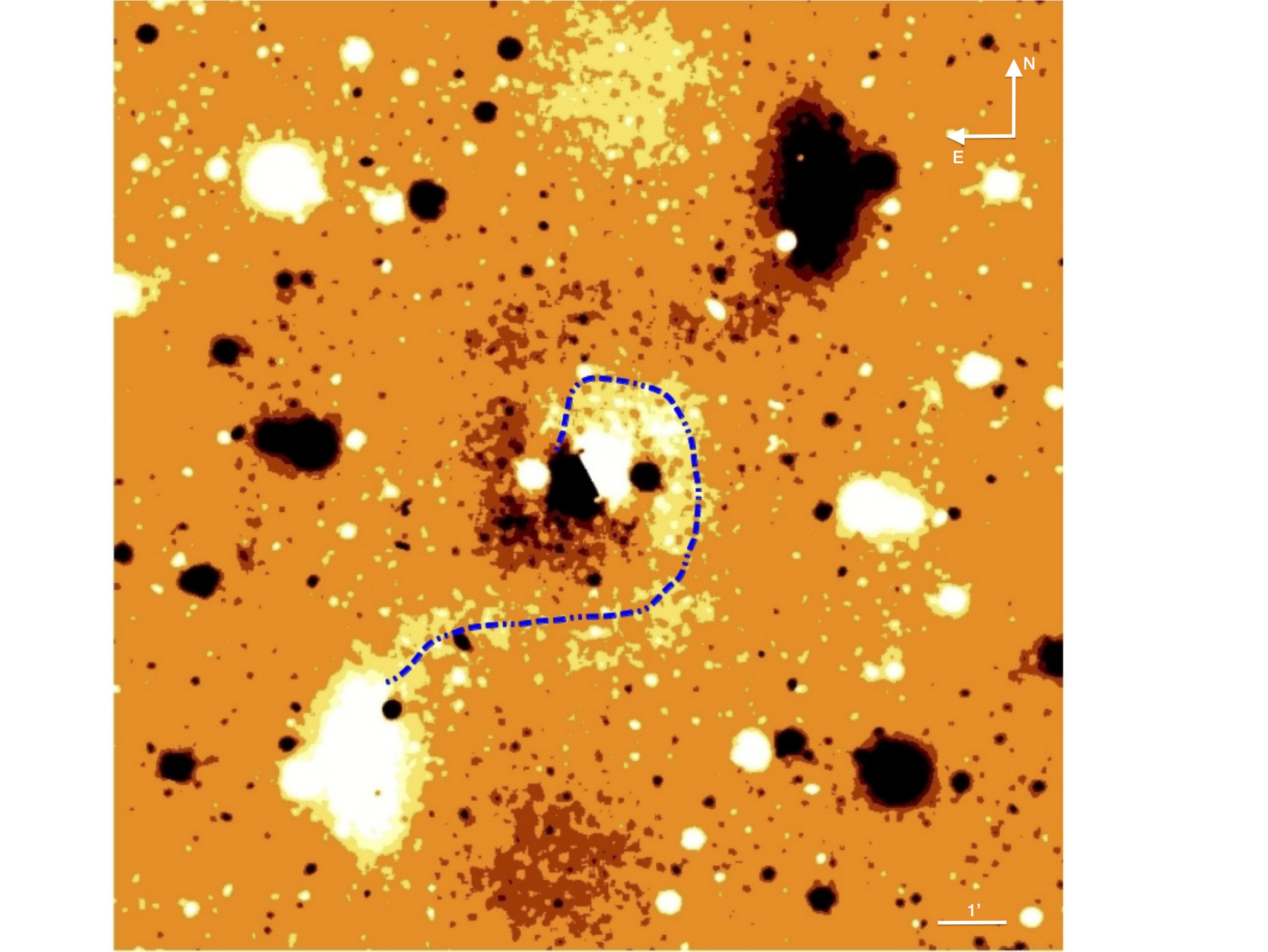}
   \caption{Long tail connecting UGC 7636 with the core of NGC 4472, from a double-folding subtraction. The image size is $15 \times\ 15$ arcmin.}
              \label{tail}
    \end{figure*}

The 2D modeling above assumed the isophotes to be homocentric and elliptical. To relax these requirements and search for asymmetric features, we rotated the original {\it g} -band image around the galaxy center by $180^{\circ}$ and then subtracted the image itself. The result is shown in Fig. \ref{tail}. In this way, we discovered the possible presence of a long tail connecting UGC 7636 to NGC 4472, twisted around the nucleus. The brightest part of this tail associated with UGC 7636 is also visible in the residual map of Fig. \ref{shell}. The tail is not shown in the {\small BMODEL} subtraction residual image, but this method is probably less sensitive to local very low surface brightness features.

\subsection{NGC 4434}
NGC 4434 (also known as VCC 1025) is an E0 galaxy where F+06 highlighted the large
nucleus, derived from a ``break'' in the surface brightness profiles around $1''$.

\begin{figure}
   \centering
   \includegraphics[width=8cm]{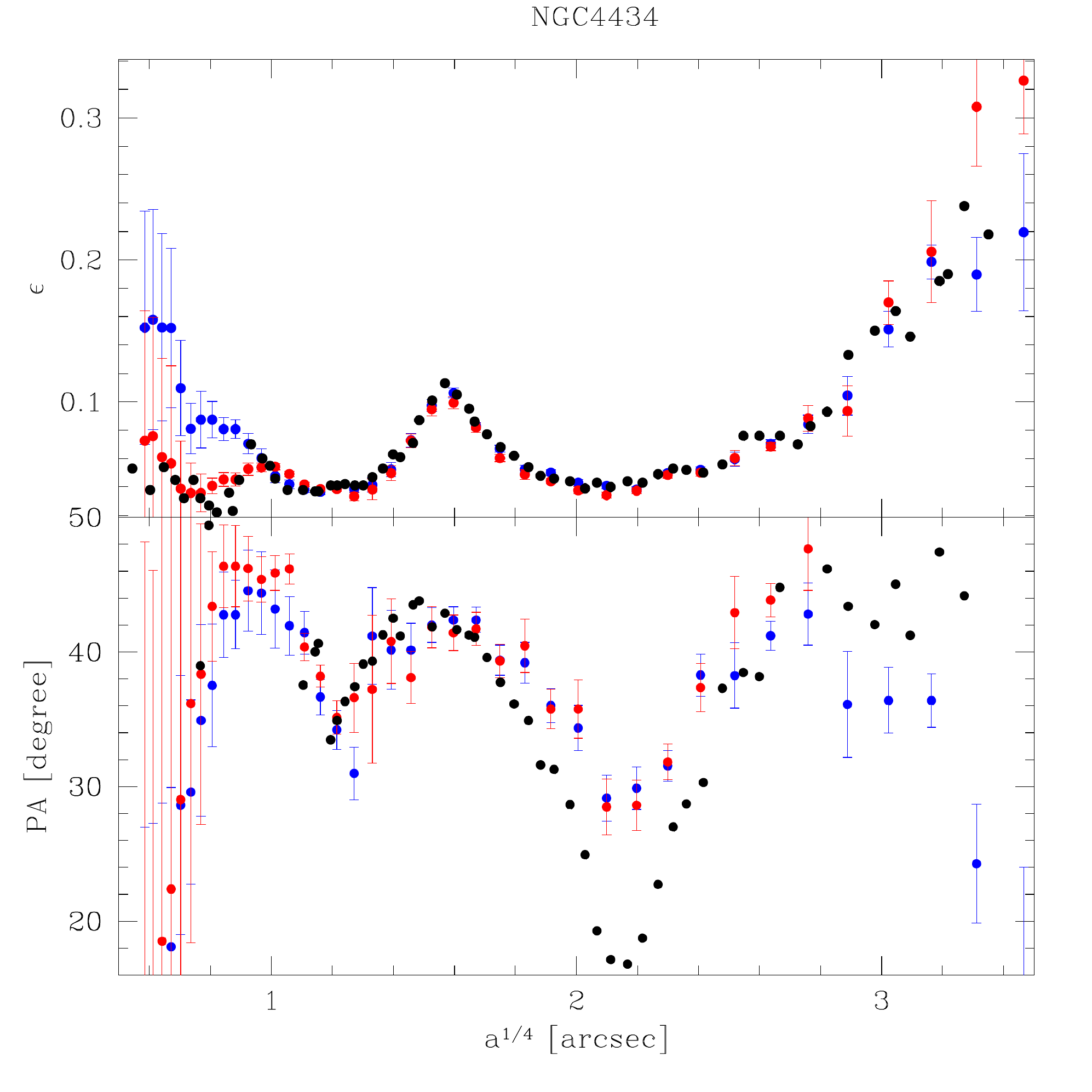}
   \includegraphics[width=8cm]{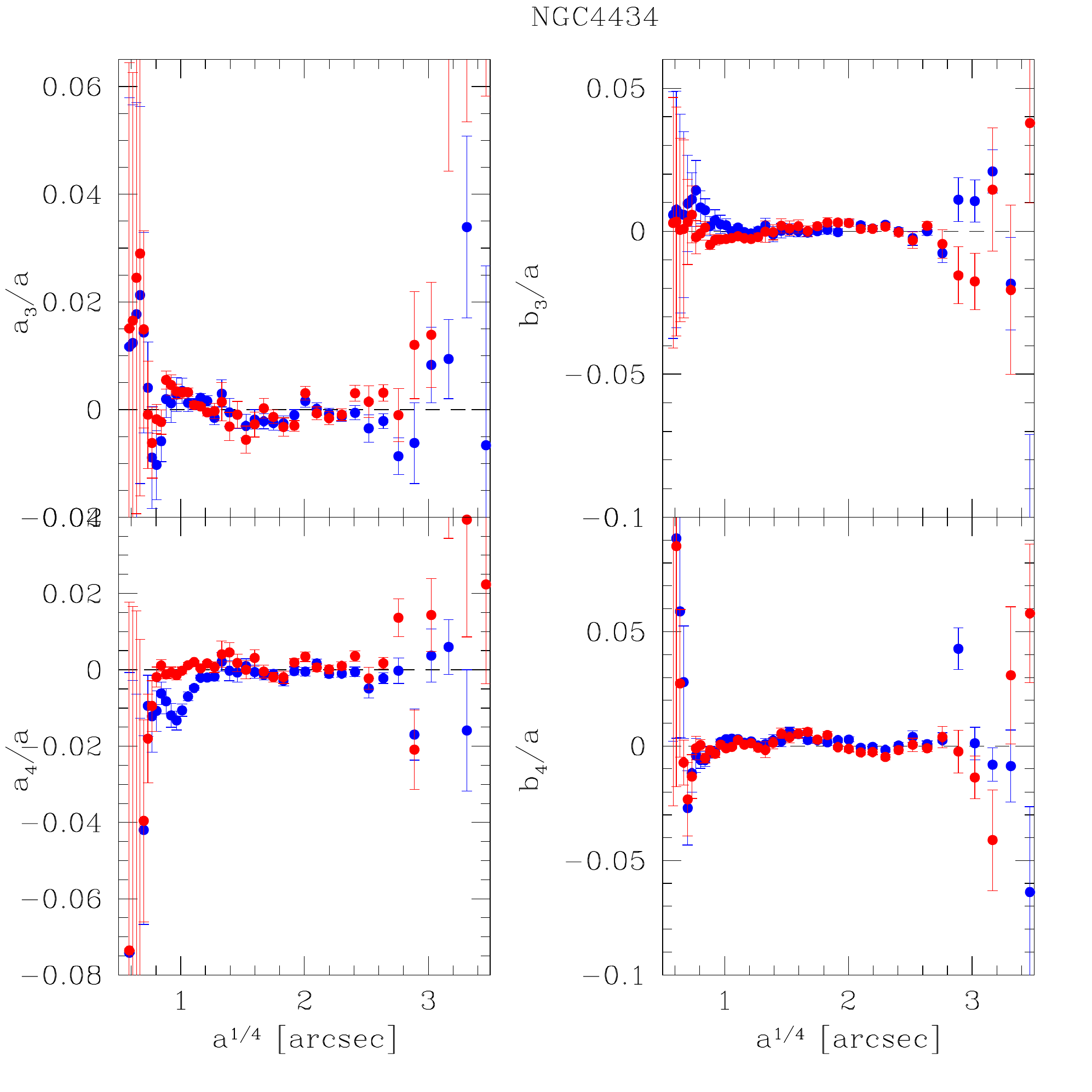}
   \caption{Same as Fig. \ref{ell_4472} for NGC 4434.}
              \label{ell4434}
\end{figure}

The ellipticity and P.A. profiles in Fig. \ref{ell4434} show strong variations within the first 20$''$
, which are not mirrored in the shape parameters ($a$ and $b$ high-order coefficients), which 
look very regular and featureless in the central $20''$, making this galaxy a quite perfect E0 system (there is a peak of 0.1 in the ellipticity 
at $a\sim6''$ while $\epsilon<0.05$ everywhere). However, the shape parameters start to show strong variation 
outside, which are difficult to comment on because of the large errors.

\begin{figure}
   \centering
   \hspace{-0cm}    \includegraphics[width=8cm]{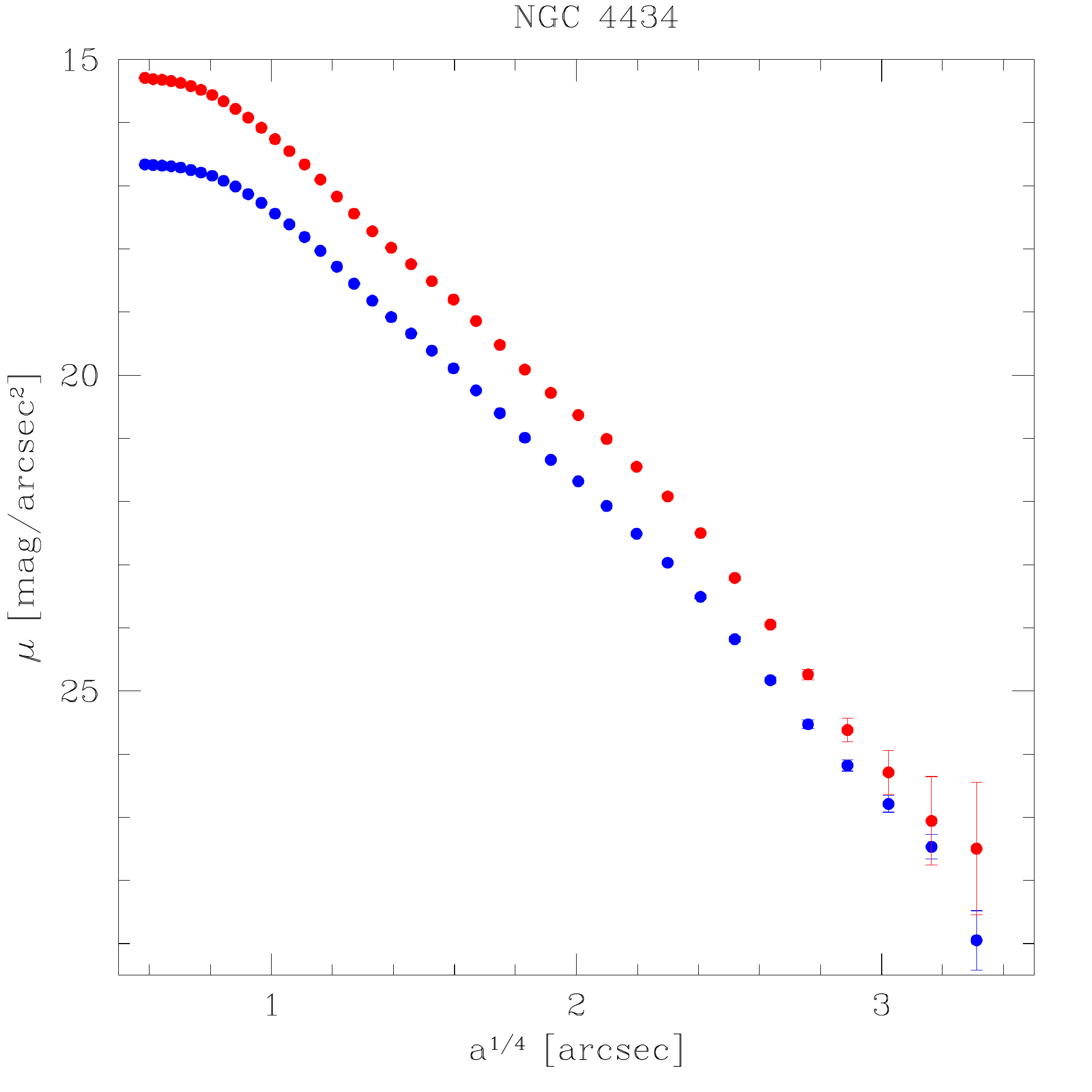}\\ 
   \includegraphics[width=8cm]{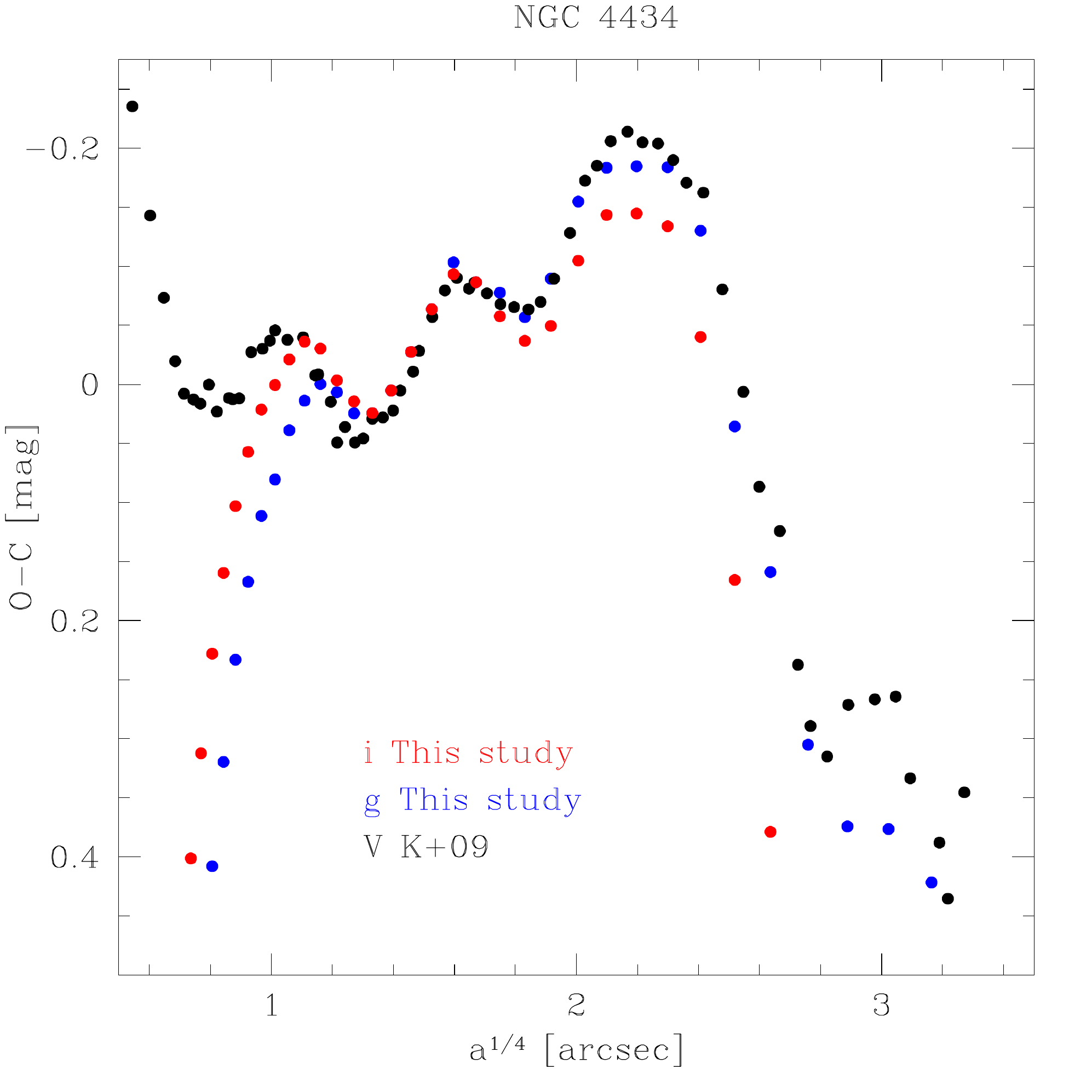}
   \includegraphics[width=8cm]{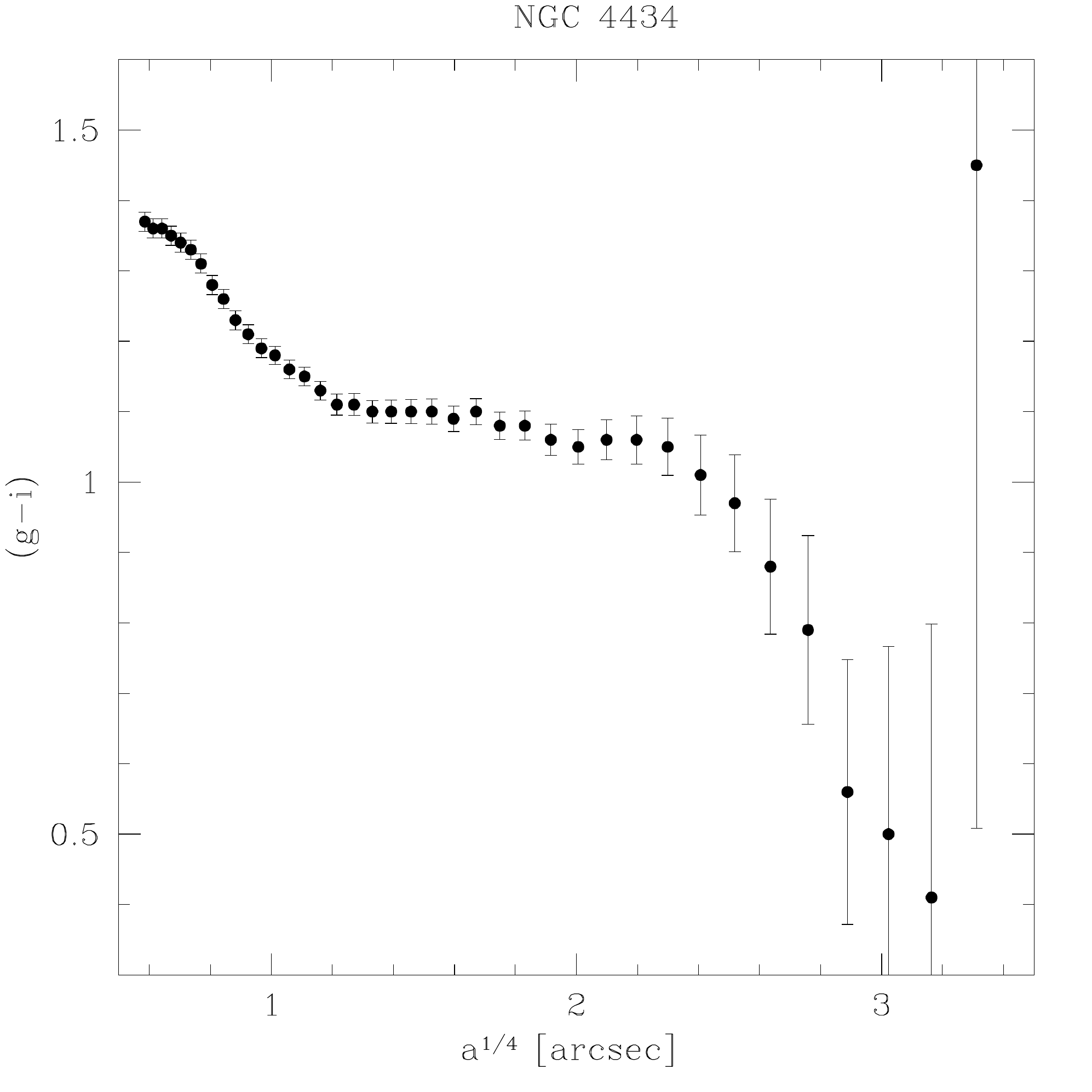}
   \caption{NGC 4434. Top panel: azimuthally averaged light profiles in the {\it g} (blue) and {\it i} (red) bands. Center panel: (O$-$C) residuals of the VEGAS profiles and the V-band profile of K+09 with respect to the best-fitting  $r^{1/4}$model (see Table \ref{tab:devauc}). The best match is obtained with the shifts listed in Table \ref{shift}. VEGAS data for $a^{1/4} < 1.2$ are affected by seeing. Bottom panel: {\it (g-i)} color profile. Again the data at $a^{1/4} < 1.2$ have not been corrected for seeing.}
              \label{prof4434}
\end{figure}

Figure \ref{prof4434} reproduces the azimuthal SB profile in {\it g} and {\it i} bands. Even deeper than for NGC 4472, reaching $\sim 28$ mag/arcsec$^2$ in {\it g} band and $\sim 27$ mag/arcsec$^2$ in {\it i} band at  $a/a_{e}\sim\ 10$, they appear regular and very similar, and both show a bump in the profiles at $a/a_{e}\sim\ 2.5$. 
This feature is evident as an excess of the residuals with respect to the best-fitting de Vaucouleurs profiles (Table \ref{tab:devauc}), which are again overplotted on the SB profiles and are better highlighted by the (O-C) curves (central panel of Fig.\ref{prof4434}). 

\begin{table}
\caption{Average colors computed as shifts giving the best match of the inner light profiles.} \label{shift}
\centering
\begin{tabular}{lcccc}
\hline\hline
\scriptsize

Name    &       (g-V) & (g-i) \\
                &     [mag] &     [mag] \\
\hline
NGC 4472 &      +0.35 &    + 1.24    \\  
NGC 4434        &       +0.20&  +1.10\\ 
NGC 4464        &       +0.39&  +1.23\\ 
NGC 4467&       +0.29&  +1.16\\ 
VCC 1199&       +0.20&  +1.24\\ 
\hline
\hline
\end{tabular}
\end{table}

The {\it (g-i)} color distribution (bottom panel of the same figure) is fairly constant out to $40''$ ($a^{1/4} \sim 2.5''$ and $a/a_{e}\sim 3.6$), while it decreases steeply immediately after the bump in the light profile. One might be tempted to blame an improper background subtraction as responsible for the effect, since the galaxy lies at the edge of the OmegaCAM field. However, the change in the slope of the {\it i} -band profile with respect to the {\it g} profile occurs at a surface brightness level where the photometric error is typically small. Moreover, just the same pattern is shown by another two galaxies of our sample (Sect. \ref{conc}).

The bump shows up lighter in the K+09 photometry, overplotted on our {\it g} band profile in Fig. \ref{prof4434}  using the same color term as applied to NGC 4472. 
Here we also see that our profile deviates from that of  {\it HST} in the very central regions ($r<1''$) as a result of the seeing 
broadening, while it remains consistent within the errors 
with  K+09 at all the other radii. 

Total luminosity and effective parameters are estimated as for NGC 4472 (Sect. \ref{mag}) and listed in Table \ref{Re}.

\subsection{NGC 4464}

\begin{figure}
   \centering
   \includegraphics[width=8cm]{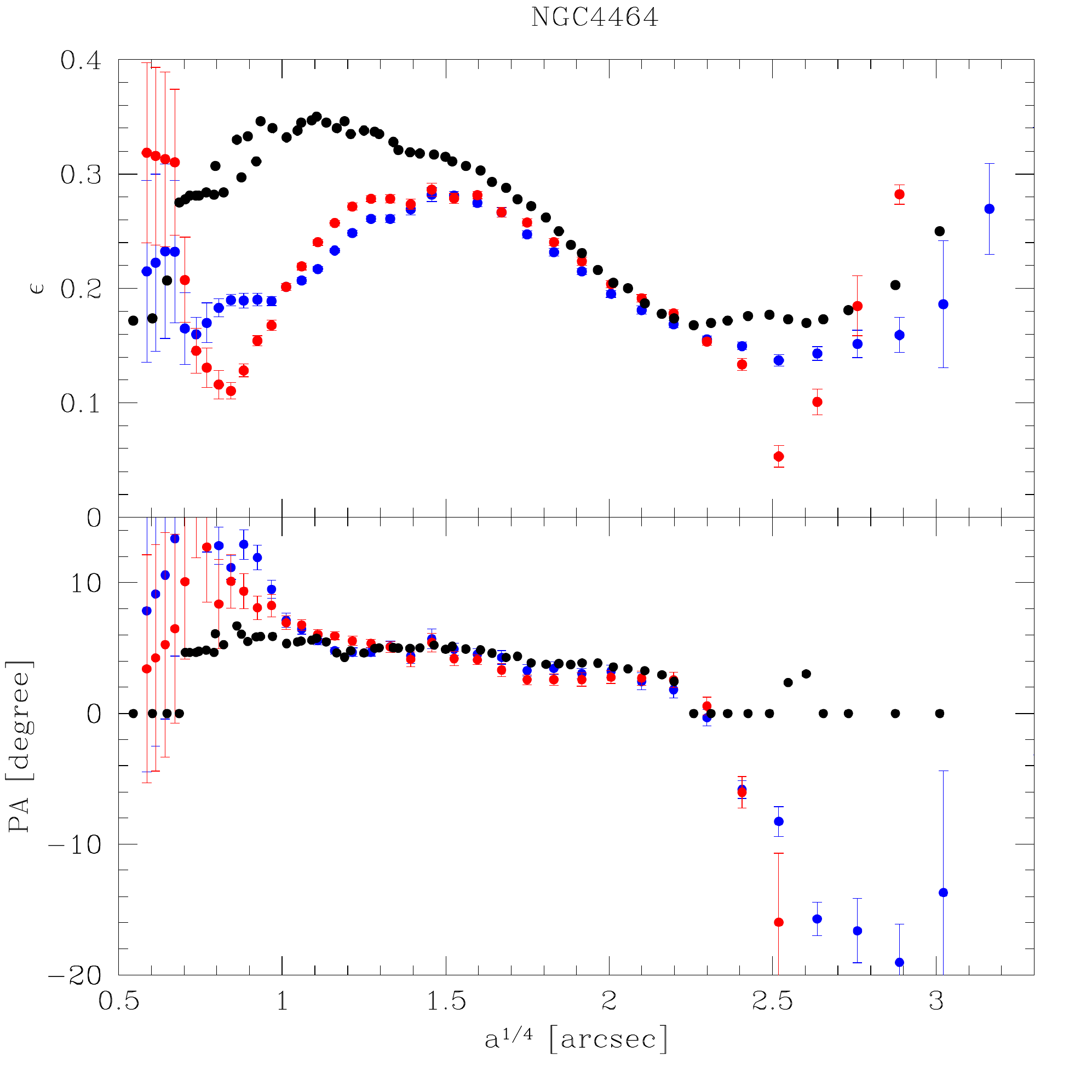}
   \includegraphics[width=8cm]{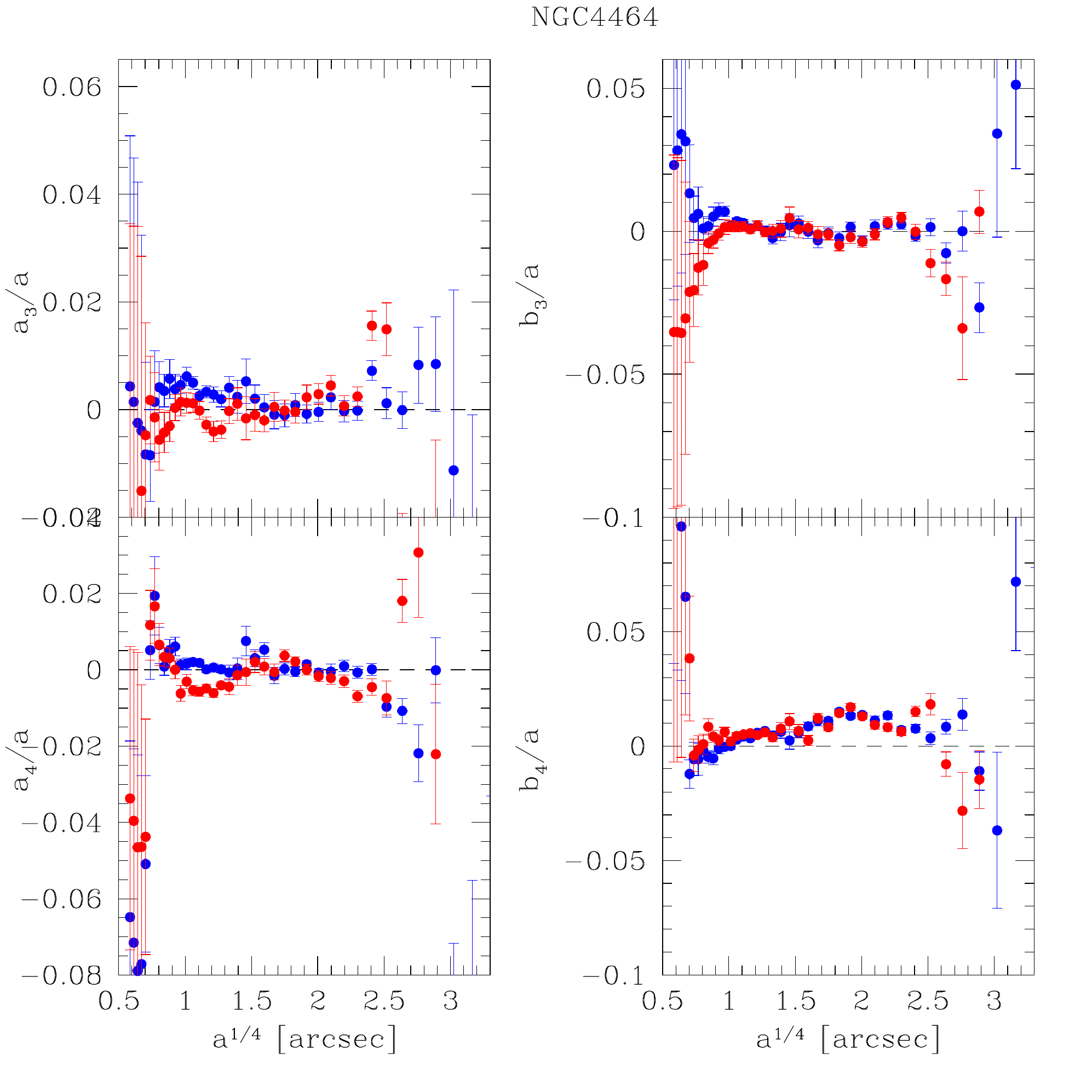}
   \caption{Same as Fig. \ref{ell_4472} for NGC 4464.}
              \label{ell4464}
    \end{figure}

\begin{figure}
   \centering
   \hspace{-0cm}    \includegraphics[width=8cm]{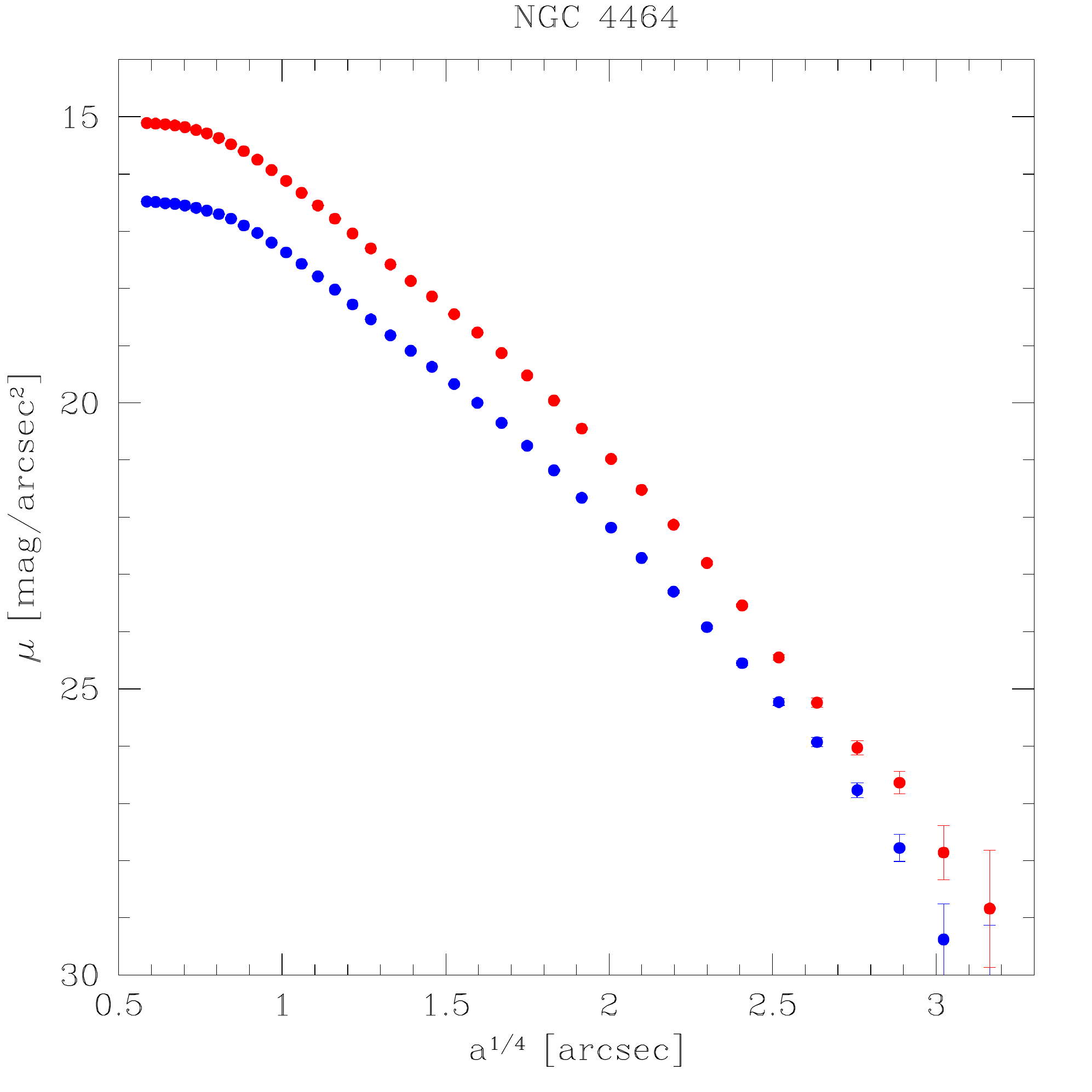}\\ 
    \includegraphics[width=8cm]{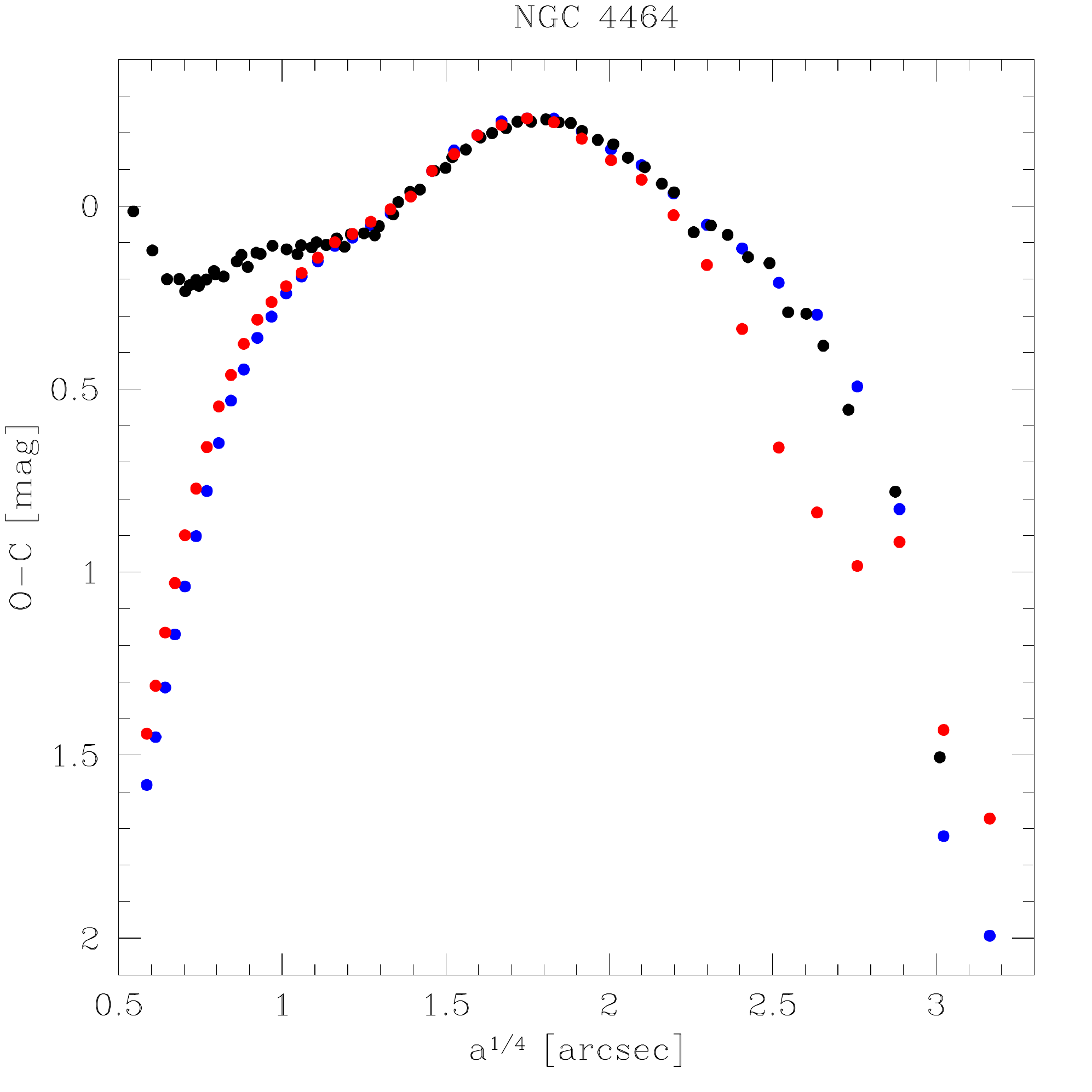}
    \includegraphics[width=8cm]{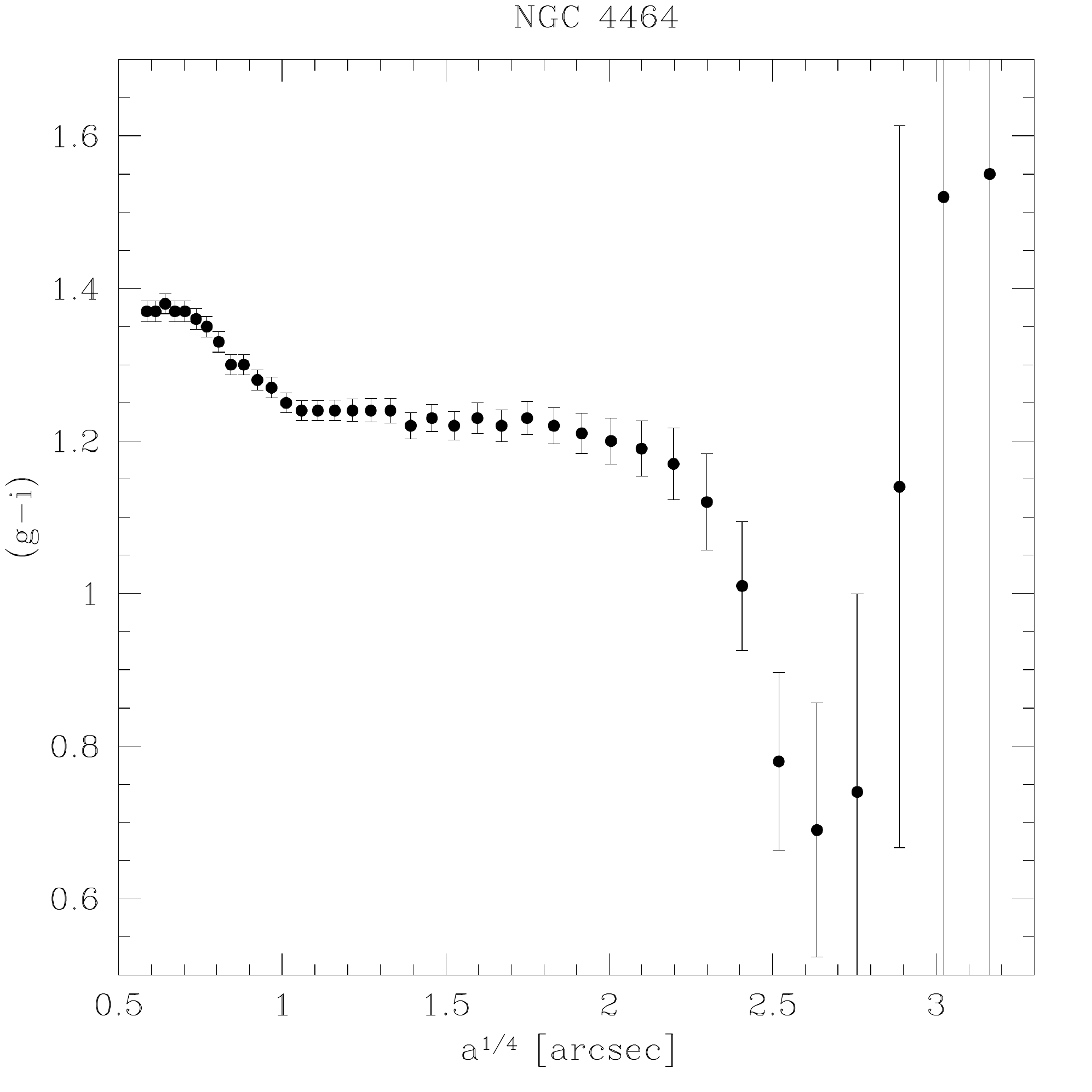}
   \caption{Same as Fig. \ref{prof4434} for NGC 4464.}
              \label{prof4464}
    \end{figure}

NGC 4464 (VCC 1178) is an E3 system.
Figure \ref{prof4464} shows the azimuthal SB profiles reaching $\sim$ 30 mag/arcsec$^2$ in {\it g} band and $\sim 29$ mag/arcsec$^2$ in {\it i} band 
at about 100$''$ ($a/a_{e}\sim 12.8$). In this case as well, the color distribution outside $1''$ is very flat over a wide radial range: $(g-i)\sim 1.2$ for $a/a_{e} < 3$. Outside, the color profile bends toward a minimum 
in correspondence of a rapid variation of the ellipticity, P.A., and shape parameters (Fig. \ref{ell4464}).
In particular, $a_4$ indicates ``disky'' isophotes both in {\it g} and {\it i} bands,
although outside $a\sim 20''$, the shape parameters are again rather noisy.

The SB profiles in both bands also show for this galaxy some hints of a substructure as
light excess with respect to the $r^{1/4}$ fit (Table \ref{tab:devauc}) shown in Fig.  \ref{prof4464}  (in this case around 
$a/a_{e} \sim 1.7$; see the (O-C) profile). 

The multiple components along the line of sight have previously been discussed 
by \citet{Halliday01} and are most likely due to the occurrence of significant
asymmetrical and symmetrical deviations of the line-of-sight velocity distribution (LOSVD)
from a Gaussian at $a \leq 10''$ along the major axis. In particular, for $a\leq 5''$, 
the measurements are consistent with the superposition of a bulge and an additional more rotationally
supported component, which agrees with our finding of flatter isophotes and $b_4>0$ in both bands.

The comparison with the K+09 photometry is shown in Fig. \ref{prof4464}. As for NGC 4434, the steep inner profile nicely follows the $r^{1/4}$ fit in Fig. \ref{prof4464}, which is a fair reproduction of the whole galaxy 
surface brightness distribution, with the caveat of the possible multicomposition as highlighted above. 

\subsection{NGC 4467}
NGC4467 is a faint-system classified dwarf elliptical (e.g., Bender et al. 1992, but see also the classification as 
an E3 galaxy by F+06). It lies at an apparent distance of $4.2'$ from NGC 4472, equivalent to 23 kpc. F+06 found that its SB profile is tidally truncated in the outer regions, where the ellipticity is also affected by the close giant companion. They also detected a small blue 
cluster within 0.1$''$ from the nucleus, a second about 0.9$''$ to the southeast. This galaxy appears to be very compact, with a nucleus brighter than galaxies of similar magnitude.

\begin{figure}
   \centering
   \includegraphics[width=8cm]{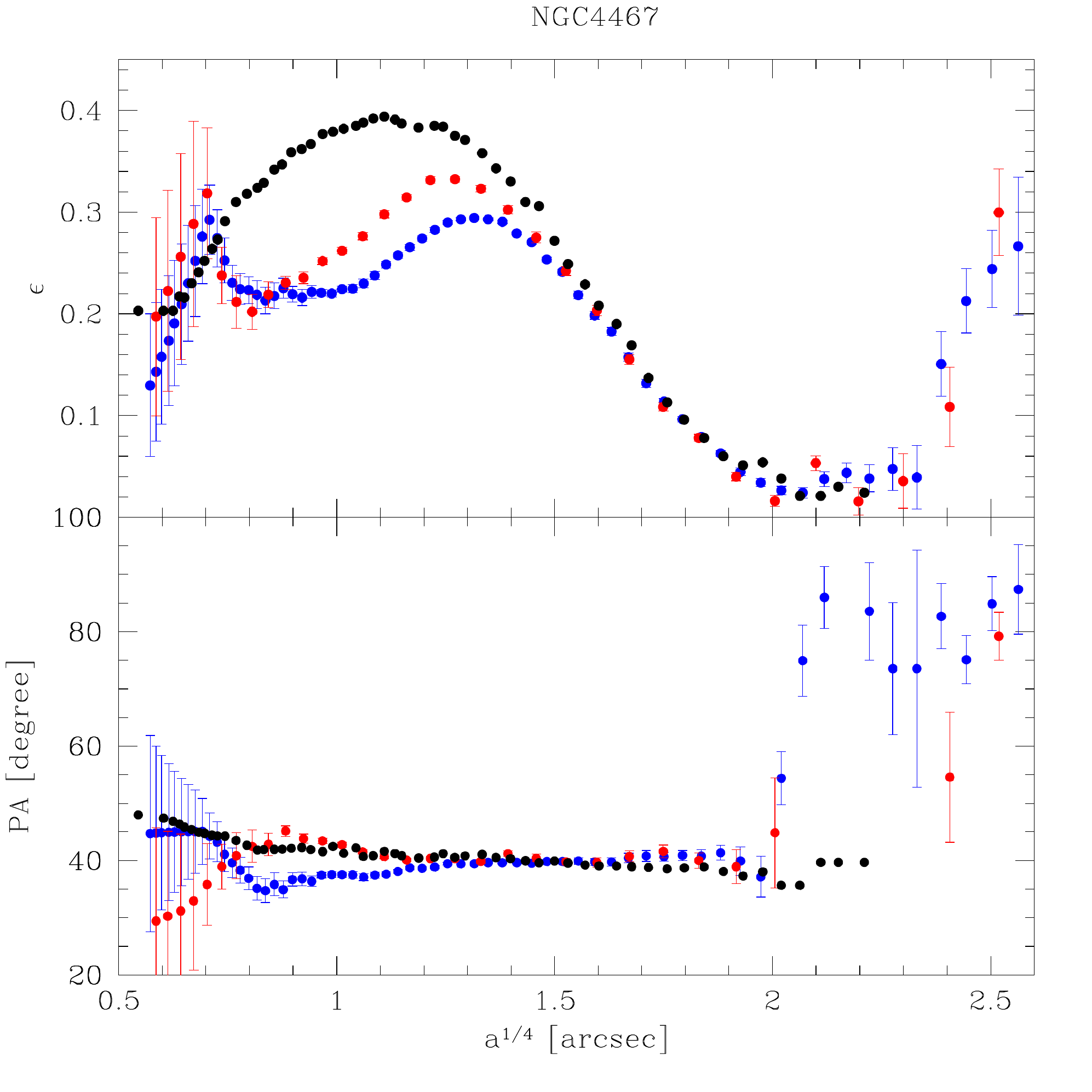}
   \includegraphics[width=8cm]{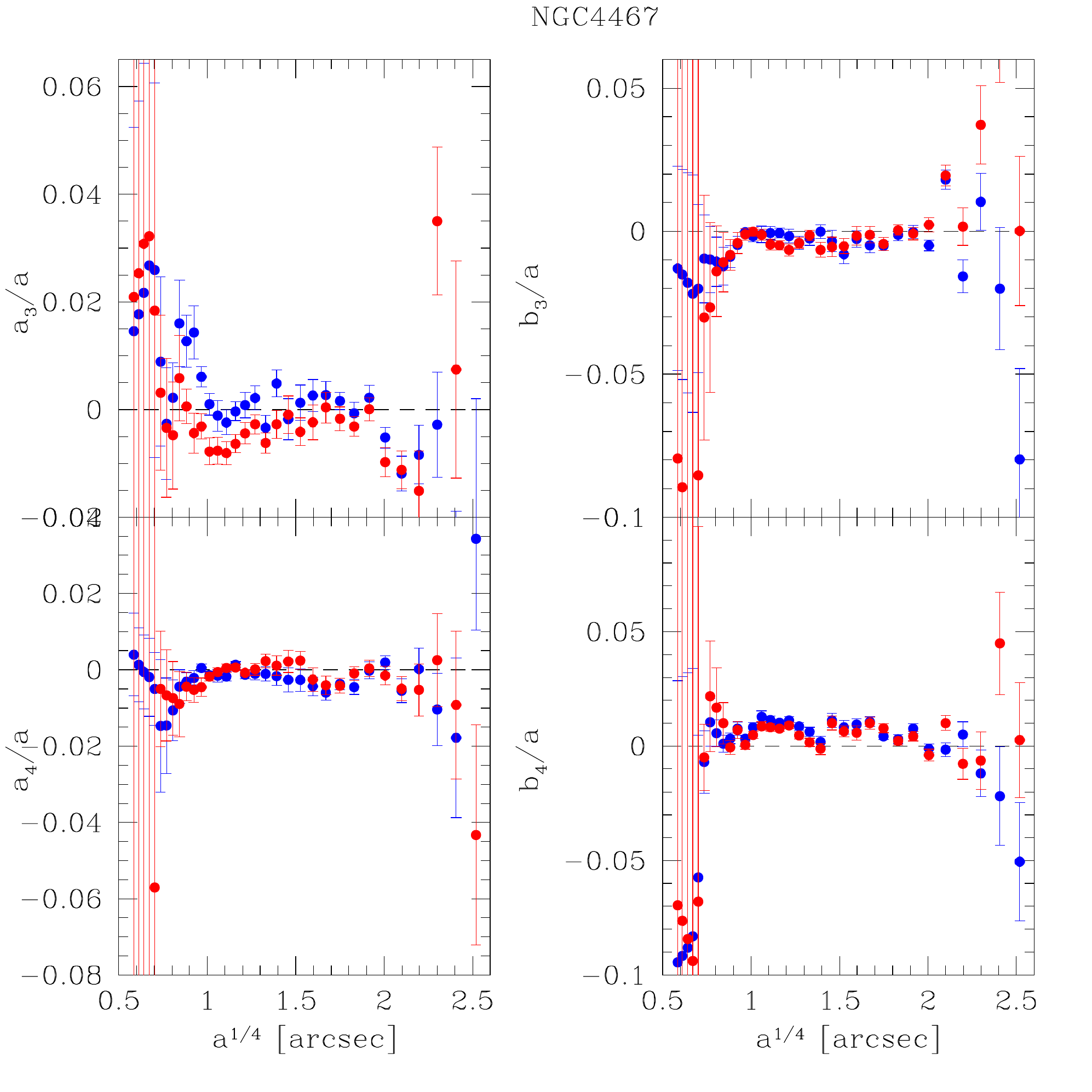}
   \caption{Same as Fig. \ref{ell_4472} for NGC 4467.}
              \label{ell4467}
    \end{figure}

 \begin{figure}
   \centering
   \hspace{-0cm}    \includegraphics[width=8cm]{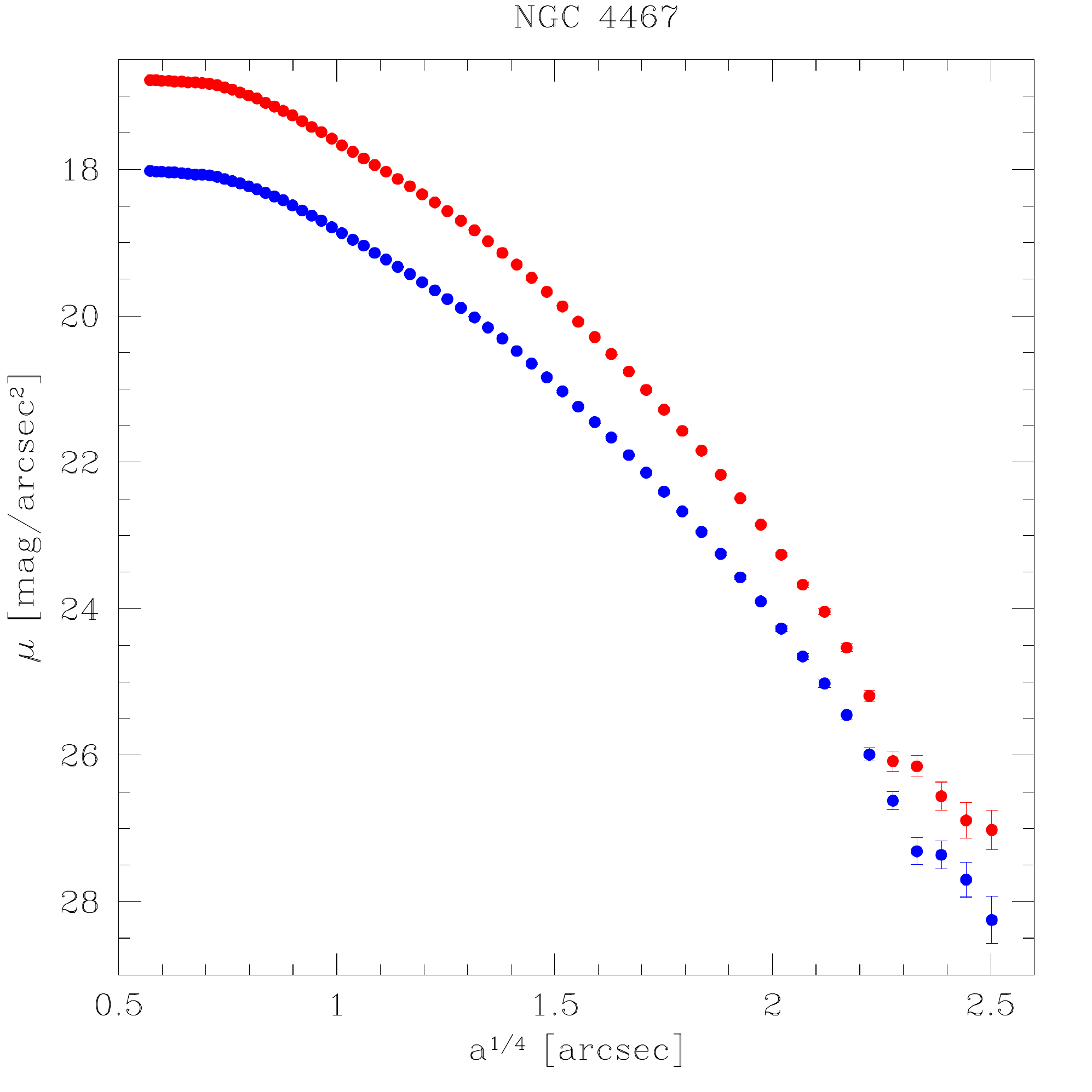}\\ 
    \includegraphics[width=8cm]{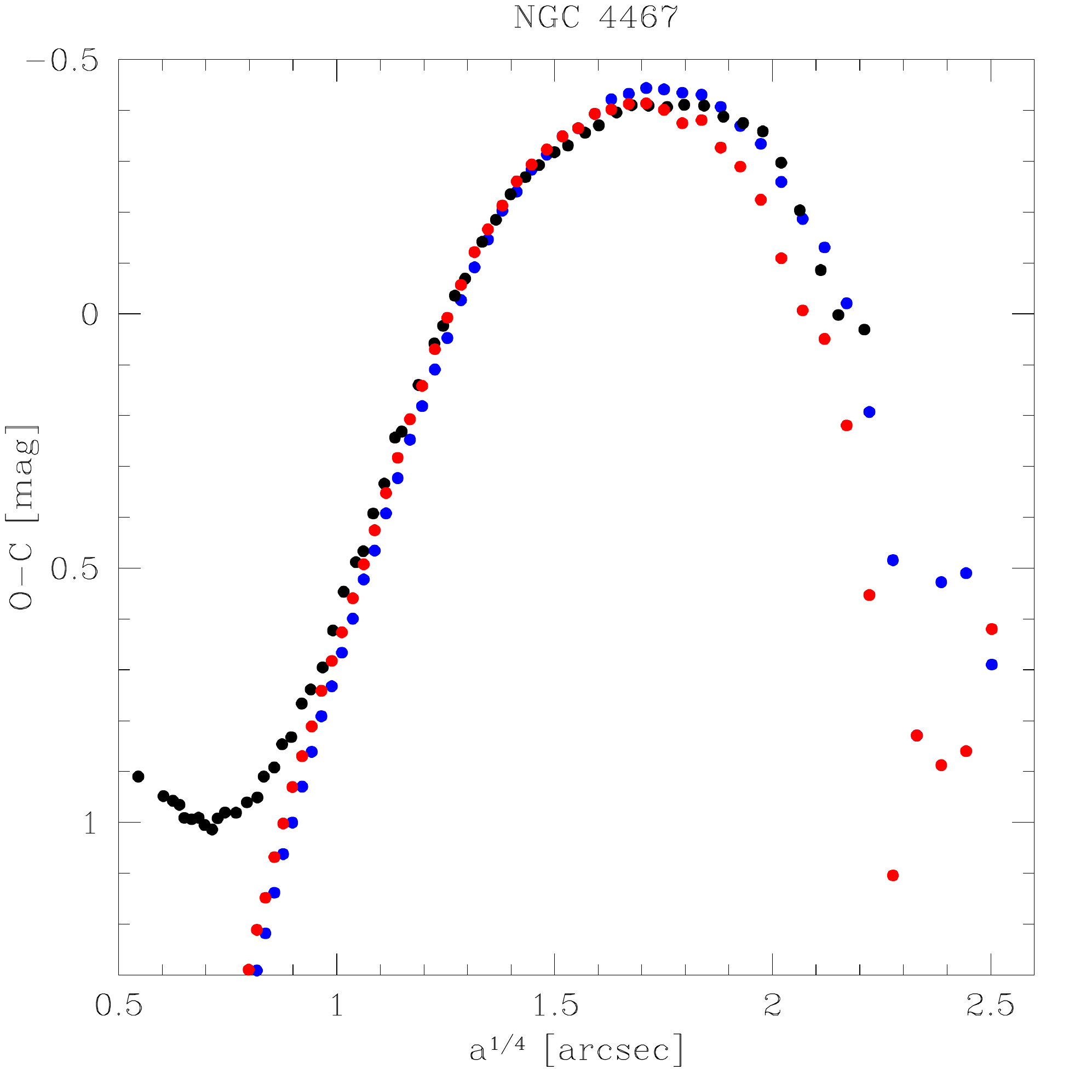}
    \includegraphics[width=8cm]{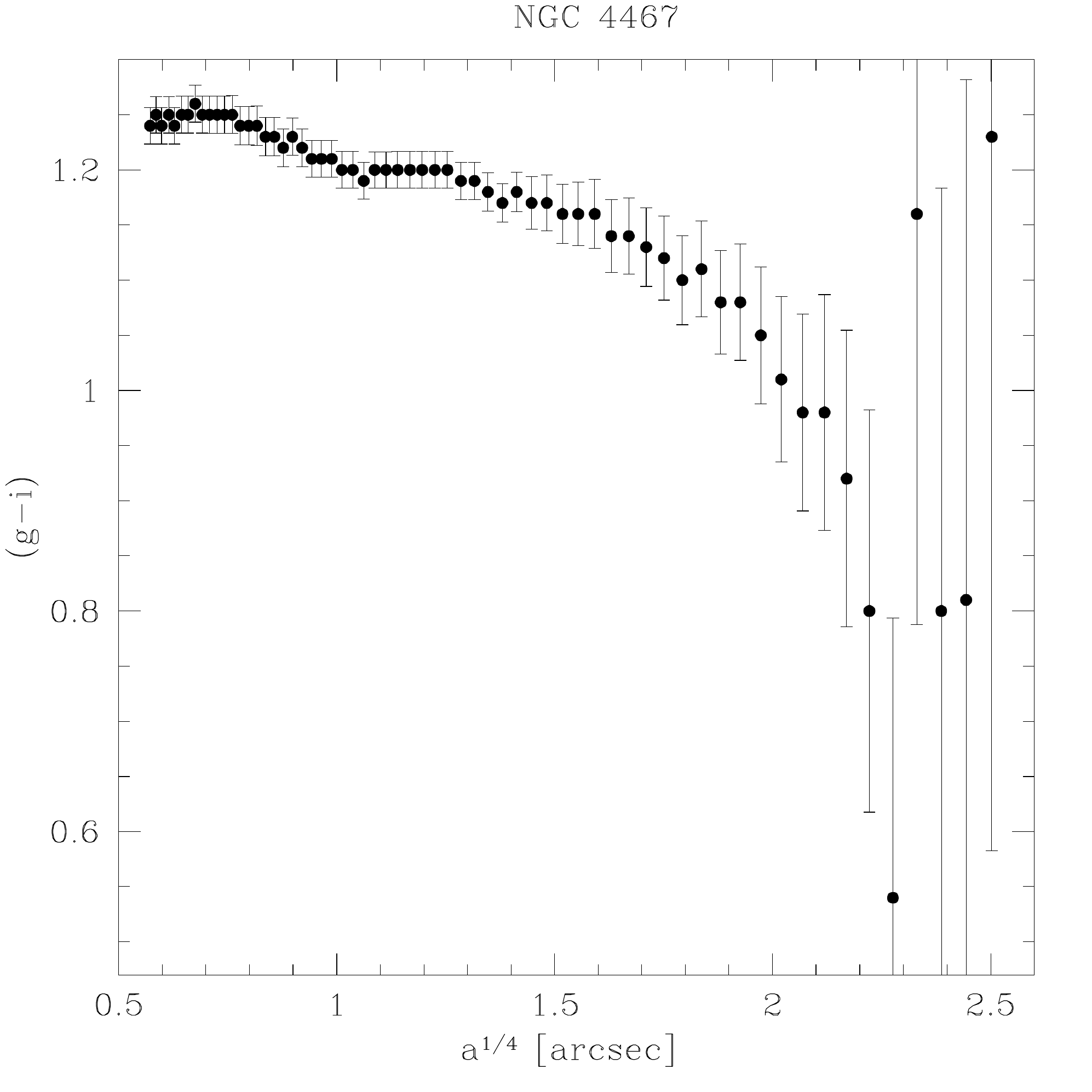}
   \caption{Same as Fig. \ref{prof4434} for NGC 4467.}
              \label{prof4467}
    \end{figure}

Our azimuthal SB profiles are shown in Fig. \ref{prof4467}. 
Despite the bright background of NGC 4472, flux could be measured out to $a/a_{e}\sim 6.8$, where $\mu_{g}\sim$ 28.3 mag/arcsec$^2$ and $\mu_{i} \sim$ 27.0 mag/arcsec$^2$. The SB profiles deviate from an $r^{1/4}$ profile at all radii, as shown by the
(O-C) profile in the same figure. 

The color distribution has a shallow gradient followed by a sharp decrease starting at $a/a_{e}\sim 2.1$. The shape parameters (in particular $b_4$, see Fig. \ref{ell4467}) 
show the emergence of disky isophotes with higher ellipticity than the center in the outer bluer regions. 
Here there is a rapid transition from boxy to 
disky isophote shapes and a quite significant twisting of the isophotes of about 20 deg at $a\sim20''$.
These are all indications of a multicomponent system.

The comparison with the K+09 photometry (Fig. \ref{prof4467}) again is quite good.

\subsection{VCC 1199}
VCC 1199 is another close companion of NGC 4472, located at $4.5'$ from its center. F+06 found that it has a surface brightness brighter than galaxies of similar luminosity and is tidally truncated in the outer regions. They also found a very thin 
edge-on disk aligned with the galaxy major axis, extending less than 1$''$, and a large-scale spiral pattern.

\begin{figure}
   \centering
   \includegraphics[width=8cm]{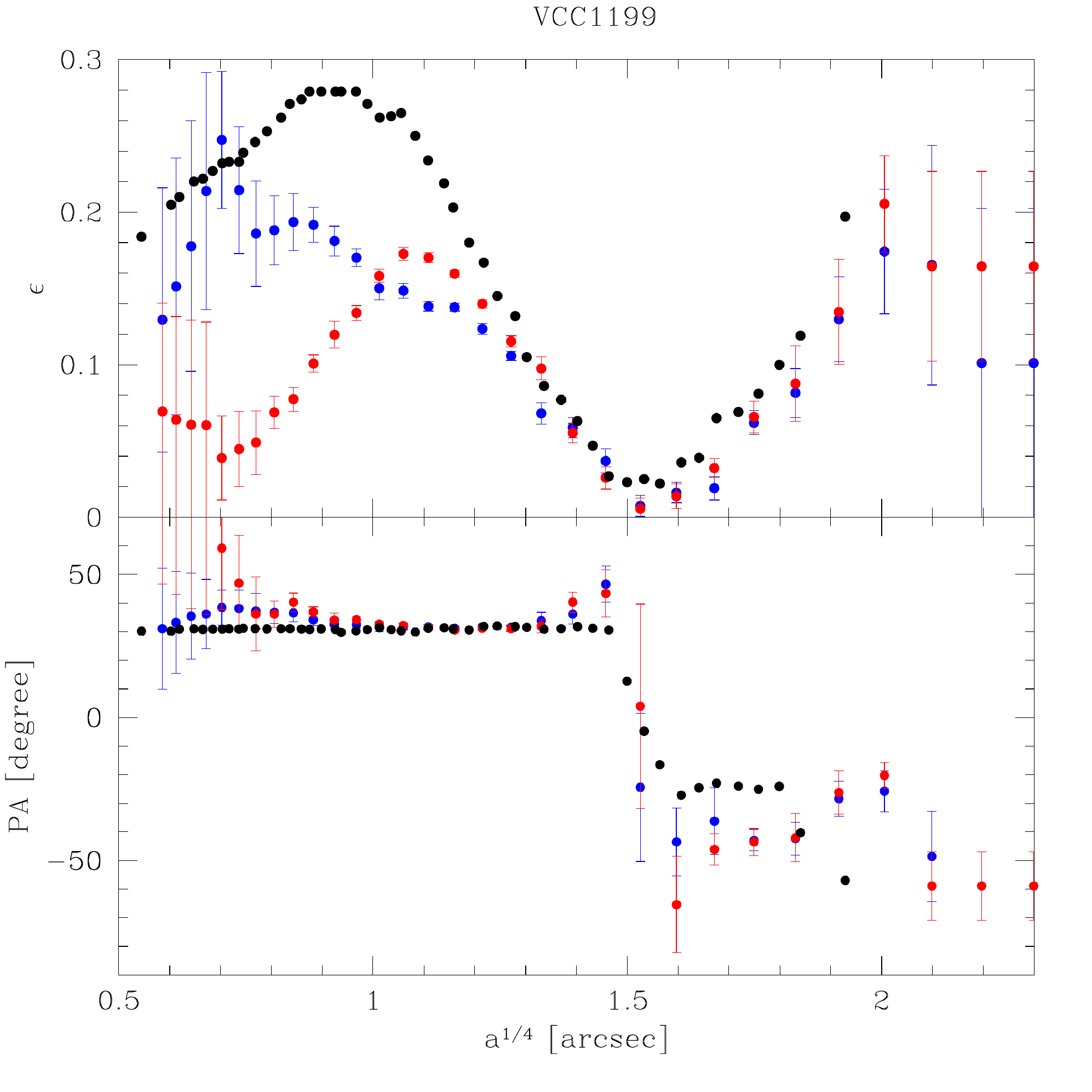}
   \includegraphics[width=8cm]{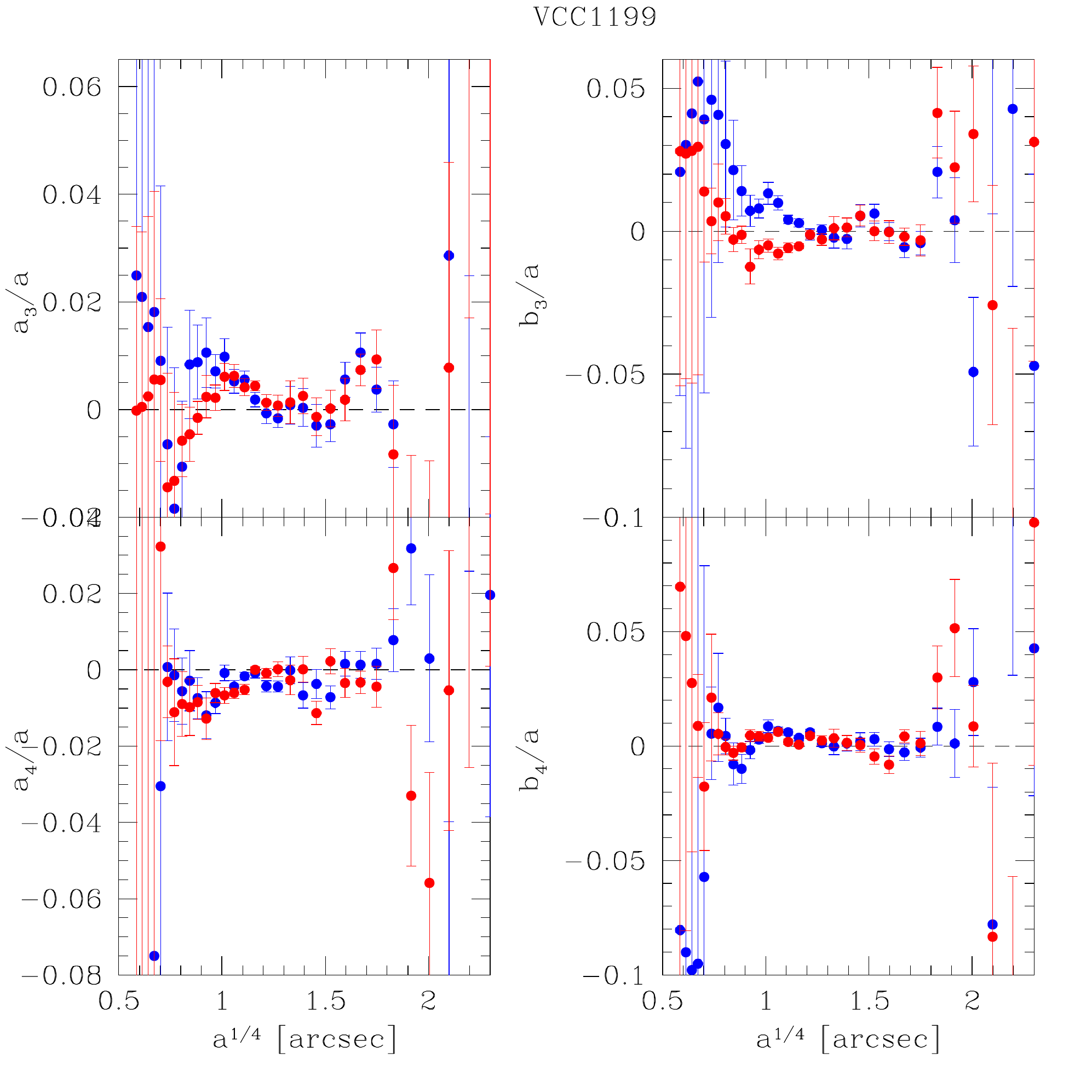}
   \caption{Same as Fig. \ref{ell_4472} for VCC 1199.}
              \label{ell1199}
    \end{figure}

\begin{figure}
   \centering
   \hspace{-0cm}    \includegraphics[width=8cm]{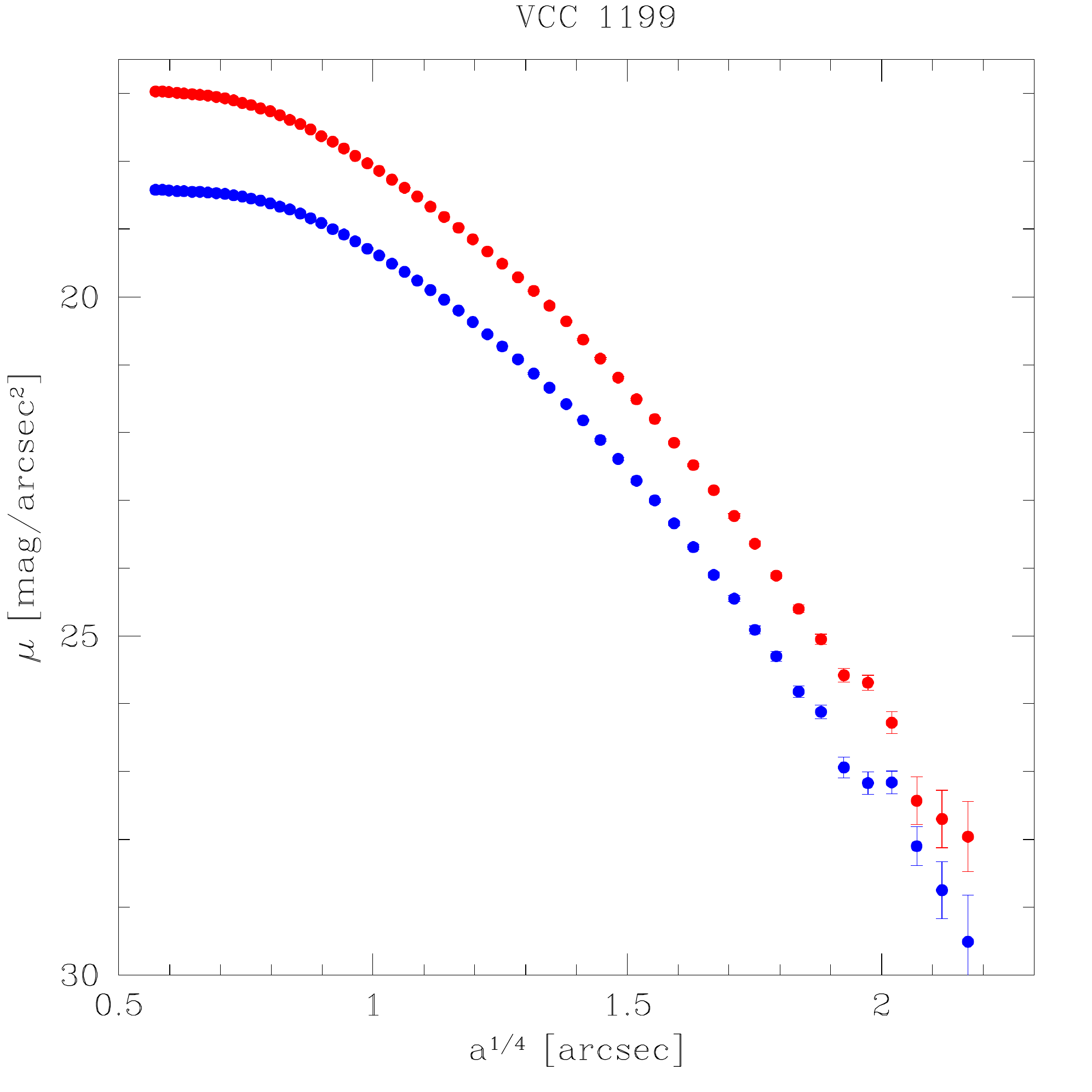}\\ 
    \includegraphics[width=8cm]{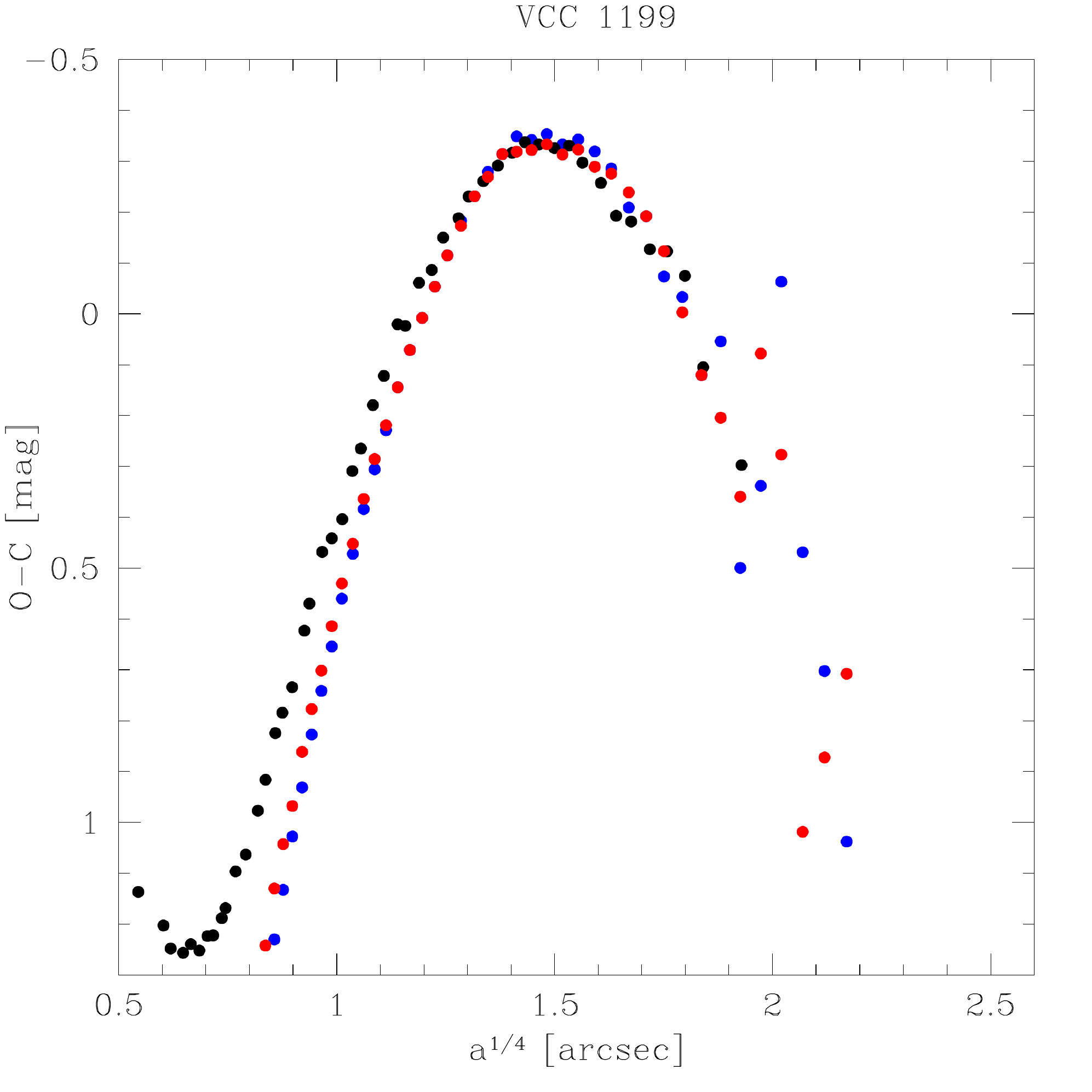}
    \includegraphics[width=8cm]{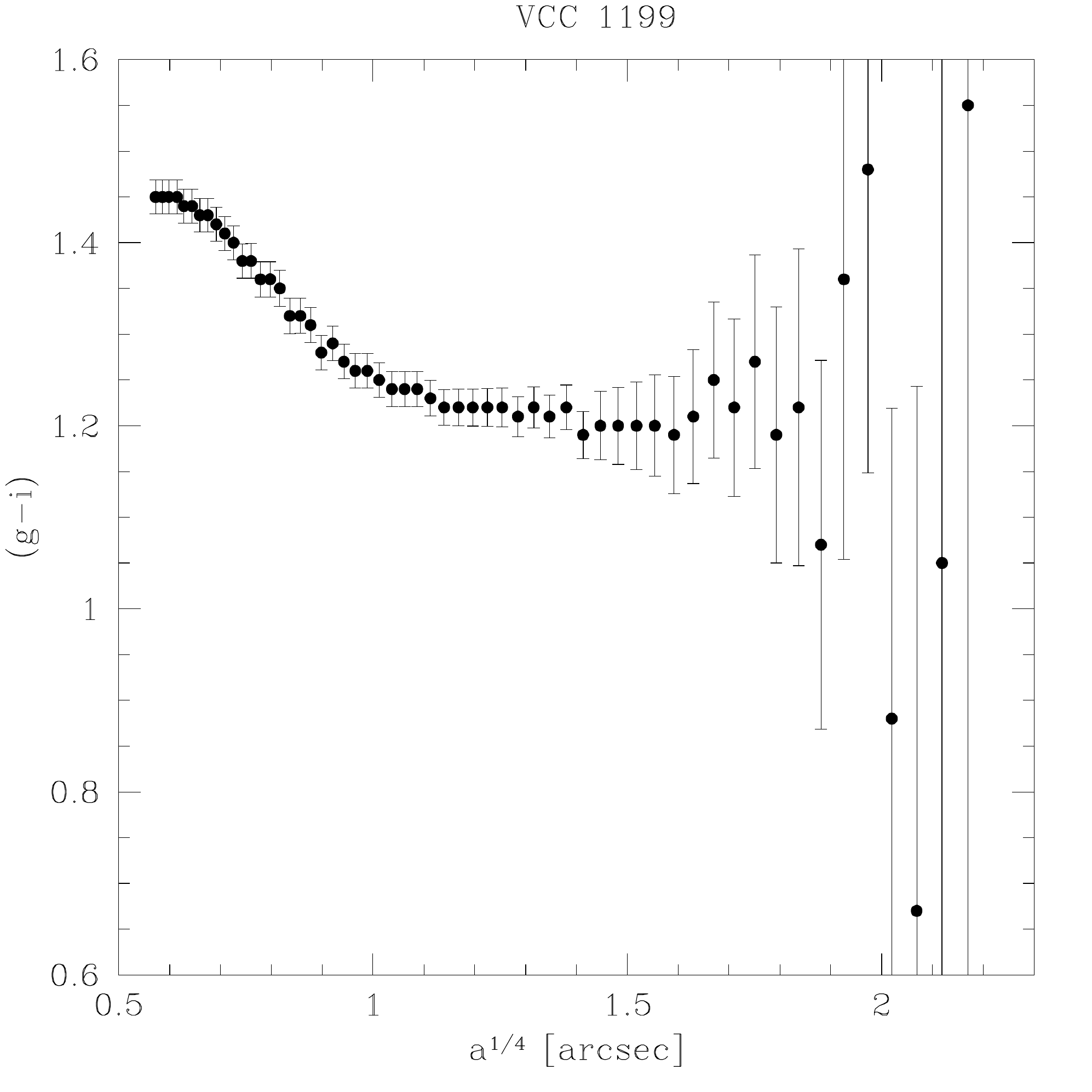}
   \caption{Same as Fig. \ref{prof4434} for VCC 1199.}
              \label{prof1199}
    \end{figure}

The {\small ELLIPSE} azimuthal SB profiles are shown in Fig. \ref{prof1199}. They provide a ({\it g-i}) color profile with almost no gradient outside $1''$, which is the reddest in our sample.
The shape parameters (see Fig. \ref{ell1199}) show a structure very similar to NGC 4467, with a
rapid and significant variation of ellipticity and position angle, and the transition from inner 
boxy isophotes to outer disky ones (although less pronounced than in NGC 4667).
This confirms the multicomponent nature of the object and the presence of an outer disk.

The comparison with the K+09 photometry (Fig. \ref{prof1199}) again is quite good.

\subsection{UGC 7636}
As a byproduct of this paper, we also analyzed the dwarf irregular UGC 7636 (VCC1249) \citep{Nilson73} located 5.6$'$
to the southeast of NGC 4472. 
This object has been extensively analyzed by \citet{Battaia12}, who studied the tidal interaction with NGC 4472 and the gas-stripping phenomena. They found an extensive series of shells and filaments, in agreement with \citet{Janowiecki10}. 
\citet{Lee00} carried out spectroscopic observations of the system and
discovered an HII region associated with this galaxy but not
spatially coincident with it, lying in the envelope of the
giant galaxy NGC 4472.

\begin{figure}
   \centering
   \includegraphics[width=8cm]{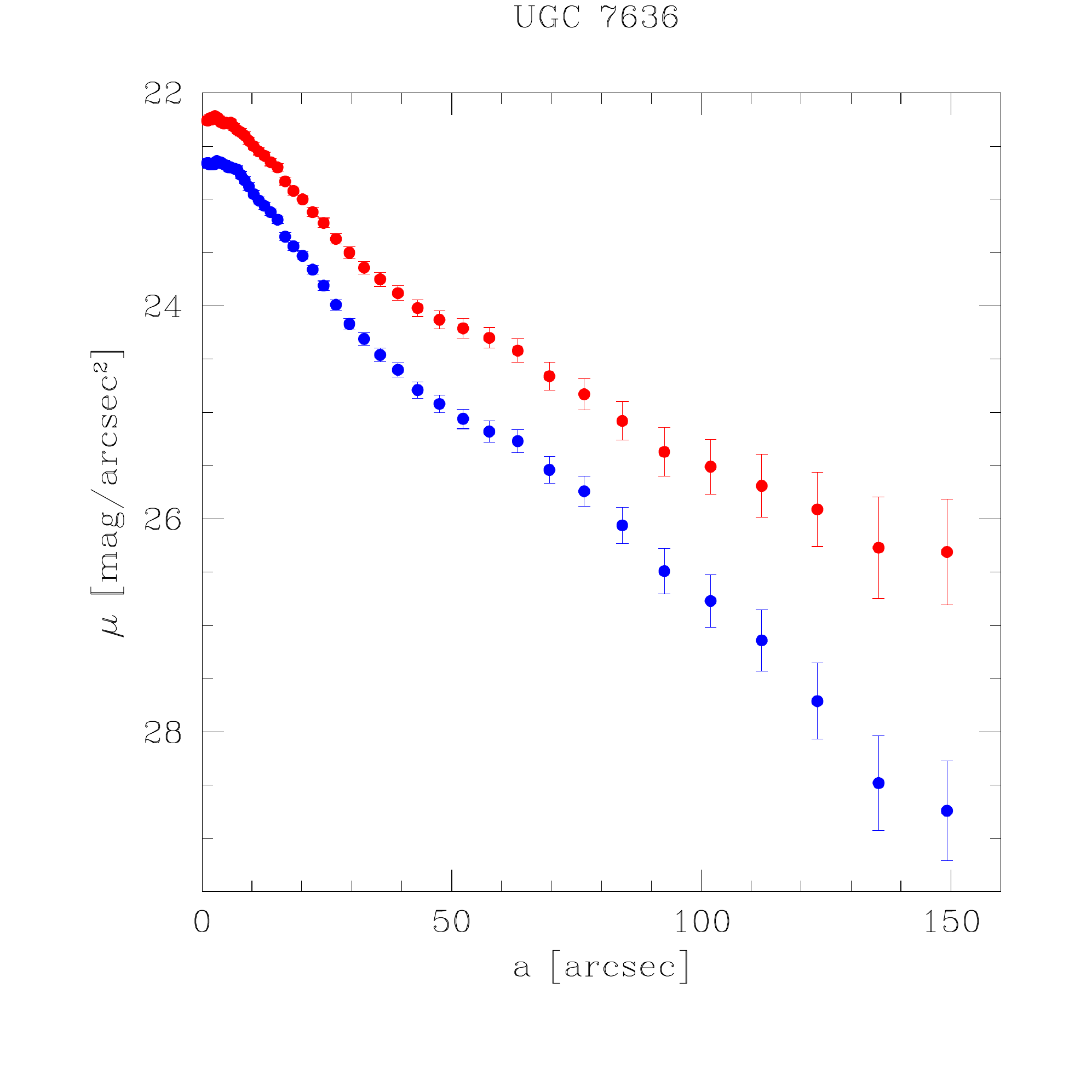}
    \includegraphics[width=8cm]{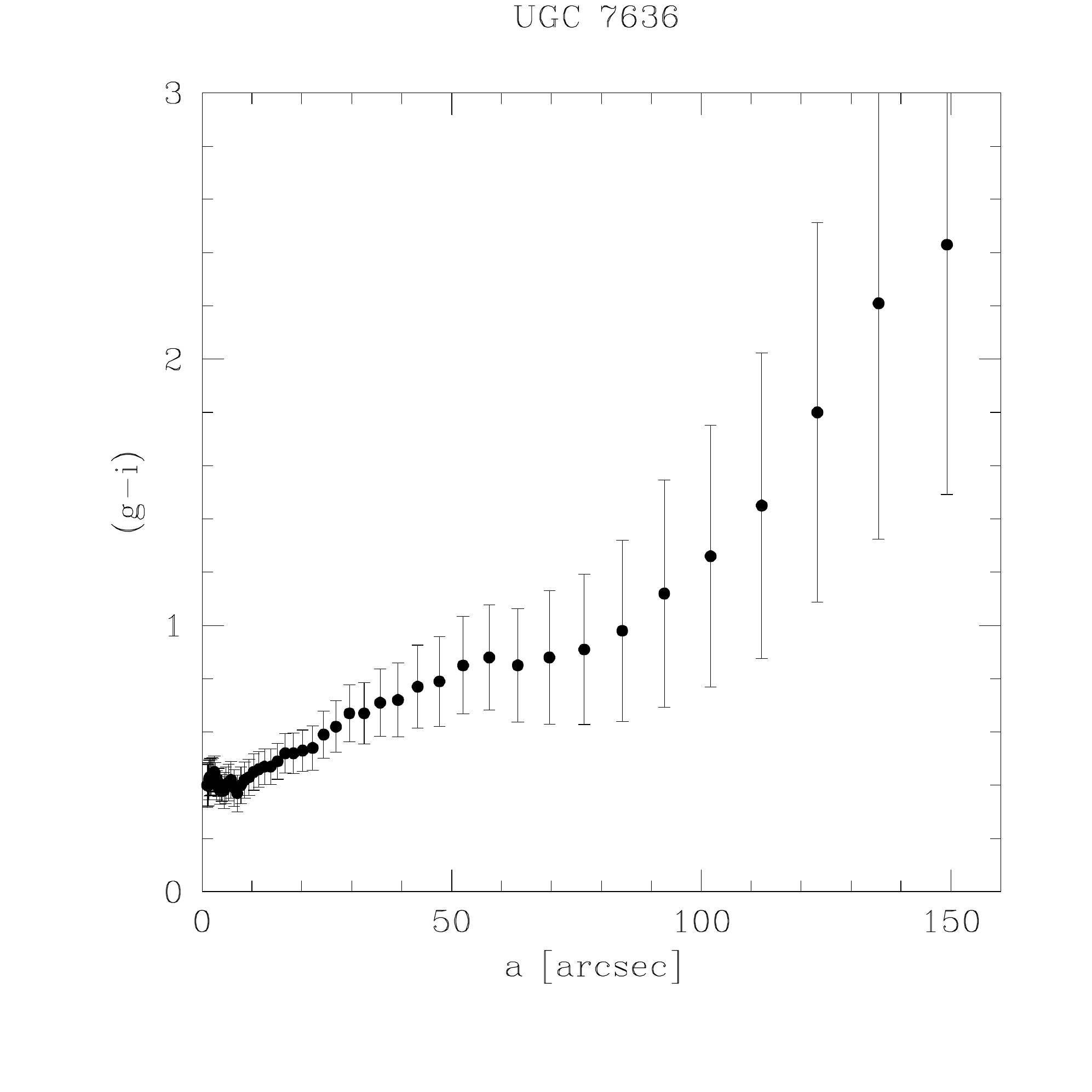}
   \caption{UGC 7636. Top panel: azimuthally averaged light profiles in the {\it g} (blue
     dots) and {\it i} (red dots) bands. Bottom panel: {\it (g-i)} color profile.}
              \label{prof_ugc}
    \end{figure}

\begin{figure}
   \includegraphics[width=9cm]{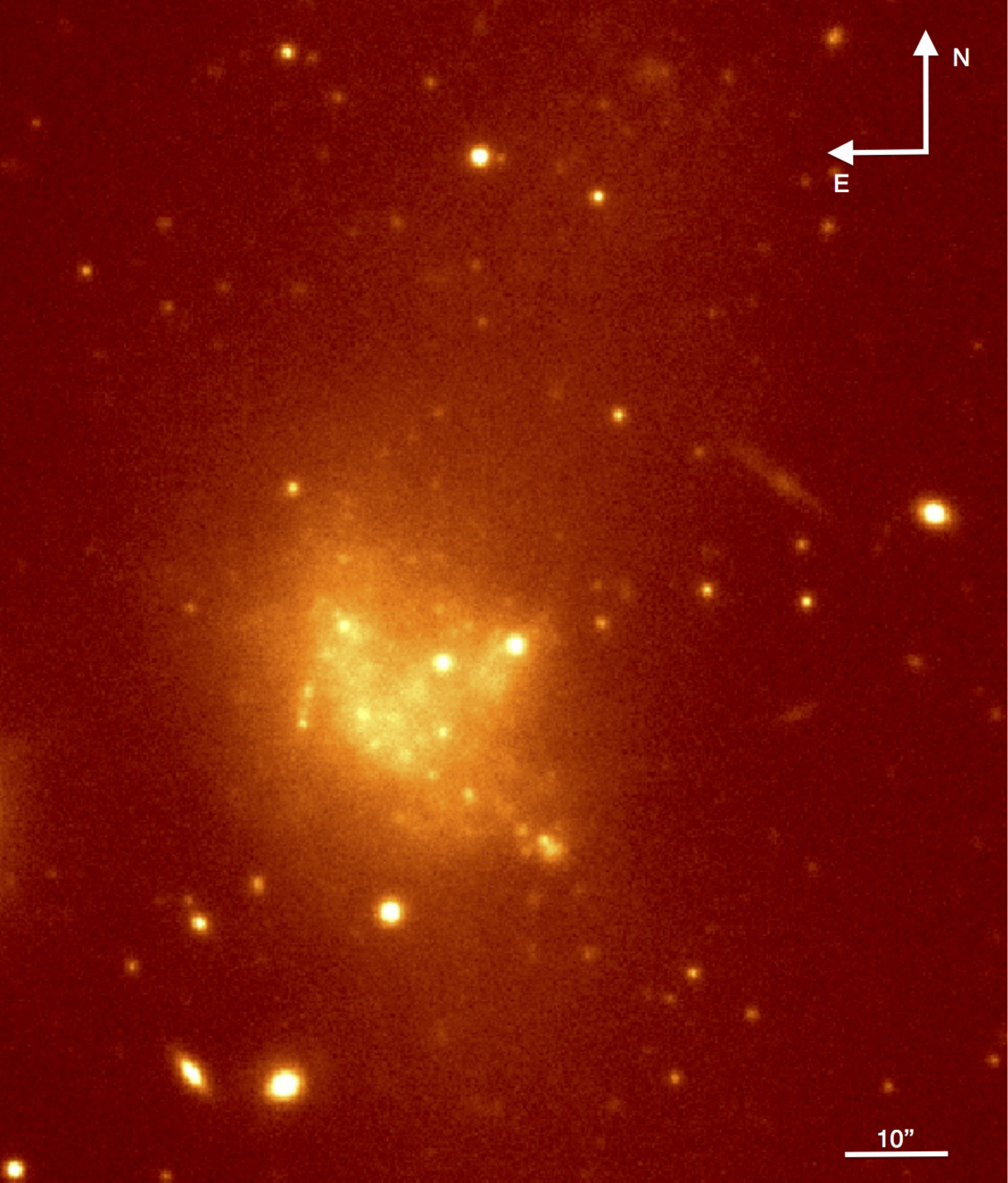}
   \caption{VST {\it g} -band image ($100 \times 117$ arcsec) of the interacting system UGC 7636.}
              \label{ugc}
    \end{figure}

Lacking the possibility of deriving a geometrical model of this very irregular object, we computed mean profiles by azimuthally averaging the background-corrected flux in annuli with orientation, flattening, and center all identical to that of the best ellipse encircling the visible boundaries of the object ($\epsilon = 0.39$, P.A. = 0 deg and center at R.A. = 12h30m01.0s Dec. = +07d55m46s).
The result is shown in Fig. \ref{prof_ugc}. The procedure is reasonably reliable because the output changes marginally by varying the input parameters within a fair range. The method is effective in providing the trend of the color with distance.
Both profiles mimic the behavior of late spiral or irregular galaxies \citep{Capaccioli73}. The temptation to fit the data with the sum of an $r^{1/4}$ bulge and an exponential disk is hampered by the complexity of the body of the object (Fig. \ref{ugc}). 

Based on Fig. 7 of \citet{Battaia12}, we judge that our light profiles agree with those of these authors, extending twice as deep, down to $\mu_{g} \sim$ 28.7 mag/arcsec$^{2}$.

We also derived the average {\it (g-i)} color profile (Fig. \ref{prof_ugc}). It reddens steadily outwards with the higher slope from $a>80$ arcsec. We note, however, that in the outer range the errors are quite large.
Our result agrees with that of \citet{Battaia12}.
Outside the main galaxy body, where the SB profiles steepen, the color becomes consistent with NGC 4472 
(Fig. \ref{prof}), indicating a continuity between the two systems, as expected in the close interaction of the dwarf irregular with the giant elliptical.

\section{Scattered light}\label{scatter}

The surface brightness levels where our estimated effects of the extended PSF are larger than twenty per cent (0.2 mag)  are $\sim 29$ mag/arcsec$^2$ in {\it g} and $\sim 28$ mag/arcsec$^2$ in {\it i}. The following is apparent from these values: 
\begin{enumerate}
\item Typically, the azimuthally averaged light profiles derived out to a surface brightness $\mu_{g} \sim 28$ mag/arcsec$^{2}$ are little affected by scattered light for all of our angularly small galaxies. This fact may explain the remarkable agreement between our results and those of Kormendy \citep{Kormendy09} because this author did not mention any correction of his light profiles, which were made using a material quite different from ours. 
\item The dip observed in the color profiles of NGC 4434, NGC 4464, and NGC 4467 (see Fig.\ref{col_all}) occurs at a surface brightness level at least two magnitudes brighter than the one where scattering becomes important.
\end{enumerate}

\section{Discussion and conclusions}\label{conc}
We have presented the VST Early-type Galaxy Survey (VEGAS) that
is  currently ongoing with VST/OmegaCAM (PI: M. Capaccioli) and aims at studying about one hundred galaxies mainly in the southern hemisphere. 
The survey is as deep as the Next Generation Virgo Survey, but has no environment constraints and is expected to provide a 
systematic coverage of the surface photometry in at least three optical bands, $g,~r,$ and {\it i}, down to 27.3, 26.8, and 26 mag 
arcsec$^{-2}$ (S/N $>3$ per arcsec$^2$), respectively, while $u$ band is foreseen for a subsample of the entire survey.
VEGAS is also expected to provide a census of the faint satellites (globular clusters, ultra-compact dwarfs, and dwarf galaxies;
see, e.g., \citealt{Cantiello15}) in the surroundings of the targeted systems, characterize their extended stellar haloes, 
and find evidence of the intracluster or group light around the giant galaxies in denser environments as well as signatures of 
merging and interactions between galaxies (e.g., tidal tails
and stellar streams) and between galaxies and the group or cluster medium.

We demonstrated the typical specifications of the survey in terms of depth and photometric accuracy and illustrated the performance of the telescope and camera as well as the data reduction and data analysis approach. To this end, we chose the field of the giant elliptical
galaxy NGC 4472 in the southern extension of the Virgo cluster. This is a well-studied system with extensive literature photometry to compare our results with. 

In particular, we presented the deep observations in two bands ({\it g} and {\it i}). The observations were collected with the VST/OmegaCAM in March, April and May of 2013. The major advantage of this wide-field dataset is the good seeing in both filters and the uniformity of the observing conditions (data are taken within one month), which are uncommon for service-mode observations. 

The surface brightness profiles of NGC 4472 reach a depth of $27.5$ mag/arcsec$^2$ in {\it g} band and $26$ mag/arcsec$^2$ in {\it i} band, which is similar to previous deep studies (see Fig. \ref{conf_prof4472}).
This depth allowed us to spot deviations from a simple de Vaucouleurs profile and in particular a change of slope at $a \sim14'.2$ (see Fig. \ref{prof})
that we have associated with a decoupled ICL component that has not been detected
in previous analyses (e.g., K+09). 
The ICL in the Virgo Cluster has been discussed before and is mainly concentrated in the cluster core. It has been detected either
through direct deep imaging \citep{Mihos05} or using planetary nebulae as stellar light tracers (ICPNe, 
e.g., \citealt{Arnaboldi02}; \citealt{Aguerri05}). In the area around NGC 4472, evidence of ICL has been obtained with 
PNe by \citet{Feldmeier04} (see also \citealt{Castro09} for a summary of ICPNe observations over a range of Virgo cluster -centric distances). However, none of these studies has addressed a detailed 2D distribution of the ICL around NGC 4472
and its connection with the giant galaxy. Here we stress that the simple inspection
of the deep SB profile of NGC 4472 clearly shows a diffuse component starting to dominate 
at $\mu_{g} \sim$26.5 mag/arcsec$^2$ (see Fig. \ref{prof}), which is compatible with the typical SB values at which \citet{Zibetti05}
have observed a change of slope induced by the ICL in a series of stacked galaxy clusters.

We note that the trend of the residuals of the luminosity profiles of NGC 4472 with respect to an $r^{1/4}$ best-fitting model has some striking analogies with the similar curve for NGC 3379 \citep{deV79}. In addition to a bright extended core, we found evidence for a wavy pattern that is possibly associated with shells of diffuse material.

We also studied the fainter ETGs in the one square degree of the OmegaCAM field: NGC 4434, NGC 4464, NGC 4467, and VCC 1199, including the dwarf irregular UGC 7636 in the proximity of the giant galaxy NGC 4472. 
For the two galaxies projected onto the bright halo of NGC 4472, NGC 4467 and VCC 1199, located at $r\sim4.1'$ and $r\sim4.5'$ from NGC 4472, we were able to estimate and subtract the galaxy background and trace the SB distribution down to 
$\mu_{g} \sim 29$ mag/arcsec$^2$ and $\mu_{i} \sim 27.5$ mag/arcsec$^2$, which is well beyond the nominal specifications of the survey.
We reached an even greater depth for the farther systems NGC 4464 and NGC 4434, which are not (deeply) affected by the extended halo of NGC 4472
and for which we have gone down to $29-30$  mag/arcsec$^2$ in {\it g} band and $\sim28$ mag/arcsec$^2$ in {\it i} band.
Together with the extremely good comparison with the V-band photometry by \citet{Kormendy09}, at least for our {\it g} band, this demonstrates that for normal galaxies the survey VEGAS provides an unprecedented view of the faint features around
early-type galaxies, with less than one night of telescope time per galaxy (in $g,~r,~i$).

For all these systems we have highlighted some substructures that were defined as deviations from a simple de Vaucouleurs (1948) best-fit profile, as done for NGC 4472. In particular, we found evidence of bumps seen in both bands for the intermediate-luminosity systems NGC 4434 and NGC 4464. These bumps are associated with strongly varying values of the ellipticity and P.A. and $a_4$ and $b_4$ parameters, hence suggesting some substructures. They are possibly also seen in their kinematics, as for NGC 4464 (\citealt{Halliday01}), but are not clearly seen in NGC 4434 (e.g., \citealt{Simien97}).

\begin{figure}
   \centering
 \includegraphics[width=8cm]{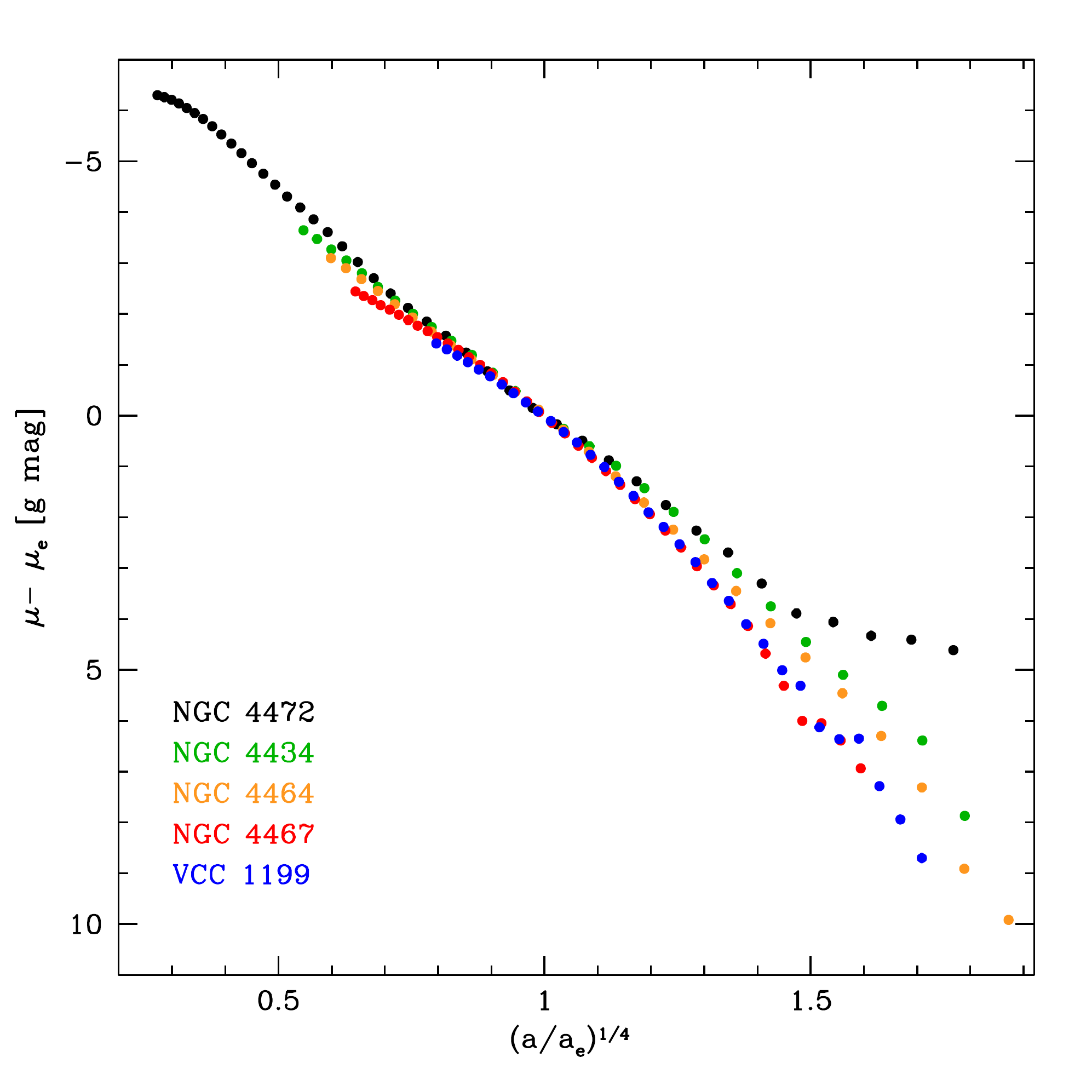}
   \includegraphics[width=8cm]{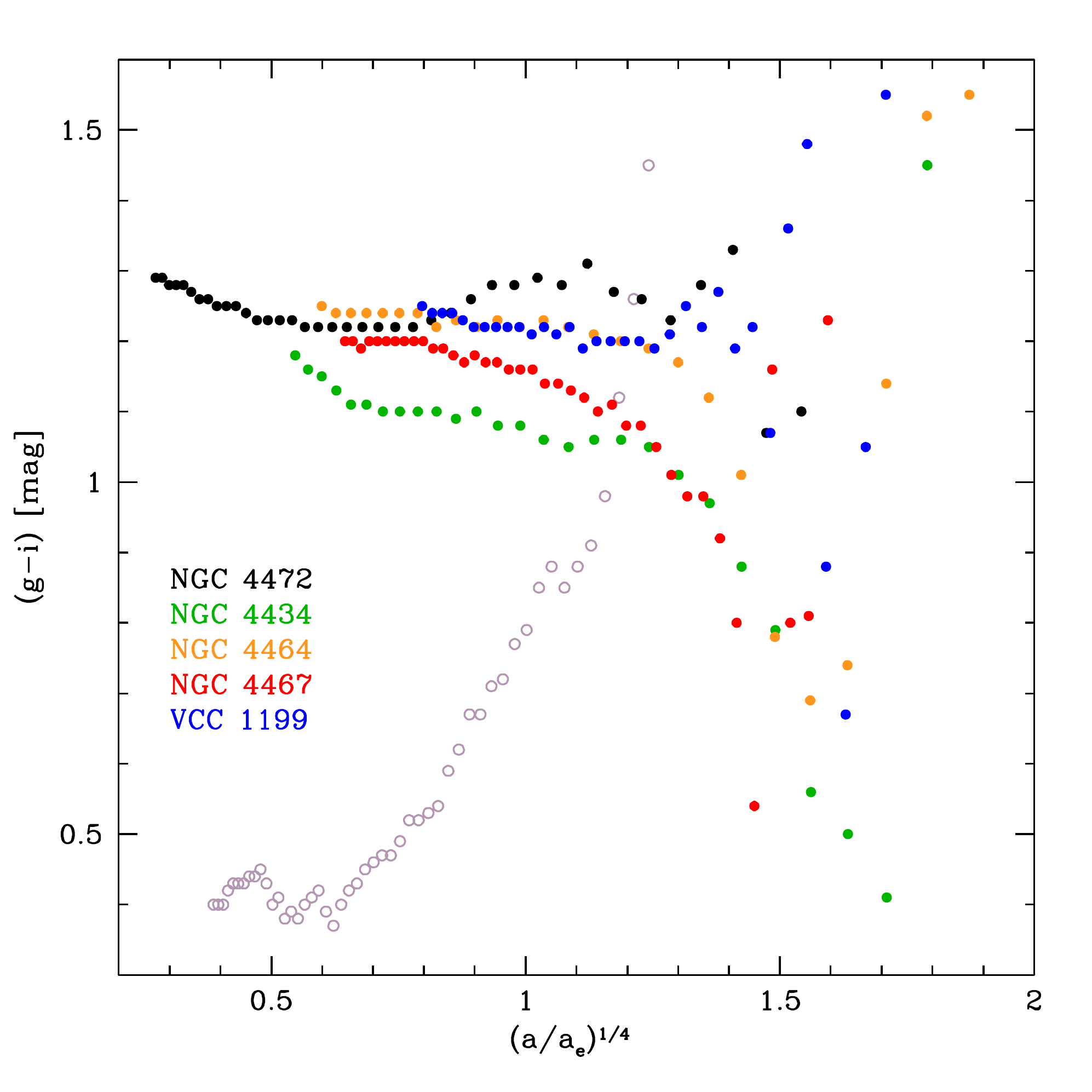}
   \caption{Top panel: azimuthally averaged light profiles in the {\it g} band for the five ETGs of this paper scaled to their effective parameters. Bottom panel: assembly of the {\it (g-i)} color profiles for the five ETGs and for the interacting system UGC 7636 (open circles).}
              \label{col_all}
    \end{figure}

The color profiles, at variance with simulations \citep{Tortora13}, do not show either the sharp decrease of the average value in the first $r_{e}$ for objects fainter than $M_{g} \sim -19$ or the pattern of the gradient as a function of the host galaxy absolute magnitude, which remains very flat with $M_{tot}$. We instead found an indication, which needs to be confirmed, that for $r >3 r_{e}$ a very negative colour gradient develops in some galaxies, which apparently vanishes at $r \simeq\ 8 r_{e}$ (see Fig.\ref{col_all}).

To conclude, we illustrated the performance and accuracy achieved with the VST/OmegaCAM to produce surface photometry of early-type galaxies also in very extreme conditions. For the case of NGC 4472 the extended halo around the giant galaxy, reaching the edge of the one-square-degree field of view, has allowed us to fully test the procedure for data reduction and background subtraction.  
The results obtained with our observations are similar in accuracy to the collection of observations from different telescopes (see K+09). In the future we expect to implement a more general surface analysis including a wider set of photometric laws (Sersic, cored Sersics, double de Vaucouleurs, etc.) to characterize the SB measurements in a larger sample of galaxies and thus discuss results in the context of galaxy formation theories. Moreover, a forthcoming paper based on the same data as were used in the present work will be devoted to the study of small stellar systems (e.g., GCs and UCDs).

\begin{acknowledgements}
     The optical imaging is collected at the ESO VLT Survey Telescope on Cerro Paranal under program ID 090.B-0414(D) and 091.B-0614(A), using the
  Italian INAF Guaranteed Time Observations.
The data reduction for
  this work was carried out with the computational infrastructures of
  the INAF-VST Center at Naples (VSTceN). The authors wish to thank
  L. Ferrarese, C. Mihos, and S. Janowiecki for the data provided, C. Tortora for illuminating discussion on colors, and R. Peletier for the ongoing discussion on the best way to remove the sky background. This research made
  use of the NASA/IPAC Extragalactic Database (NED), which is operated
  by the Jet Propulsion Laboratory, California Institute of
  Technology, under contract with the National Aeronautics and Space
  Administration, and has been partly supported by the PRIN-INAF ``Galaxy evolution with the VLT Survey Telescope (VST)'' (PI A. Grado). M. Cantiello acknowledges support from Progetto FSE Abruzzo ``Sapere e Crescita''.
\end{acknowledgements}



\begin{appendix} 
\section{Data reduction}\label{app}
\subsection{Overscan correction and master bias}
For each exposure, the median value of an overscan region is computed and then subtracted, row by row. 
Then a masterbias, created as a sigma-clipped ($5\sigma$) average of (at least) ten bias frames, is subtracted from all the other scientific and technical exposures for full 2D bias removal.

\subsection{Flat-fielding}
The conversion from photons to ADUs, called gain, varies over the whole camera frame, owing to the optical design, pixel response, and electronics behavior. 
In principle, an exposure of a uniformly illuminated field is sufficient to build a gain-variation map. 
We use exposures of the sky at twilight. 
Because of the wide field of the instrument, these twilight flat fields may suffer from illumination variations amounting to some percent units on a degree scale. 
This undesired effect is mitigated by the illumination-correction procedure described below (Sect. \ref{IC}). 

A master flat-field is created by averaging a set of twilight flat-fields (typically five); a sigma-clipping rejection procedure helps removing  non-stationary  features. 
The method tracks the gain variations at high spatial frequencies well, but sometimes fails at low frequencies.  
The reason may be the color and flux mismatch between twilight and science exposures. 
In this case, the twilight flat-fields are combined with some science images taken during the same night with exposure times similar to those of the images under correction. 
This type of frame combination has been used to process the NGC 4472 exposures. Specifically, we applied the formula
\begin{equation}\label{mflat}
MasterFlat^i = \frac{MFlat^i}{<MFlat^i>} \times Gain^i \times IC^i,
\label{MF1}
\end{equation} 
where\begin{equation}
MFlat ^i= \frac{SFlat_{low}^i}{<SFlat_{low}^i>} \times \frac{TFlat^i}{TFlat_{low}^i}.
\label{MF2}
\end{equation}
The superscript indicates the $i^{th}$ CCD, the subscript {\it low} is for the low-frequency spatial component  obtained by applying a low-pass spatial  filter in the Fourier space. 
The  master  twilight (TFlat) and master skyflat (SFlat) are  produced  using  a sigma-clipped average of overscan- and bias-corrected twilight frames and sky frames, respectively. 
The choice of the exposures used to produce the master skyflat requires special care. 
The dithering pattern of the exposures must be wider than the largest structure in the images (such as galaxies or the halo of bright stars) to avoid fictitious gain variations. Moreover, all the bright features in the science images (galaxies stars, halos, etc.) are accurately masked.
In all these formulas, chevron brackets indicate medians done on a $1000\times 2000$ pixel central spot in the CCDs.

The terms $Gain^i \times IC^i$ accounts  for the average CCD gain and for the illumination correction and is described in Sect. \ref{IC}.

\subsection{Defringing}

The  {\it i} -band images need a correction for the fringe pattern caused by thin-film interference of sky emission lines in the detector. This is an additive component and, as such, it must be subtracted. The first step of the defringing is determining the fringing pattern by the formula

\begin{equation}
{\it frP} = \frac{SFlat}{TFlat}\times <TFlat>-{\it imsurfit}(\frac{SFlat}{TFlat}\times <TFlat>),
\end{equation} 
where {\it imsurfit} indicates a fifth-order surface Chebyshev polynomial fit.

Once the pattern is found, it must be subtracted from the science image, 
\begin{equation}
Im_{defring}=Im_{fring}- fr_{scale} \times frP
\end{equation}
using a scale factor, $fr_{scale}$, that is derived as follows. 
We assume that the fringe-pattern features are quite stable in time. We have then a priori determined the regions in the OmegaCAM frame where they clearly stand out. The best scale factor minimizes within these regions the absolute differences between peak and valley values in the fringe-corrected image.

\subsection{Gain harmonization} \label{GH}
The gain harmonization procedure sets the photometric zero point over the whole OmegaCAM mosaic. We derive the relative gain coefficients that minimize the background differences in adjacent CCDs. First we select a set of auxiliary scientific images belonging to the same night and having approximately the same exposure time as the science image to be calibrated. 

Each such image is heavily clipped around the median pixel level to flag out all the sources; holes created by the procedure are filled up in a subsequent step.
After overscan and bias correction, the auxiliary images are properly scaled and sigma-clipped combined. The scaling factor is calculated as the median over the scientific image divided by the median of the medians. All the holes surviving the stacking procedure are filled by interpolated values. 
The resulting image, corrected for the master twilight flat-frame, is then fitted with a third-order polynomial surface. This is used to compute 32 median values over  
subregions of $1000\times 2000$ pixels centered on each CCD. 
These values, normalized to the median of all the CCDs medians, are the relative gain corrections. The gain harmonization correction typically ranges from 0.9 to 1.17. 

\subsection{Illumination correction} \label{IC}
Another effect to be considered is the scattered light in the telescope and in the camera that is due to insufficient baffling, which produces an uncontrolled redistribution of light. 
In the presence of this additive contribution to the signal, the flat field is no longer an accurate model of the spatial detector response. 
Indeed, after flat-fielding, the image background appears perfectly flat, but the photometric response is position dependent \citep{Andersen95}. 
This bias in the flat field can be mitigated by applying the illumination correction (IC) map. 
We determine such a map by comparing our magnitude measurements of stars observed in equatorial fields with the corresponding SDSS DR8 psf magnitudes. 

\begin{figure*}
   \centering
   \includegraphics[width=4.5cm, angle=-90]{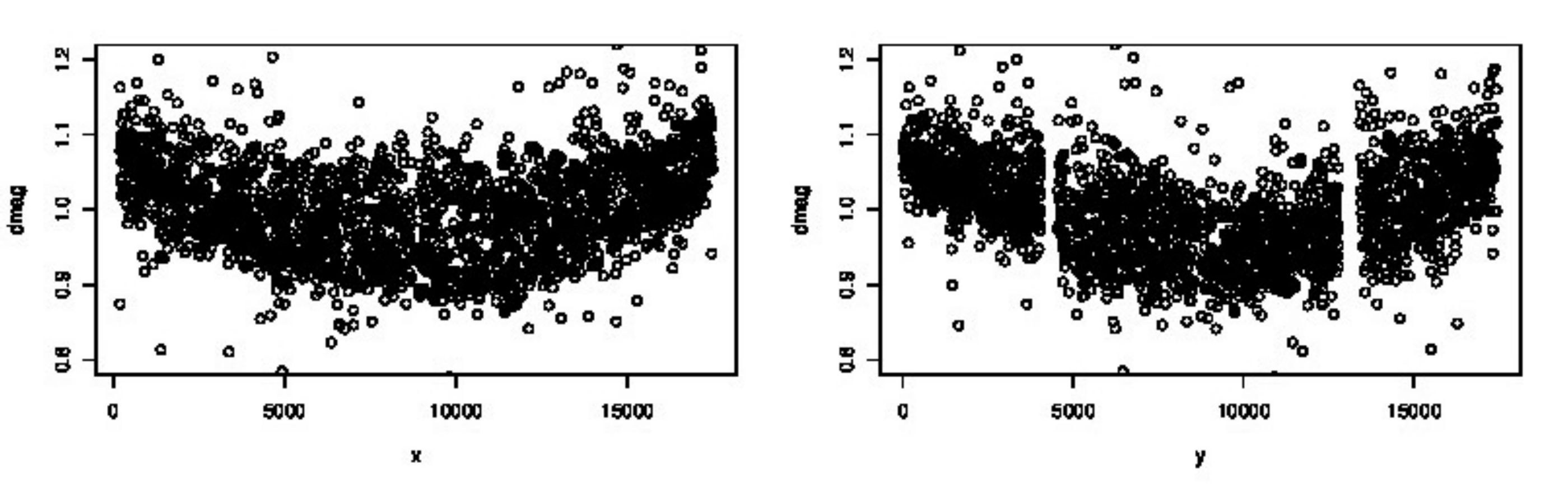}
   \caption{Differences of magnitude among observed and SDSS DR8 equatorial stars as a function of $x$ and $y$ pixel coordinates.}
              \label{Fig1}
    \end{figure*}

\begin{figure*}
   \centering
   \includegraphics[width=10cm, angle=-90]{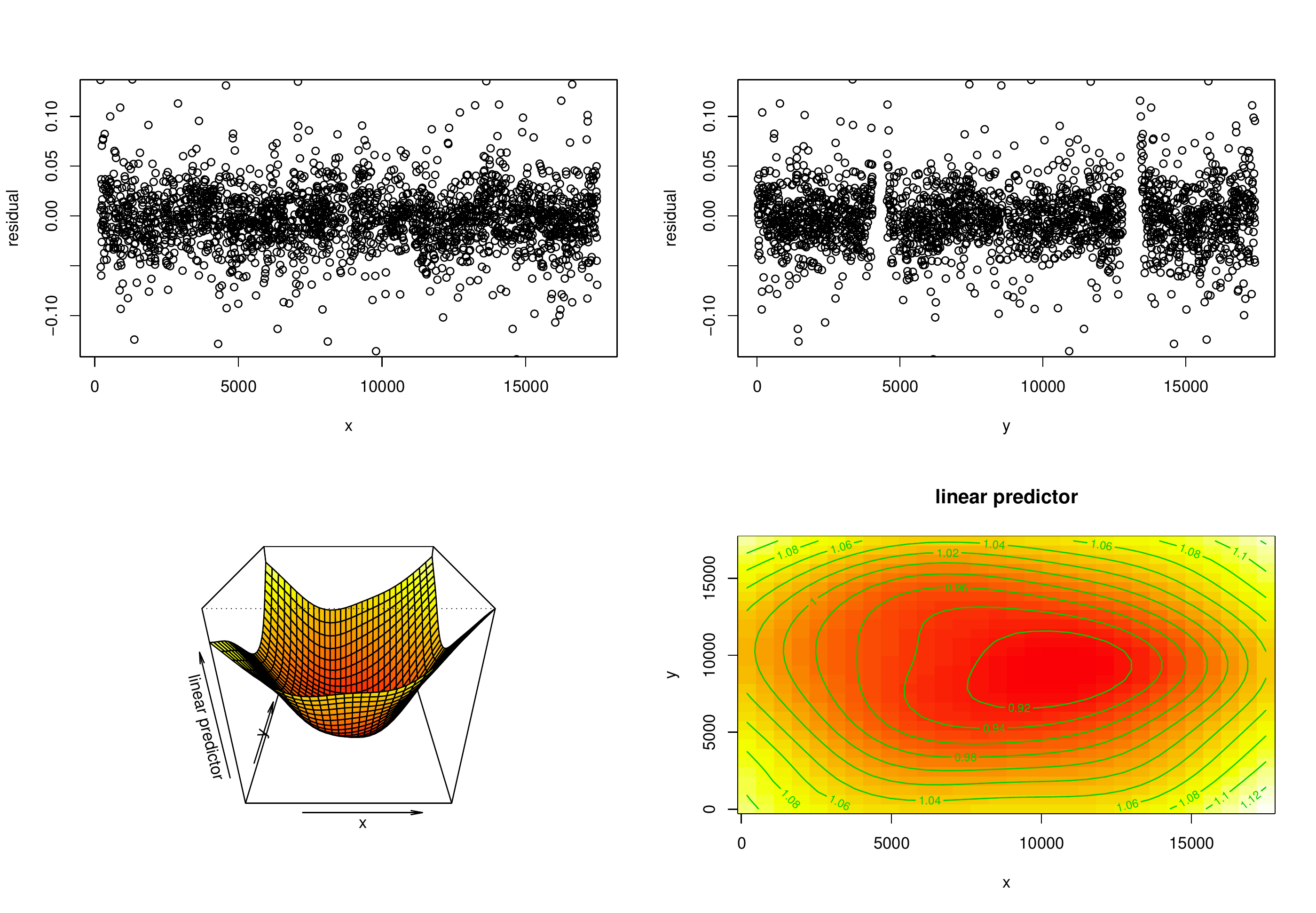}
   \caption{Differences of magnitude among observed and SDSS DR8 equatorial stars as a function of $x$ and $y$ pixel coordinates after applying the surface fit (top). {\it Bottom right panel}: contour
     plot of the IC image. {\it Bottom left panel}: IC 3D view.}
              \label{IC_map}
    \end{figure*}

The differences of magnitudes, $\Delta m(x, y)$, as a function of the position are fitted with a generalized additive model (GAM) \cite{Wood11} to derive a surface used to correct the science images during the pre-reduction stage. 
GAM also provides a well-behaved surface when the standard stars do not sample the field of view uniformly, and in general the resulting image has a smoother behavior at the frame edges than
do simple polynomial fits.  
Figures \ref{Fig1} and \ref{IC_map} illustrate the position dependency of the zero point before and after the IC application and the IC shape. The statistics on the differences in magnitude between the reference photometric catalog and the magnitude of sources before and after the illumination correction are the following:  STD = 0.09 and MAD = 0.084 before and STD 0.05 and MAD 0.026 after the correction. The IC was created using  2189 sources.
As shown in Eq. \ref{MF1}, the IC is embedded in the master flat field.
In this way, the images have a uniform zero point all over the field, but the background does not appear flat. 
To have a flat background, the properly rescaled
IC\ surface is also subtracted from the images.

\begin{table}
\caption{\label{pho}Absolute photometric calibration for NGC 4472.} \centering
\begin{tabular}{lccccc}
\hline\hline
Band & Zero Point & Color term {\it (g-i)} & Extinction \\
\hline
{\it g} & 24.864 $\pm$ 0.006 & 0.027 $\pm$ 0.006 & 0.180 $\pm$ 0.0 \\
{\it i} & 24.160 $\pm$ 0.006 & -0.004 $\pm$ 0.005 & 0.043 $\pm$ 0.0 \\
\hline
\end{tabular}
\end{table}

\subsection{Photometric and astrometric calibration}\label{calib}
In VST-tube, the absolute photometric calibration is performed by observing standard star fields each night and comparing their OmegaCAM magnitudes with SDSS DR8 photometry. For the data analyzed in this work, the absolute photometric calibration was derived using 4392 sources in the {\it g} and 4489 in the {\it i} band.
For each night and band, the zero point (ZP) and color term were obtained using the tool Photcal provided by Mario Radovich \citep{mario}. 
The extinction coefficient was derived from the extinction curve M.OMEGACAM.2011-12-01T16:15:04.474 provided by ESO.
Table \ref{pho} lists the fitted values for the zero points and color terms obtained for the nights used for the absolute photometric calibration. 

Relative photometric correction among the exposures was obtained by minimizing the quadratic sum of magnitude differences between overlapping detections. 
The tool used for this task was SCAMP \citep{Bertin06}. 
The final coadded images were then normalized to an exposure time of one second of time and a ZP of 30 magnitudes.

The absolute and relative astrometric calibrations were performed using SCAMP. 
For the absolute astrometric calibration we refer to the 2MASS catalog. Compared to this catalog, the rms of the residuals after the astrometric correction has been applied  is 0.28''. The rms on the residuals of the differences between coordinates of overlapping detections, that is, the internal astrometric accuracy,  is  0.09''.
The image resampling for the application of the astrometric solution and final image coaddition is made with the program SWARP \citep{Bertin02}. 

\end{appendix}

\begin{appendix} 

\section{Convolution by the scattering profile of the point spread function}\label{PSF1}
To evaluate the contribution of the scattered light, which is indeed a reason of concern for the surface photometry of galaxy outskirts, we first derived an extended stellar point spread function (PSF) by combining the unsaturated azimuthally averaged light profiles of stars of different luminosities, properly shifted in magnitudes. Our interest is not in the seeing profile, that is, in the inner few arcseconds of the PSF, but instead in the wings produced by the scattering in the mirror and in the atmosphere \citep{Capaccioli83}.

\begin{figure*}
   \centering
   \includegraphics[width=10cm]{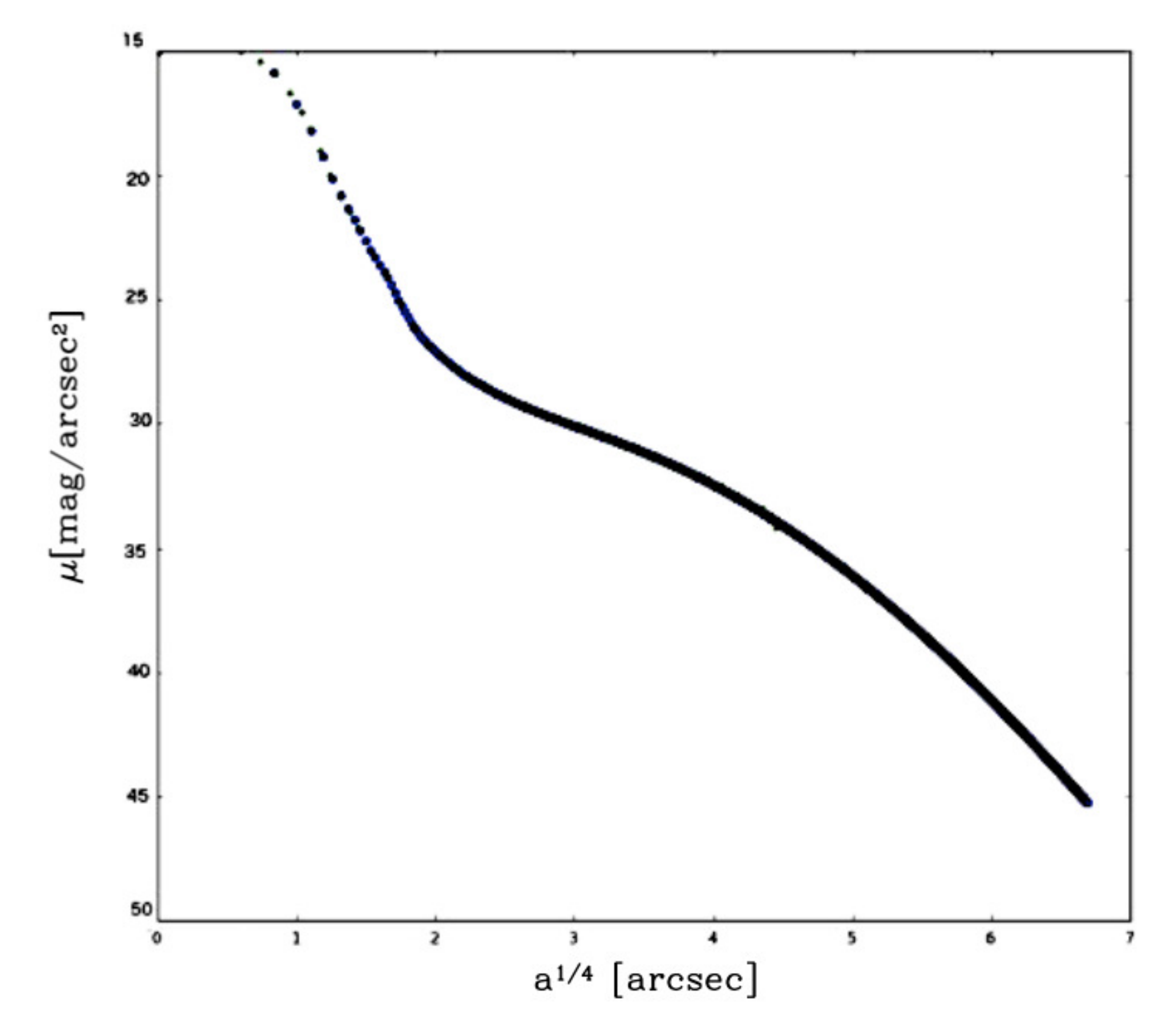}
   \caption{Adopted extended PSF for VST.}
              \label{psf}
    \end{figure*}

The measured PSF profiles for the {\it g} and {\it i} band are shown in Fig. \ref{psf}, normalized to unity up to the last observed point. Although the inner PSF has an average behavior that is
uncorrelated with the actual seeing of each of the images contributing to the final mosaic, it could not be used for deconvolving the inner regions of the galaxy. Nonetheless, it must be kept just for providing a way to normalize the PSF itself.

To extend the PSF beyond the observational limits, we adopted the polynomial expansion of \citet{Capaccioli83}, which was used to interpolate the total PSF profile (see Fig. \ref{psf}):
\begin{equation}
PSF = c_{0} + \sum\limits_{i=1}^3 c_{i} (log\ r)^{i},
\label{psf_form}
\end{equation} where $c_{0} = 2.187 \times 10^{-6}$ (mag/arcsec$^{2}$), $c_{1} = 1.725 \times 10^{-5}$, $c_{2} = -8.559 \times 10^{-6}$ and $c_{3} = 1.570 \times 10^{-6}$ for the {\it g} band and $c_{0} = -9.497 \times 10^{-4}$ (mag/arcsec$^{2}$), $c_{1} = 2.109 \times 10^{-3}$, $c_{2} = 1.157 \times 10^{-3}$ and $c_{3} = 2.134 \times 10^{-4}$ for the {\it i} band.
As expected, the {\it g} PSF spans a wider range than in the {\it i} band. The total integrated energy included in the inner regions, $r^{1/4} \leq 2.3$, is 94 $\%$ of the total flux from the stars for the {\it g} band.

The expressions \ref{psf_form} were used to estimate the effect of the scattered light in the outskirts of the galaxies of this study, which have quite different sizes. It is indeed expected that the effect will be quite different at the same surface brightness level between angularly large and small galaxies.

Two methods were employed.
The first method is a plain numerical convolution of each galaxy modeled through azimuthally averaged light profiles under the assumptions that the isophotes are ellipses of average flattening and no twisting. At first order, the difference between the model and its convolution provides an estimate of the excess of light in the observed galaxy caused by the broad smearing of the extended PSF. 

Another method consists of a straightforward deconvolution of the noiseless model of the galaxy by the extended PSF. To this end, we used the {\small IRAF} task {\small LUCY}. The two methods provide very consistent results that will be illustrated in a forthcoming paper (Spavone et al., in preparation). 

One additional comment is in order about the effect of the background interpolation on the partial removal of the excess of light that
is due to PSF scattering. It is expected and verified numerically that small galaxies will be widely broadened by the PSF wings. If this causes the outer light profile to become much flatter, one may expect that the background interpolation procedure is capable of removing part of it, if not all. This is precisely what our numerical experiments show (Spavone et al., in preparation).

\end{appendix}

\begin{appendix}
\section{Online tables}
\longtab{1}{
\setlength{\tabcolsep}{2.5pt}
\begin{longtable}{lcccccccccccccc} 
\caption{\label{Tab_vegas} The VEGAS sample.}\\
\hline\hline
\tiny
Name&R.A.&Dec.&Type&T-type&P.A.&$\mu_{e}$&$B_{t}$&(B-V)&$\sigma$&v&ag&$B_{tc}$&$M_{B}$\\
          & [deg]&[deg]&        &           & [deg]& [mag/arcsec$^{2}$]& [mag]& [mag] &[km/s]& [km/s]&[mag] &[mag] & [mag]\\
\hline
\endfirsthead
\caption{continued.}\\
\hline\hline
Name&R.A.&Dec.&Type&T-type&P.A.&$\mu_{e}$&$B_{t}$&(B-V)&$\sigma$&v&ag&$B_{tc}$&$M_{B}$\\
          & [deg]&[deg]&        &           & [deg]& [mag/arcsec$^{2}$]& [mag]& [mag] &[km/s]& [km/s]&[mag] &[mag] & [mag]\\
\hline
\endhead
\hline
\endfoot
ESO075-028&21.8495056&-71.4132778&E&-4.8&19.64&21.154&13.389&1.08&205.49&3928&0.117&13.213&-20.425\\
ESO137-010&16.263925&-60.8030833&E-S0&-3.3&167.73&22.841&12.571&-&215.79&3407.5&1.028&11.491&-21.839\\
ESO137-045&16.8507634&-60.8087988&E&-4.9&27.47&21.485&13.452&-&210.4&3292.2&1.028&12.374&-20.9\\
ESO138-005&16.8981651&-58.7781169&E-S0&-3&143.57&21.021&12.838&-&349.44&2632.8&0.577&12.221&-20.549\\
ESO183-030&18.9488541&-54.5456659&E-S0&-3.3&23.25&20.42&12.619&0.97&172.81&2724.8&0.367&12.211&-20.628\\
ESO194-021&0.4949153&-51.5206693&E-S0&-3&100.31&20.817&13.612&-&228.59&3439&0.069&13.492&-19.838\\
ESO270-014&13.4742477&-44.1719654&E-S0&-3.1&33.29&21.474&13.932&1.14&171.8&3871.9&0.535&13.339&-20.323\\
ESO322-038&12.6384452&-41.5033989&E-S0&-3&84.22&20.721&13.787&-&216.39&3126.3&0.576&13.164&-20.024\\
ESO322-051&12.681667&-41.6066322&E-S0&-2.8&51.1&20.906&14.026&1.15&205.69&3237.1&0.625&13.353&-19.893\\
ESO323-015&12.8621611&-38.8291944&E-S0&-3&153.97&21.802&13.91&-&153.54&3058.5&0.362&13.502&-19.626\\
ESO423-024&5.5780811&-29.2322714&S0&-2.1&-9&20.712&13.213&-&176.01&3937.3&0.111&13.043&-20.619\\
ESO428-011&7.2586941&-29.3589361&E-S0&-3.1&23.25&21.197&12.94&1.05&170.16&2115.7&0.812&12.097&-20.157\\
ESO499-023&9.9404727&-26.0949238&E-S0&-3.2&102.03&20.674&12.925&-&213.6&2512&0.307&12.58&-20.118\\
ESO507-021&12.8412924&-26.8425635&E-S0&-2.5&27.57&20.964&13.396&-&202.71&3167.6&0.333&13.015&-20.231\\
ESO507-025&12.8588463&-26.4521027&E-S0&-3&99.37&21.186&12.612&1&260.24&3235.3&0.387&12.177&-21.125\\
ESO567-051&10.3340084&-21.5186201&E&-4.5&170.55&21.145&14.159&0.96&151.81&3706.1&0.234&13.87&-19.72\\
IC1459&22.9529445&-36.4621765&E&-4.8&43.1&20.62&10.949&0.98&306.1&1794.5&0.071&10.852&-21.064\\
IC2311&8.3127802&-25.3695888&E&-4.8&-9&21.324&12.494&1.01&224.35&1844&0.622&11.844&-20.124\\
IC2552&10.1794778&-34.8447266&E-S0&-2.8&89.29&21.803&13.085&1.03&159.5&3113.9&0.518&12.521&-20.638\\
IC2586&10.5173341&-28.7165771&E&-4.9&85.13&21.389&13.564&1&346.02&3673.3&0.272&13.237&-20.304\\
IC2594&10.6011578&-24.3230111&E-S0&-3&111.53&21.696&13.499&-&216.74&3546.9&0.266&13.18&-20.31\\
IC2597&10.6298222&-27.0812518&E&-3.9&7.99&21.841&12.924&1&257.97&2995.2&0.306&12.573&-20.525\\
IC3370&12.4603126&-39.3379207&E&-4.9&52.86&21.747&11.997&0.97&204.49&2937.5&0.401&11.552&-21.484\\
IC3896&12.9453372&-50.3467167&E&-4.8&6.36&21.903&12.172&1.17&203.29&2052.5&0.914&11.226&-20.936\\
IC4197&13.1345203&-23.7969418&E-S0&-3.1&162.33&21.164&13.546&1.07&185.99&3013.2&0.628&12.873&-20.286\\
IC4296&13.6108474&-33.9658219&E&-4.9&41.57&21.785&11.576&1.01&332.81&3781.4&0.276&11.244&-22.395\\
IC4421&14.4753451&-37.5835389&E&-4.7&164.35&21.367&13.391&1.02&202.28&3636.4&0.347&12.989&-20.551\\
IC4797&18.9415806&-54.3058612&E&-3.9&148.1&20.467&12.263&1.01&212.13&2715.3&0.336&11.887&-20.953\\
IC4889&19.7542178&-54.3442167&E&-4.7&2.03&20.529&12.058&0.95&181.88&2554.2&0.229&11.791&-20.907\\
IC4943&20.1078543&-48.3756503&E&-4.9&-9&21.043&13.622&0.95&165.25&2913.6&0.216&13.363&-19.642\\
IC5011&20.4760662&-36.0272354&S0&-2&17.77&20.043&12.683&-&212.81&2289.9&0.184&12.465&-20.042\\
IC5063&20.8673137&-57.0688469&S0-a&-1.2&119.98&21.984&12.962&1.02&160.91&3383.1&0.267&12.644&-20.686\\
IC5181&22.2226958&-46.0176284&S0&-2&73.1&19.507&12.464&0.97&-&1980.4&0.086&12.349&-19.766\\
IC5250A&22.7881584&-65.0608292&E-S0&-2.8&-9&21.897&12.706&-&190&3581&0.141&12.512&-20.926\\
IC5267&22.953758&-43.3960525&S0-a&-1.1&137.55&21.612&11.391&0.89&-&1714.9&0.054&11.311&-20.443\\
IC5328&23.5545553&-45.0160034&E&-4.2&40.87&21.144&12.268&0.96&195.43&3138&0.064&12.157&-21.002\\
NGC0474&1.3351857&3.4153385&S0&-2&-9&21.979&12.383&0.86&157.12&2371.6&0.15&12.197&-20.463\\
NGC0584&1.5224182&-6.8680509&E&-4.7&104.48&20.364&11.326&0.96&199.34&1849.8&0.183&11.114&-20.953\\
NGC0636&1.6518172&-7.5125923&E&-4.9&5.84&20.801&12.351&0.95&165.06&1854.7&0.109&12.214&-19.853\\
NGC0720&1.8834808&-13.7385764&E&-4.9&141.9&20.921&11.148&0.98&241.14&1717.3&0.069&11.053&-20.81\\
NGC0731&1.915624&-9.0108753&E&-4.1&155.5&21.032&13.069&0.93&157.22&3881.2&0.094&12.916&-20.793\\
NGC0936&2.4603827&-1.1559362&S0-a&-1.2&132.45&21.368&11.195&0.97&179.33&1368.2&0.153&11.022&-20.367\\
NGC1052&2.6846621&-8.2558045&E&-4.7&109.01&21.055&11.454&0.94&213.31&1483.8&0.115&11.316&-20.201\\
NGC1162&2.982225&-12.3985278&E&-4.9&-9&-9&13.806&-&194.85&3936.2&0.205&13.542&-20.168\\
NGC1199&3.0606951&-15.6138377&E&-4.8&47.67&21.69&12.388&1.02&203.87&2681.6&0.24&12.109&-20.731\\
NGC1201&3.0688995&-26.0696616&E-S0&-2.6&8.08&20.67&11.703&0.94&163.82&1680.3&0.067&11.61&-20.087\\
NGC1209&3.1008348&-15.6112629&E&-4.5&83.25&20.623&12.35&0.96&229.75&2641&0.16&12.151&-20.654\\
NGC1316&3.3782565&-37.2082112&S0&-1.8&49.65&-9&9.409&0.89&224.54&1788.3&0.091&9.292&-22.517\\
NGC1332&3.4381105&-21.3352765&E-S0&-2.9&121.49&20.127&11.198&0.96&320.86&1526.2&0.14&11.035&-20.483\\
NGC1340&3.4721166&-31.0681243&E&-4&167.2&-9&11.218&0.88&166.02&1183.1&0.079&11.121&-19.662\\
NGC1380&3.6076629&-34.9761257&S0&-2.3&6.23&-9&10.896&0.94&211.01&1874.1&0.073&10.795&-21.121\\
NGC1387&3.6158538&-35.5066275&E-S0&-2.8&-9&20.549&11.772&0.99&170.22&1260.5&0.054&11.699&-19.252\\
NGC1395&3.6415799&-23.0274642&E&-4.9&104.5&21.371&10.601&0.96&244.67&1701.4&0.1&10.476&-21.278\\
NGC1399&3.641408&-35.4506257&E&-4.6&-9&-9&10.426&0.96&336.04&1425.7&0.055&10.35&-20.902\\
NGC1400&3.6585692&-18.6881481&E&-3.7&42.93&21.09&11.933&0.96&251.52&589.9&0.28&11.644&-20.436\\
NGC1404&3.6477546&-35.5942446&E&-4.8&163.3&-9&10.891&0.97&228.35&1946.3&0.049&10.812&-21.206\\
NGC1407&3.6699637&-18.5803554&E&-4.5&-9&22.053&10.701&1.03&270.65&1791.4&0.296&10.378&-21.485\\
NGC1426&3.7136417&-22.1083611&E&-4.9&112.49&21.291&12.271&0.9&150.74&1444.7&0.071&12.179&-19.21\\
NGC1427&3.7053874&-35.3927754&E&-4&77.85&-9&11.777&0.91&155.99&1388.2&0.051&11.705&-19.476\\
NGC1439&3.7472206&-21.9207205&E&-4.8&-9&21.971&12.29&0.88&150.85&1667.9&0.125&12.141&-19.557\\
NGC1453&3.7742559&-3.968888&E&-4.7&19&21.556&12.586&1.05&331.89&3906.6&0.456&12.071&-21.638\\
NGC1537&4.2279746&-31.6452899&E&-3.6&102.31&20.539&11.472&0.89&159.3&1406.8&0.104&11.347&-19.905\\
NGC1549&4.2625382&-55.5922874&E&-4.3&138.25&21.092&10.678&0.93&202.69&1243.4&0.053&10.606&-20.177\\
NGC1550&4.3272&2.4098611&E&-4.1&27.75&22.156&13.163&1.08&307.97&3785.2&0.576&12.53&-21.132\\
NGC1553&4.2695796&-55.780024&S0&-2.3&150.1&20.906&10.285&0.88&177.25&1148.4&0.064&10.203&-20.398\\
NGC1587&4.5110844&0.6616683&E&-4.8&72.82&21.531&12.721&1.02&227.14&3671.7&0.311&12.355&-21.235\\
NGC1588&4.5121711&0.6647302&E&-4.8&164.55&21.39&13.893&0.98&152.53&3484.9&0.309&13.532&-19.932\\
NGC1596&4.460582&-55.027671&S0&-2&19.23&19.872&12.009&0.94&172.02&1511.4&0.042&11.944&-19.377\\
NGC1700&4.9489833&-4.8659744&E&-4.7&86.97&20.421&12.03&0.97&238.71&3891.4&0.187&11.784&-21.925\\
NGC1726&4.9949816&-7.7553421&S0&-2.4&0.5&21.531&12.679&0.98&247.72&3991.3&0.303&12.316&-21.439\\
NGC1993&5.5904361&-17.8152222&E-S0&-3.2&77.35&21.134&13.621&-&150.51&3140.3&0.295&13.279&-19.909\\
NGC2073&5.7649753&-21.9990983&E-S0&-3&-9&21.017&13.454&-&152.78&2983.8&0.128&13.282&-19.785\\
NGC2089&5.7976106&-17.6024234&E-S0&-2.9&41.75&20.44&13.04&-&206.01&3001.4&0.287&12.708&-20.359\\
NGC2217&6.3610332&-27.2336632&S0-a&-0.6&-9&21.966&11.633&1&220.83&1617&0.187&11.422&-20.157\\
NGC2271&6.7147194&-23.4759167&E-S0&-3.1&78.49&21.947&13.201&-&228.16&2596&0.507&12.654&-20.079\\
NGC2293&6.7951824&-26.7539001&S0-a&-1&129.8&22.023&12.233&1.07&262.62&2037&0.517&11.685&-20.477\\
NGC2305&6.8103583&-64.2733333&E&-4.9&137.24&21.08&12.757&1.02&246.92&3570&0.332&12.372&-21.013\\
NGC2325&7.04455&-28.69725&E&-4.8&4.2&22.772&12.298&-&183.57&2157.2&0.498&11.768&-20.53\\
NGC2380&7.3985&-27.5291&S0&-2&-9&21.364&12.291&1.05&191&1782&1.331&10.934&-20.929\\
NGC2434&7.5808894&-69.2842123&E&-4.8&139.56&21.553&12.338&1.07&186.89&1449.7&1.069&11.247&-19.934\\
NGC2663&8.7522889&-33.79475&E&-4.9&110.6&22.45&11.89&-&292&2156&1.559&10.299&-21.999\\
NGC2695&8.9075195&-3.0670221&S0&-2.2&160.55&20.946&12.843&0.95&199.85&1834.3&0.076&12.74&-19.327\\
NGC2822&9.2305596&-69.6448861&E&-4.5&93.8&21.485&11.557&-&156.32&1614.9&0.414&11.119&-20.335\\
NGC2865&9.3917154&-23.1616091&E&-4.2&154.64&20.2&12.446&0.91&171.96&2722.1&0.361&12.045&-20.828\\
NGC2887&9.3900028&-63.8125833&E-S0&-3.1&79.87&21.791&12.763&1.1&281.95&2874.5&0.983&11.737&-21.17\\
NGC2902&9.5146999&-14.7358814&S0&-2&21.78&20.766&13.237&-&-&1993.1&0.283&12.923&-19.285\\
NGC2904&9.5047205&-30.3849959&E-S0&-3&89.13&21.362&13.52&1.06&234.45&2371.4&0.566&12.919&-19.628\\
NGC2974&9.709242&-3.698761&E&-4.3&43.99&21.095&11.876&1&237.13&1891.4&0.235&11.613&-20.549\\
NGC2983&9.7280895&-20.4772017&S0-a&-0.8&86.42&-9&12.736&0.985&173.16&2029.9&0.261&12.446&-19.763\\
NGC2986&9.7377826&-21.2781478&E&-4.7&-9&21.479&11.692&0.97&259.17&2326.1&0.252&11.405&-21.141\\
NGC3078&9.973478&-26.9262608&E&-4.8&177.2&21.084&12.067&1.01&252.65&2563.6&0.31&11.719&-21.015\\
NGC3082&9.9814223&-30.3577716&E-S0&-2.8&28.7&20.239&13.576&-&191.73&2805.9&0.345&13.189&-19.751\\
NGC3087&9.9857443&-34.2251812&E&-4.3&46.05&20.623&12.63&1.05&184&2636.7&0.464&12.127&-20.643\\
NGC3091&10.0039505&-19.6362038&E&-4.8&144.43&21.742&12.126&1&321.37&3809&0.188&11.881&-21.757\\
NGC3100&10.0113312&-31.6642967&S0&-2&148.22&21.846&12.031&0.89&199.92&2583.6&0.323&11.67&-21.064\\
NGC3108&10.0414028&-31.6774686&S0-a&-1.1&58.71&21.713&12.756&1.03&203.62&2669.1&0.343&12.373&-20.432\\
NGC3115&10.0872139&-7.7185556&E-S0&-2.9&42.51&19.537&10.082&0.97&261.08&648.6&0.201&9.871&-19.939\\
NGC3136&10.0967083&-67.3779722&E&-4.9&27.24&21.442&11.697&1.01&228.88&1706&1.037&10.634&-21.005\\
NGC3136B&10.17025&-67.0050833&E&-3.7&23.25&21.844&12.774&0.98&172.82&1782.9&0.862&11.885&-19.869\\
NGC3250&10.4422974&-39.9439207&E&-4.9&139.67&20.971&12.162&1.05&267.08&2816.6&0.455&11.664&-21.243\\
NGC3258&10.4815547&-35.6053845&E&-4.3&67.34&21.863&12.502&1.01&263.27&2800.1&0.349&12.111&-20.796\\
NGC3260&10.4851259&-35.5951366&E&-4.9&9.08&20.741&13.733&1.06&204.46&2427.7&0.366&13.33&-19.255\\
NGC3268&10.5001969&-35.3255622&E&-4.3&70.82&22.048&12.324&1.05&230.41&2798.6&0.444&11.838&-21.069\\
NGC3271&10.5072837&-35.3586217&S0&-1.8&109.24&21.003&12.822&1.09&255.26&3800.7&0.472&12.294&-21.321\\
NGC3305&10.6032811&-27.1622212&E&-4.9&-9&20.729&13.748&1.02&221.96&3933.7&0.339&13.35&-20.359\\
NGC3308&10.6062169&-27.4378684&E-S0&-3&34.4&22.229&13.39&1.03&189.66&3555.2&0.338&12.999&-20.491\\
NGC3311&10.6118836&-27.5278788&E-S0&-3.3&-9&23.867&12.799&1&185.17&3835.5&0.344&12.397&-21.265\\
NGC3315&10.6220067&-27.1912561&E-S0&-2.9&-9&21.866&14.361&1.05&179.47&3755.8&0.353&13.951&-19.663\\
NGC3497&11.1216884&-19.4715273&S0&-1.8&56.67&-9&12.996&-&224.46&3701&0.173&12.767&-20.823\\
NGC3557&11.166028&-37.539245&E&-4.9&32.4&21.105&11.405&1.03&267.72&3056.9&0.436&10.924&-22.205\\
NGC3585&11.2214178&-26.754864&E&-4.8&103.97&20.928&10.819&0.97&205.67&1373.4&0.276&10.522&-20.8\\
NGC3606&11.2710034&-33.8274951&E&-4.8&-9&20.571&13.399&-&208.18&3000.6&0.322&13.032&-20.066\\
NGC3640&11.3519162&3.234786&E&-4.9&97.5&21.095&11.331&0.92&192.27&1315.2&0.185&11.127&-20.327\\
NGC3706&11.4956778&-36.3912997&E-S0&-3.2&78.38&21.116&12.352&1.04&270.27&2979.6&0.406&11.902&-21.166\\
NGC3818&11.6992674&-6.1555814&E&-4.6&95.93&21.452&12.709&0.96&194.78&1696.1&0.156&12.528&-19.44\\
NGC3904&11.8203344&-29.2768138&E&-4.8&11.35&20.692&11.795&0.98&205.48&1576.1&0.31&11.461&-20.178\\
NGC3923&11.8504901&-28.8059748&E&-4.8&48&21.553&10.767&1&256.61&1550&0.35&10.394&-21.245\\
NGC3962&11.9111368&-13.9749309&E&-4.8&10&21.333&11.608&0.95&232.98&1854.5&0.193&11.388&-20.727\\
NGC4024&11.9753387&-18.3468954&E-S0&-3&63.78&20.882&12.699&0.94&150.57&1690.8&0.182&12.492&-19.372\\
NGC4105&12.11134&-29.7602517&E&-4.7&136.43&21.548&11.567&0.95&261.81&1873.4&0.269&11.27&-20.797\\
NGC4179&12.2144722&1.2996667&S0&-1.9&146.98&20.656&11.839&0.92&168.93&1261.7&0.143&11.677&-19.712\\
NGC4373&12.4216498&-39.7596561&E-S0&-3.3&53.87&21.529&11.904&0.98&247.52&3421.8&0.331&11.522&-21.863\\
NGC4546&12.5915281&-3.7932771&E-S0&-2.7&89.68&20.654&11.353&0.98&195.22&1054.2&0.146&11.191&-19.839\\
NGC4636&12.7137983&2.6876184&E&-4.8&149.65&-9&10.429&0.94&200.04&925.3&0.124&10.291&-20.492\\
NGC4643&12.7222582&1.9782772&S0-a&-0.6&131.66&20.484&11.674&0.96&150.1&1329.1&0.132&11.52&-19.998\\
NGC4645&12.7361061&-41.7499317&E&-3.9&46.21&21.089&12.891&1.06&188.92&2613.3&0.591&12.261&-20.509\\
NGC4645B&12.7253256&-41.3624578&S0&-2&157.9&21.285&13.48&-&183.84&2308.2&0.815&12.631&-19.836\\
NGC4696&12.8136889&-41.3111281&E&-3.8&87.79&-9&11.654&-&256.08&2973.9&0.483&11.127&-21.941\\
NGC4696B&12.789364&-41.2375561&E-S0&-2.9&40.15&20.559&13.786&1.08&243.62&2831.8&0.503&13.241&-19.699\\
NGC4697&12.8099705&-5.8006924&E&-4.5&83.07&21.664&10.25&0.91&168.12&1240.5&0.128&10.104&-21.218\\
NGC4751&12.8807793&-42.6600013&E-S0&-2.8&174.58&21.386&13.095&-&349.18&2101.9&0.522&12.541&-19.712\\
NGC4767&12.8980417&-39.7143333&E&-4&141&21.184&12.52&1.04&212.57&3060&0.464&12.01&-21.118\\
NGC4783&12.9101591&-12.5583986&E&-4.9&133.5&21.474&12.825&-&265.16&3992.9&0.23&12.535&-21.266\\
NGC4830&12.9577444&-19.6913076&E-S0&-2.8&164.26&22.237&13.086&1&180.08&3352&0.351&12.684&-20.727\\
NGC4831&12.9601913&-27.2922281&E-S0&-3.2&176.83&21.338&13.497&-&155.31&3243.3&0.421&13.027&-20.275\\
NGC4915&13.0244861&-4.5463022&E&-4.7&-9&20.043&12.89&0.89&211.21&3135.5&0.125&12.718&-20.585\\
NGC4936&13.0713803&-30.5262311&E&-4.8&6.45&22.208&11.976&-&285.45&3095&0.359&11.57&-21.618\\
NGC4946&13.0915&-43.5910833&E&-3.8&135.06&21.44&13.397&1.05&198.44&3112.6&0.499&12.851&-20.307\\
NGC4958&13.096928&-8.0201732&S0&-1.9&7.18&19.531&11.49&0.92&156.08&1294.7&0.204&11.267&-20.122\\
NGC4976&13.14372&-49.5063&E&-4.5&161.2&21.005&10.97&1.01&161.24&1411.5&0.813&10.136&-21.186\\
NGC4984&13.1492258&-15.516305&S0-a&-0.8&45&19.79&12.198&0.92&-&1214.8&0.267&11.913&-19.268\\
NGC4993&13.1632522&-23.3838872&E-S0&-3&173.2&20.399&13.452&-&169.48&2951.7&0.534&12.874&-20.254\\
NGC5011&13.2144109&-43.0961268&E&-4.8&153.97&21.663&12.396&1.02&256.28&3127&0.439&11.91&-21.278\\
NGC5017&13.2151385&-16.7657761&E&-4.2&28.5&20.482&13.544&0.97&176.85&2529.4&0.358&13.148&-19.657\\
NGC5018&13.2169453&-19.518201&E&-4.4&99.13&20.342&11.687&0.92&208.19&2843.2&0.412&11.232&-21.804\\
NGC5044&13.256655&-16.385291&E&-4.8&41&22.813&11.589&1&241.81&2692.9&0.303&11.245&-21.695\\
NGC5061&13.3014199&-26.8370999&E&-4.3&108.55&20.444&11.209&0.92&185.8&2031&0.297&10.882&-21.372\\
NGC5077&13.3254466&-12.6565178&E&-4.8&4.18&21.123&12.328&1.03&257.63&2828.7&0.212&12.074&-20.993\\
NGC5084&13.3380211&-21.8272255&S0&-2&80.4&22.64&11.599&1.11&202.77&1721.3&0.506&11.067&-20.85\\
NGC5087&13.3402759&-20.6110916&E-S0&-2.9&5.94&20.651&12.228&1.01&282.8&1814.2&0.447&11.754&-20.313\\
NGC5090&13.3535811&-43.7046333&E&-4.9&105.91&22.041&12.562&-&268.68&2946&0.614&11.904&-21.132\\
NGC5114&13.4004778&-32.3438288&E-S0&-3.1&77.51&20.788&13.641&-&200.47&3674.3&0.207&13.379&-20.186\\
NGC5140&13.4393684&-33.8685154&E-S0&-3&46.93&21.731&12.851&-&194.95&3866.7&0.247&12.546&-21.139\\
NGC5193&13.5315347&-33.2339269&E&-4.1&-9&21.472&12.579&0.93&205.67&3733.9&0.246&12.276&-21.338\\
NGC5266&13.7172545&-48.1693386&S0&-2.5&97.52&21.951&12.053&-&201.36&3006.3&0.384&11.624&-21.443\\
NGC5304&13.833746&-30.5784422&E-S0&-3.2&140.02&21.65&13.619&-&211.01&3774.7&0.241&13.322&-20.317\\
NGC5493&14.1914944&-5.0436389&S0&-2.1&-9&20.442&12.289&0.87&204.05&2695.5&0.154&12.094&-20.911\\
NGC5576&14.3510283&3.2710301&E&-4.8&89.66&20.585&11.789&0.89&184.27&1506.7&0.133&11.634&-20.23\\
NGC5638&14.4945621&3.2332952&E&-4.8&154.71&21.429&12.156&0.94&161.92&1657.4&0.141&11.99&-20.028\\
NGC5791&14.9795028&-19.2668765&E&-4.2&171.63&21.133&12.698&1.01&252.01&3346.8&0.392&12.256&-21.156\\
NGC5796&14.9900088&-16.6239335&E&-4.7&89.73&21.094&12.729&1.07&273.17&2910.3&0.451&12.234&-20.894\\
NGC5812&15.0154681&-7.4572376&E&-4.8&73.37&21.219&12.192&1.03&199.61&1917.7&0.374&11.789&-20.509\\
NGC5813&15.0198046&1.7020095&E&-4.9&142.5&22.337&11.517&0.99&238.27&1955.8&0.246&11.242&-21.142\\
NGC5831&15.0685994&1.2199386&E&-4.8&128.71&21.585&12.437&0.97&164.07&1630.9&0.253&12.159&-19.859\\
NGC5838&15.0906396&2.0994875&E-S0&-2.7&38.75&20.767&11.786&0.98&276.8&1252.1&0.23&11.537&-19.917\\
NGC5846&15.1081241&1.6062912&E&-4.7&-9&22.042&11.091&1.01&236.81&1749&0.241&10.824&-21.338\\
NGC5846A&15.108091&1.5949712&E&-4.3&111.67&-9&12.721&-&221.98&2309.9&0.242&12.445&-20.253\\
NGC5869&15.163722&0.4701069&S0&-2.2&110.7&21.624&13.148&0.96&167.78&2074&0.236&12.881&-19.626\\
NGC5898&15.3037785&-24.0980507&E&-4.3&50.75&21.236&12.438&1.06&206.88&2127.4&0.626&11.78&-20.646\\
NGC5903&15.3101444&-24.0685833&E&-4.8&164.34&22.059&12.215&1.01&206.54&2561.6&0.655&11.521&-21.318\\
NGC6305&17.300247&-59.1719975&E-S0&-2.9&136.7&20.401&13.22&1.07&155.12&2677.2&0.44&12.74&-20.065\\
NGC6758&19.2311953&-56.3095216&E&-4.2&110.85&21.325&12.606&1.04&241.89&3484.6&0.286&12.268&-21.143\\
NGC6799&19.5379184&-55.9080055&E-S0&-3.5&106.7&21.536&13.509&-&150.64&3413.5&0.267&13.192&-20.166\\
NGC6851&20.0595516&-48.2845444&E&-4.5&160.87&20.553&12.689&0.91&228.49&3049.5&0.203&12.439&-20.689\\
NGC6861&20.122079&-48.3702995&E-S0&-2.7&133.11&20.455&12.077&1.01&414&2823.9&0.234&11.8&-21.14\\
NGC6868&20.1650169&-48.3794926&E&-4.9&80.83&21.451&11.671&1.01&255.04&2949.3&0.24&11.386&-21.65\\
NGC6875&20.2201365&-46.1617223&E-S0&-2.6&22.06&20.394&12.984&0.95&-&3121.4&0.171&12.767&-20.392\\
NGC6876&20.3052897&-70.8590603&E&-4.9&110.19&22.036&11.943&-&233.44&3868.9&0.194&11.691&-21.923\\
NGC6893&20.3471223&-48.2393344&S0&-2&8.63&20.909&12.747&0.98&-&3056&0.173&12.528&-20.6\\
NGC6903&20.3958003&-19.3254017&E-S0&-3.3&-9&22.062&12.9&-&226.61&3292.3&0.289&12.562&-20.822\\
NGC6920&20.7325936&-80.0008073&S0&-2&131.52&20.861&13.193&1.22&234.96&2634.8&1.032&12.121&-20.577\\
NGC6942&20.6771732&-54.3030482&S0-a&-0.3&134.28&22.719&12.969&-&-&3274&0.205&12.688&-20.558\\
NGC6958&20.8118337&-37.9973321&E&-3.7&97.42&20.842&12.284&0.91&187.77&2719.5&0.195&12.048&-20.826\\
NGC6964&20.7900844&0.3008814&E&-4.5&164.5&20.686&13.954&1.01&206.61&3791.5&0.426&13.472&-20.261\\
NGC7020&21.1889168&-64.0253426&S0-a&-1.2&165.27&22.096&12.693&0.95&195.43&3153&0.17&12.475&-20.653\\
NGC7029&21.1978029&-49.2837797&E&-4.6&67.61&20.71&12.395&0.89&185.01&2804.1&0.16&12.193&-20.715\\
NGC7041&21.2756582&-48.3635931&E-S0&-3&83.61&20.731&12.102&0.91&225.52&1945.6&0.169&11.904&-20.163\\
NGC7049&21.3167278&-48.5608333&S0&-1.9&63.72&21.086&11.676&1.05&246.48&2261.8&0.243&11.399&-21.026\\
NGC7079&21.5431359&-44.0675693&S0&-1.8&76.69&20.763&12.485&0.87&158.9&2653.5&0.137&12.308&-20.496\\
NGC7097&21.6702463&-42.5394127&E&-4.9&17.17&20.724&12.611&0.91&221.85&2602.1&0.086&12.486&-20.283\\
NGC7135&21.8294475&-34.876334&E-S0&-2.9&44.87&22.791&12.794&0.99&-&2718&0.123&12.63&-20.243\\
NGC7144&21.8784659&-48.2539237&E&-4.8&-9&21.371&11.687&0.92&174.03&1924.4&0.089&11.569&-20.449\\
NGC7166&22.0091491&-43.389778&E-S0&-2.9&10.21&20.573&13.033&0.99&-&2451.6&0.066&12.93&-19.693\\
NGC7168&22.0353944&-51.7431111&E&-4.8&69.4&20.964&12.834&0.93&180.17&2846&0.1&12.691&-20.249\\
NGC7173&22.034261&-31.9737228&E&-4.3&140.54&22.144&12.905&0.92&203.86&2497.2&0.115&12.752&-19.945\\
NGC7176&22.0356785&-31.9900967&E&-4.8&75.1&21.549&12.434&-&253.84&2515.4&0.115&12.282&-20.416\\
NGC7192&22.1139243&-64.3161871&E&-4&-9&21.44&12.202&0.95&178.6&2959.5&0.147&12.011&-20.962\\
NGC7196&22.0985626&-50.1190894&E&-4.8&58.52&20.734&12.366&0.94&278.92&2886.5&0.095&12.227&-20.746\\
NGC7200&22.1193075&-49.995513&E&-3.7&44.89&20.759&13.774&0.91&194.45&2907.9&0.084&13.647&-19.326\\
NGC7216&22.2099498&-68.6619852&E&-4.2&127.41&21.384&13.535&0.95&172.58&3529.7&0.152&13.33&-20.055\\
NGC7302&22.5399237&-14.1205203&E-S0&-2.8&95.4&20.788&13.151&0.97&150.61&2671.5&0.309&12.802&-20.105\\
NGC7391&22.8433537&-1.5447783&E&-4.9&96.32&21.331&13.117&1.07&243.84&3045.4&0.416&12.655&-20.591\\
NGC7484&23.1180401&-36.2753442&E&-4.8&-9&21.908&12.981&-&193.19&2738.1&0.074&12.866&-20.007\\
NGC7507&23.2021016&-28.5396671&E&-4.5&-9&20.384&11.378&0.98&222.37&1606.6&0.213&11.14&-20.557\\
NGC7585&23.3003716&-4.6504582&S0-a&-1.1&94.86&21.323&12.445&0.9&214.25&3458&0.232&12.162&-21.328\\
NGC7600&23.3149747&-7.5805507&E-S0&-2.9&62.34&21.187&12.908&0.96&210.2&3441.5&0.141&12.716&-20.748\\
NGC7676&23.4838058&-59.7167044&E-S0&-3.3&86.18&20.301&13.539&0.98&198.52&3367.2&0.068&13.42&-19.854\\
NGC7702&23.5913627&-56.0122925&S0-a&-1.1&118.07&21.423&13.067&0.92&-&3227.3&0.071&12.948&-20.241\\
NGC7744&23.7497847&-42.9096325&E-S0&-2.5&108.55&20.634&12.797&0.95&-&3091.8&0.063&12.687&-20.441\\
NGC7796&23.9832784&-55.4583824&E&-4&177.96&21.203&12.44&0.97&258.84&3342.4&0.045&12.345&-20.93\\
\hline
\end{longtable}
\tablefoot {Column1: Object name. Column 2: Right ascension. Column 3: Declination. Column 4: Morphological type. Column 5: Morphological type code. Column 6: Position angle. Column 7: Mean effective surface brightness. Column 8: Total B-magnitude. Column 9: Total (B-V) color. Column 10: Central velocity dispersion. Column 11: Mean Heliocentric radial velocity (cz). Column 12: Galactic extinction in B-band. Column 13: Total B-magnitude. Column 14: Absolute B-band magnitude.}
} 

\begin{table*}
\caption{Surface photometry of NGC 4472.}\label{Tab_NGC4472}
\begin{tabular}{lcccccccccc}
\hline\hline
$a^{1/4}_{g}$ & $\mu_{g}$ & $\sigma(\mu_{g})$ & $\epsilon_{g}$ &
$P.A._{g}$ & $a^{1/4}_{i}$ & $\mu_{i}$ & $\sigma(\mu_{i})$ &
$\epsilon_{i}$ & $P.A._{i}$ \\

[arcsec] & [mag/arcsec$^{2}$] & [mag/arcsec$^{2}$] & & [deg] & [arcsec] & [mag/arcsec$^{2}$] & [mag/arcsec$^{2}$] & & [deg] \\
\hline
0.000&16.64&0.013&0.202&0.49&0.000&15.34&0.009&0.145&-88.83\\
0.586&16.64&0.007&0.202&0.49&0.586&15.34&0.006&0.145&-88.83\\
0.613&16.65&0.007&0.204&0.75&0.613&15.35&0.006&0.146&-88.53\\
0.642&16.65&0.007&0.205&1.05&0.642&15.35&0.006&0.149&-88.00\\
0.672&16.65&0.007&0.206&1.34&0.672&15.35&0.006&0.153&-87.54\\
0.703&16.65&0.007&0.098&2.83&0.703&15.35&0.006&0.054&-80.71\\
0.736&16.65&0.007&0.111&10.19&0.736&15.35&0.006&0.044&-31.71\\
0.769&16.65&0.007&0.143&10.31&0.769&15.36&0.006&0.084&-21.02\\
0.806&16.66&0.007&0.170&10.93&0.806&15.36&0.006&0.132&-21.19\\
0.843&16.66&0.007&0.132&16.65&0.843&15.37&0.006&0.117&-17.28\\
0.883&16.67&0.007&0.133&12.79&0.883&15.38&0.006&0.150&-9.12\\
0.924&16.69&0.006&0.104&3.51&0.924&15.39&0.006&0.125&-7.76\\
0.967&16.71&0.006&0.099&-2.08&0.967&15.41&0.006&0.113&-12.77\\
1.012&16.73&0.006&0.092&-7.91&1.012&15.44&0.006&0.107&-13.68\\
1.059&16.77&0.006&0.076&-15.60&1.059&15.48&0.006&0.102&-13.98\\
1.109&16.82&0.006&0.073&-14.93&1.109&15.54&0.006&0.093&-16.18\\
1.161&16.89&0.006&0.065&-14.46&1.161&15.61&0.006&0.086&-15.95\\
1.215&16.98&0.006&0.061&-13.29&1.215&15.70&0.006&0.075&-15.39\\
1.271&17.08&0.006&0.061&-14.13&1.271&15.81&0.006&0.072&-15.47\\
1.331&17.20&0.007&0.048&-18.52&1.331&15.94&0.006&0.057&-20.40\\
1.393&17.34&0.007&0.053&-19.67&1.393&16.08&0.006&0.061&-18.07\\
1.458&17.50&0.007&0.060&-16.03&1.458&16.25&0.006&0.061&-20.55\\
1.526&17.68&0.007&0.067&-19.40&1.526&16.43&0.006&0.072&-22.06\\
1.597&17.87&0.007&0.074&-18.72&1.597&16.62&0.006&0.083&-20.66\\
1.671&18.07&0.007&0.093&-18.65&1.671&16.83&0.006&0.099&-21.89\\
1.749&18.27&0.007&0.109&-20.60&1.749&17.04&0.006&0.110&-20.69\\
1.831&18.49&0.007&0.123&-19.96&1.831&17.26&0.006&0.126&-20.61\\
1.916&18.72&0.007&0.142&-20.04&1.916&17.49&0.006&0.145&-20.81\\
2.006&18.94&0.007&0.163&-21.18&2.006&17.71&0.006&0.162&-21.34\\
2.099&19.17&0.007&0.175&-21.11&2.099&17.95&0.007&0.174&-21.12\\
2.197&19.42&0.007&0.175&-20.81&2.197&18.20&0.007&0.176&-20.81\\
2.299&19.70&0.007&0.172&-20.92&2.299&18.48&0.007&0.175&-21.63\\
2.407&20.01&0.008&0.164&-20.88&2.407&18.79&0.007&0.166&-21.34\\
2.519&20.33&0.008&0.160&-21.45&2.519&19.11&0.008&0.160&-22.34\\
2.636&20.63&0.009&0.164&-21.61&2.636&19.41&0.009&0.163&-22.57\\
2.759&20.91&0.010&0.170&-22.30&2.759&19.69&0.010&0.168&-23.01\\
2.888&21.18&0.011&0.175&-22.94&2.888&19.96&0.011&0.175&-22.77\\
3.023&21.46&0.011&0.174&-23.41&3.023&20.23&0.013&0.177&-22.51\\
3.164&21.79&0.013&0.170&-23.03&3.164&20.55&0.015&0.169&-21.18\\
3.311&22.16&0.015&0.161&-23.60&3.311&20.90&0.019&0.155&-21.48\\
3.466&22.54&0.017&0.162&-24.89&3.466&21.26&0.025&0.154&-21.89\\
3.627&22.88&0.020&0.180&-26.01&3.627&21.60&0.032&0.164&-23.47\\
3.796&23.20&0.023&0.197&-26.10&3.796&21.91&0.040&0.171&-24.85\\
3.973&23.52&0.028&0.212&-26.81&3.973&22.24&0.052&0.177&-22.88\\
4.159&23.91&0.034&0.206&-26.82&4.159&22.60&0.070&0.169&-17.53\\
4.353&24.32&0.041&0.214&-29.74&4.353&23.05&0.101&0.149&-14.40\\
4.556&24.79&0.052&0.220&-32.35&4.556&23.53&0.151&0.131&-7.06\\
4.768&25.29&0.069&0.234&-35.05&4.768&24.06&0.238&0.112&0.38\\
4.990&25.72&0.091&0.270&-39.20&4.990&24.44&0.332&0.076&-72.85\\
5.223&26.33&0.133&0.270&-46.22&5.223&25.00&0.543&0.076&-72.85\\
5.467&26.92&0.194&0.307&-53.26&5.467&25.85&1.154&0.076&-72.85\\
5.722&27.09&0.218&0.417&-54.40&5.722&25.99&1.304&0.076&-72.85\\
5.988&27.36&0.265&0.505&-56.16&&&&&\\
6.268&27.44&0.284&0.582&-60.02&&&&&\\
6.560&27.64&0.328&0.644&-63.09&&&&&\\
\hline
\end{tabular}
\tablefoot {Column1: Semi-major axis in the {\it g} band. Column 2: Azimuthally averaged surface brightness in the {\it g} band. Column 3: Error on the surface brightness in the {\it g} band. Column 4: Ellipticity in the {\it g} band. Column 5: Position angle in the {\it g} band. Column 6: Semi-major axis in the {\it i} band. Column 7: Azimuthally averaged surface brightness in the {\it i} band. Column 8: Error on the surface brightness in the {\it i} band. Column 9: Ellipticity in the {\it i} band. Column 10: Position angle in the {\it i} band.}
\end{table*}

\begin{table*}
\caption{Surface photometry of NGC 4434.}\label{Tab_NGC4434}
\begin{tabular}{lcccccccccc}
\hline\hline
$a^{1/4}_{g}$ & $\mu_{g}$ & $\sigma(\mu_{g})$ & $\epsilon_{g}$ &
$P.A._{g}$ & $a^{1/4}_{i}$ & $\mu_{i}$ & $\sigma(\mu_{i})$ &
$\epsilon_{i}$ & $P.A._{i}$\\

[arcsec] & [mag/arcsec$^{2}$] & [mag/arcsec$^{2}$] & & [deg] & [arcsec] & [mag/arcsec$^{2}$] & [mag/arcsec$^{2}$] & & [deg] \\
\hline
0.000&16.62&0.012&0.152&9.65&0.000&15.23&0.006&0.073&9.35\\
0.586&16.66&0.007&0.152&9.65&0.586&15.29&0.006&0.073&9.35\\
0.613&16.67&0.007&0.158&11.41&0.613&15.31&0.006&0.076&12.02\\
0.642&16.68&0.007&0.152&14.85&0.642&15.32&0.006&0.061&18.51\\
0.672&16.69&0.007&0.152&18.09&0.672&15.34&0.006&0.057&22.38\\
0.703&16.71&0.007&0.110&28.62&0.703&15.37&0.006&0.039&29.03\\
0.736&16.75&0.007&0.081&29.59&0.736&15.42&0.006&0.036&36.16\\
0.769&16.79&0.007&0.087&34.91&0.769&15.48&0.006&0.036&38.35\\
0.806&16.84&0.007&0.087&37.51&0.806&15.56&0.006&0.041&43.38\\
0.843&16.92&0.007&0.081&42.76&0.843&15.66&0.006&0.045&46.36\\
0.883&17.01&0.007&0.081&42.76&0.883&15.78&0.006&0.045&46.36\\
0.924&17.13&0.007&0.071&44.55&0.924&15.92&0.006&0.053&46.19\\
0.967&17.27&0.007&0.060&44.37&0.967&16.08&0.006&0.054&45.39\\
1.012&17.44&0.007&0.048&43.19&1.012&16.26&0.006&0.054&45.86\\
1.059&17.61&0.007&0.042&41.94&1.059&16.45&0.006&0.049&46.16\\
1.109&17.81&0.007&0.038&41.43&1.109&16.66&0.006&0.042&40.36\\
1.161&18.03&0.007&0.036&36.65&1.161&16.90&0.006&0.038&38.19\\
1.215&18.28&0.007&0.040&34.21&1.215&17.17&0.006&0.039&35.13\\
1.271&18.55&0.008&0.038&30.97&1.271&17.44&0.006&0.033&36.60\\
1.331&18.82&0.008&0.042&41.18&1.331&17.72&0.006&0.038&37.22\\
1.393&19.08&0.008&0.052&40.14&1.393&17.98&0.006&0.050&40.78\\
1.458&19.34&0.009&0.073&40.14&1.458&18.24&0.006&0.073&38.09\\
1.526&19.61&0.009&0.097&41.99&1.526&18.51&0.006&0.095&41.85\\
1.597&19.89&0.009&0.106&42.37&1.597&18.80&0.006&0.099&41.42\\
1.671&20.24&0.009&0.083&42.37&1.671&19.14&0.006&0.082&41.71\\
1.749&20.60&0.010&0.067&39.39&1.749&19.52&0.006&0.060&39.33\\
1.831&20.99&0.010&0.052&39.20&1.831&19.91&0.006&0.049&40.45\\
1.916&21.34&0.011&0.050&36.01&1.916&20.28&0.006&0.044&35.75\\
2.006&21.68&0.012&0.043&34.35&2.006&20.63&0.006&0.038&35.75\\
2.099&22.07&0.014&0.041&29.14&2.099&21.01&0.007&0.034&28.48\\
2.197&22.51&0.017&0.038&29.87&2.197&21.45&0.007&0.037&28.61\\
2.299&22.97&0.020&0.050&31.53&2.299&21.92&0.009&0.049&31.82\\
2.407&23.51&0.028&0.052&38.28&2.407&22.50&0.012&0.050&37.35\\
2.519&24.18&0.034&0.060&38.24&2.519&23.21&0.021&0.060&42.92\\
2.636&24.83&0.048&0.071&41.20&2.636&23.95&0.040&0.069&43.85\\
2.759&25.53&0.067&0.084&42.81&2.759&24.74&0.083&0.088&47.65\\
2.888&26.18&0.094&0.104&36.10&2.888&25.62&0.185&0.093&58.63\\
3.023&26.79&0.133&0.151&36.39&3.023&26.29&0.346&0.170&62.97\\
3.164&27.47&0.194&0.199&36.39&3.164&27.06&0.702&0.206&65.79\\
3.311&28.95&0.471&0.190&24.26&3.311&27.50&1.047&0.308&65.79\\        
\hline
\end{tabular}
\tablefoot {Column1: Semi-major axis in the {\it g} band. Column 2: Azimuthally averaged surface brightness in the {\it g} band. Column 3: Error on the surface brightness in the {\it g} band. Column 4: Ellipticity in the {\it g} band. Column 5: Position angle in the {\it g} band. Column 6: Semi-major axis in the {\it i} band. Column 7: Azimuthally averaged surface brightness in the {\it i} band. Column 8: Error on the surface brightness in the {\it i} band. Column 9: Ellipticity in the {\it i} band. Column 10: Position angle in the {\it i} band.}
\end{table*}

\begin{table*}
\caption{Surface photometry of NGC 4464.}\label{Tab_NGC4464}
\begin{tabular}{lcccccccccc}
\hline\hline
$a^{1/4}_{g}$ & $\mu_{g}$ & $\sigma(\mu_{g})$ & $\epsilon_{g}$ &
$P.A._{g}$ & $a^{1/4}_{i}$ & $\mu_{i}$ & $\sigma(\mu_{i})$ &
$\epsilon_{i}$ & $P.A._{i}$\\

[arcsec] & [mag/arcsec$^{2}$] & [mag/arcsec$^{2}$] & & [deg] & [arcsec] & [mag/arcsec$^{2}$] & [mag/arcsec$^{2}$] & & [deg] \\
\hline
0.000&16.43&0.012&0.215&7.85&0.000&15.06&0.008&0.319&3.41\\
0.586&16.48&0.007&0.215&7.85&0.586&15.11&0.006&0.319&3.41\\
0.613&16.49&0.007&0.223&9.13&0.613&15.12&0.006&0.316&4.25\\
0.642&16.51&0.007&0.233&10.58&0.642&15.13&0.006&0.313&5.27\\
0.672&16.52&0.007&0.232&13.37&0.672&15.15&0.006&0.310&6.48\\
0.703&16.55&0.007&0.165&22.47&0.703&15.18&0.006&0.208&10.08\\
0.736&16.59&0.007&0.160&20.53&0.736&15.23&0.006&0.145&16.19\\
0.769&16.64&0.007&0.170&15.65&0.769&15.29&0.006&0.131&12.73\\
0.806&16.70&0.007&0.183&12.83&0.806&15.37&0.006&0.116&8.37\\
0.843&16.78&0.007&0.190&11.16&0.843&15.48&0.006&0.111&10.10\\
0.883&16.90&0.007&0.189&12.93&0.883&15.60&0.006&0.128&9.35\\
0.924&17.03&0.007&0.190&11.92&0.924&15.75&0.006&0.154&8.08\\
0.967&17.20&0.007&0.189&9.49&0.967&15.93&0.006&0.168&8.25\\
1.012&17.37&0.007&0.201&7.12&1.012&16.12&0.006&0.201&6.93\\
1.059&17.57&0.007&0.207&6.44&1.059&16.33&0.006&0.219&6.77\\
1.109&17.79&0.007&0.217&5.57&1.109&16.55&0.006&0.240&6.04\\
1.161&18.02&0.007&0.233&4.80&1.161&16.78&0.006&0.257&5.93\\
1.215&18.28&0.007&0.248&4.67&1.215&17.04&0.006&0.272&5.55\\
1.271&18.54&0.008&0.261&4.67&1.271&17.30&0.007&0.279&5.36\\
1.331&18.82&0.008&0.261&5.10&1.331&17.58&0.007&0.279&5.09\\
1.393&19.09&0.008&0.269&4.40&1.393&17.87&0.007&0.274&4.14\\
1.458&19.37&0.009&0.282&5.69&1.458&18.14&0.007&0.286&5.42\\
1.526&19.67&0.009&0.281&4.92&1.526&18.45&0.007&0.279&4.19\\
1.597&20.00&0.010&0.275&4.53&1.597&18.77&0.008&0.282&4.10\\
1.671&20.35&0.010&0.266&4.28&1.671&19.13&0.008&0.267&3.32\\
1.749&20.75&0.011&0.247&3.28&1.749&19.52&0.008&0.258&2.58\\
1.831&21.18&0.012&0.232&3.45&1.831&19.96&0.008&0.240&2.59\\
1.916&21.66&0.013&0.215&3.04&1.916&20.45&0.009&0.224&2.59\\
2.006&22.18&0.015&0.195&3.22&2.006&20.98&0.010&0.204&2.77\\
2.099&22.71&0.018&0.181&2.45&2.099&21.52&0.012&0.192&2.70\\
2.197&23.30&0.024&0.169&1.82&2.197&22.13&0.015&0.178&2.57\\
2.299&23.92&0.032&0.155&-0.32&2.299&22.80&0.021&0.154&0.56\\
2.407&24.55&0.042&0.150&-5.82&2.407&23.54&0.030&0.134&-6.03\\
2.519&25.23&0.058&0.137&-8.26&2.519&24.45&0.049&0.053&-15.97\\
2.636&25.93&0.083&0.143&-15.71&2.636&25.24&0.079&0.101&-46.77\\
2.759&26.77&0.130&0.152&-16.62&2.759&26.03&0.131&0.185&-54.86\\
2.888&27.78&0.237&0.159&-19.04&2.888&26.64&0.198&0.282&-65.85\\
3.023&29.38&0.625&0.186&-13.70&3.023&27.86&0.478&0.282&-65.85\\
3.164&30.39&1.256&0.270&-35.14&3.164&28.84&1.025&0.282&-65.85\\
\hline
\end{tabular}
\tablefoot {Column1: Semi-major axis in the {\it g} band. Column 2: Azimuthally averaged surface brightness in the {\it g} band. Column 3: Error on the surface brightness in the {\it g} band. Column 4: Ellipticity in the {\it g} band. Column 5: Position angle in the {\it g} band. Column 6: Semi-major axis in the {\it i} band. Column 7: Azimuthally averaged surface brightness in the {\it i} band. Column 8: Error on the surface brightness in the {\it i} band. Column 9: Ellipticity in the {\it i} band. Column 10: Position angle in the {\it i} band.}
\end{table*}

\begin{table*}
\caption{Surface photometry of NGC 4467.}\label{Tab_NGC4467}
\begin{tabular}{lcccccccccc}
\hline\hline
$a^{1/4}_{g}$ & $\mu_{g}$ & $\sigma(\mu_{g})$ & $\epsilon_{g}$ &
$P.A._{g}$ & $a^{1/4}_{i}$ & $\mu_{i}$ & $\sigma(\mu_{i})$ &
$\epsilon_{i}$ & $P.A._{i}$\\

[arcsec] & [mag/arcsec$^{2}$] & [mag/arcsec$^{2}$] & & [deg] & [arcsec] & [mag/arcsec$^{2}$] & [mag/arcsec$^{2}$] & & [deg] \\
\hline
0.000&17.96&0.021&0.130&44.73&0.000&16.74&0.013&0.185&29.39\\
0.572&18.02&0.008&0.130&44.73&0.572&16.78&0.007&0.185&29.39\\
0.586&18.03&0.008&0.143&44.82&0.586&16.78&0.007&0.195&30.39\\
0.598&18.03&0.008&0.158&44.88&0.598&16.79&0.007&0.210&30.66\\
0.615&18.04&0.008&0.174&44.94&0.615&16.79&0.007&0.225&31.26\\
0.628&18.04&0.008&0.191&44.98&0.628&16.80&0.007&0.240&31.82\\
0.644&18.05&0.008&0.210&45.02&0.644&16.80&0.007&0.257&32.38\\
0.659&18.06&0.008&0.230&45.05&0.659&16.81&0.007&0.274&32.98\\
0.675&18.07&0.008&0.252&45.07&0.675&16.81&0.007&0.293&33.45\\
0.692&18.07&0.008&0.276&45.08&0.692&16.82&0.007&0.314&33.91\\
0.709&18.08&0.008&0.293&44.27&0.709&16.83&0.007&0.318&37.46\\
0.726&18.10&0.008&0.274&43.24&0.726&16.85&0.007&0.255&39.60\\
0.743&18.13&0.009&0.253&41.08&0.743&16.88&0.007&0.219&39.77\\
0.760&18.16&0.009&0.230&39.61&0.760&16.91&0.007&0.212&40.61\\
0.779&18.19&0.009&0.224&38.30&0.779&16.95&0.007&0.210&41.28\\
0.798&18.23&0.009&0.223&36.90&0.798&16.99&0.007&0.202&42.07\\
0.817&18.27&0.009&0.219&35.15&0.817&17.03&0.007&0.207&43.50\\
0.836&18.32&0.009&0.213&34.72&0.836&17.09&0.007&0.217&42.85\\
0.857&18.37&0.009&0.218&35.82&0.857&17.14&0.007&0.224&44.22\\
0.877&18.42&0.009&0.225&34.92&0.877&17.20&0.007&0.230&44.90\\
0.898&18.49&0.009&0.220&36.68&0.898&17.26&0.007&0.236&43.68\\
0.921&18.56&0.009&0.216&36.81&0.921&17.34&0.007&0.236&43.70\\
0.943&18.63&0.008&0.222&36.38&0.943&17.42&0.007&0.244&44.20\\
0.965&18.70&0.008&0.221&37.44&0.965&17.49&0.007&0.253&43.49\\
0.989&18.79&0.008&0.220&37.54&0.989&17.58&0.007&0.254&42.99\\
1.012&18.87&0.008&0.224&37.53&1.012&17.67&0.007&0.263&42.82\\
1.037&18.96&0.008&0.225&37.50&1.037&17.76&0.007&0.270&42.42\\
1.062&19.04&0.008&0.230&37.11&1.062&17.85&0.007&0.277&41.32\\
1.087&19.14&0.008&0.238&37.47&1.087&17.94&0.007&0.290&41.13\\
1.113&19.23&0.008&0.248&37.61&1.113&18.03&0.007&0.301&40.43\\
1.140&19.33&0.008&0.258&38.08&1.140&18.13&0.007&0.310&40.42\\
1.168&19.43&0.008&0.265&38.72&1.168&18.23&0.007&0.316&40.17\\
1.196&19.54&0.008&0.274&38.65&1.196&18.34&0.007&0.322&40.13\\
1.225&19.65&0.008&0.283&38.86&1.225&18.45&0.007&0.329&40.29\\
1.254&19.77&0.008&0.290&39.45&1.254&18.57&0.007&0.332&40.32\\
1.285&19.89&0.008&0.293&39.42&1.285&18.70&0.007&0.329&40.33\\
1.316&20.02&0.009&0.294&39.45&1.316&18.83&0.007&0.324&40.04\\
1.347&20.16&0.009&0.293&39.65&1.347&18.98&0.007&0.315&39.80\\
1.380&20.31&0.009&0.291&39.62&1.380&19.14&0.007&0.309&40.22\\
1.413&20.48&0.009&0.279&39.64&1.413&19.30&0.007&0.298&39.99\\
1.447&20.65&0.012&0.270&39.55&1.447&19.48&0.009&0.283&39.94\\
1.482&20.84&0.013&0.253&39.82&1.482&19.67&0.009&0.266&39.60\\
1.518&21.03&0.013&0.241&39.82&1.518&19.87&0.009&0.248&40.10\\
1.554&21.24&0.014&0.219&39.92&1.554&20.08&0.010&0.227&39.61\\
1.592&21.45&0.016&0.199&39.76&1.592&20.29&0.011&0.204&38.47\\
1.630&21.66&0.016&0.183&39.76&1.630&20.52&0.011&0.182&39.37\\
1.670&21.90&0.017&0.158&40.45&1.670&20.76&0.012&0.160&40.69\\
1.710&22.14&0.018&0.132&40.79&1.710&21.01&0.012&0.135&38.78\\
1.751&22.40&0.019&0.114&40.68&1.751&21.28&0.013&0.111&40.11\\
1.793&22.67&0.020&0.097&40.91&1.793&21.57&0.014&0.083&38.47\\
1.837&22.95&0.022&0.079&40.77&1.837&21.84&0.015&0.078&38.24\\
1.881&23.25&0.024&0.063&41.35&1.881&22.17&0.016&0.057&40.44\\
1.926&23.57&0.026&0.045&39.91&1.926&22.49&0.018&0.044&42.22\\
1.973&23.90&0.031&0.034&37.14&1.973&22.85&0.022&0.033&35.67\\
2.020&24.27&0.038&0.026&54.38&2.020&23.26&0.027&0.017&61.34\\
2.069&24.65&0.045&0.024&74.94&2.069&23.67&0.032&0.019&85.44\\
2.119&25.02&0.053&0.037&85.99&2.119&24.04&0.040&0.029&-82.29\\
2.170&25.45&0.067&0.044&-83.88&2.170&24.53&0.054&0.035&-63.69\\
2.222&25.99&0.091&0.038&83.57&2.222&25.19&0.080&0.003&-84.78\\
2.276&26.62&0.127&0.048&73.55&2.276&26.08&0.141&0.001&-174.80\\
2.331&27.31&0.186&0.039&73.55&2.331&26.15&0.148&0.105&76.97\\
2.387&27.36&0.192&0.151&82.69&2.387&26.56&0.195&0.168&87.41\\
2.444&27.70&0.236&0.213&75.10&2.444&26.89&0.245&0.168&87.41\\
2.503&28.25&0.324&0.244&84.88&2.503&27.02&0.271&0.168&87.41\\
\hline
\end{tabular}
\tablefoot {Column1: Semi-major axis in the {\it g} band. Column 2: Azimuthally averaged surface brightness in the {\it g} band. Column 3: Error on the surface brightness in the {\it g} band. Column 4: Ellipticity in the {\it g} band. Column 5: Position angle in the {\it g} band. Column 6: Semi-major axis in the {\it i} band. Column 7: Azimuthally averaged surface brightness in the {\it i} band. Column 8: Error on the surface brightness in the {\it i} band. Column 9: Ellipticity in the {\it i} band. Column 10: Position angle in the {\it i} band.}
\end{table*}

\begin{table*}
\caption{Surface photometry of VCC 1199.}\label{Tab_VCC1199}
\begin{tabular}{lcccccccccc}
\hline\hline
$a^{1/4}_{g}$ & $\mu_{g}$ & $\sigma(\mu_{g})$ & $\epsilon_{g}$ &
$P.A._{g}$ & $a^{1/4}_{i}$ & $\mu_{i}$ & $\sigma(\mu_{i})$ &
$\epsilon_{i}$ & $P.A._{i}$\\

[arcsec] & [mag/arcsec$^{2}$] & [mag/arcsec$^{2}$] & & [deg] & [arcsec] & [mag/arcsec$^{2}$] & [mag/arcsec$^{2}$] & & [deg] \\
\hline
0.000&18.37&0.025&0.112&30.72&0.000&16.90&0.014&0.059&77.81\\
0.572&18.42&0.009&0.112&30.72&0.572&16.97&0.007&0.059&77.81\\
0.586&18.42&0.009&0.139&30.72&0.586&16.97&0.007&0.057&76.75\\
0.598&18.43&0.009&0.139&34.27&0.598&16.98&0.007&0.056&75.66\\
0.615&18.44&0.009&0.151&34.03&0.615&16.99&0.007&0.054&74.53\\
0.628&18.44&0.009&0.167&33.85&0.628&17.00&0.007&0.053&73.38\\
0.644&18.45&0.009&0.182&34.97&0.644&17.01&0.007&0.053&72.21\\
0.659&18.45&0.009&0.198&35.99&0.659&17.02&0.007&0.053&71.03\\
0.675&18.46&0.009&0.217&36.85&0.675&17.03&0.007&0.057&69.97\\
0.692&18.47&0.009&0.238&37.58&0.692&17.05&0.007&0.054&65.10\\
0.709&18.48&0.009&0.237&39.69&0.709&17.07&0.007&0.044&64.72\\
0.726&18.50&0.009&0.215&38.95&0.726&17.10&0.007&0.044&53.69\\
0.743&18.52&0.009&0.192&38.09&0.743&17.14&0.007&0.042&41.89\\
0.760&18.55&0.009&0.187&38.03&0.760&17.17&0.007&0.052&38.01\\
0.779&18.58&0.010&0.188&37.64&0.779&17.22&0.007&0.052&38.01\\
0.798&18.62&0.010&0.191&37.35&0.798&17.26&0.007&0.063&36.03\\
0.817&18.67&0.010&0.192&36.88&0.817&17.32&0.007&0.062&35.63\\
0.836&18.71&0.010&0.188&37.46&0.836&17.39&0.007&0.068&41.37\\
0.857&18.77&0.010&0.197&36.61&0.857&17.45&0.007&0.092&34.81\\
0.877&18.84&0.010&0.199&33.76&0.877&17.53&0.007&0.099&36.07\\
0.898&18.91&0.010&0.188&33.70&0.898&17.63&0.007&0.100&37.36\\
0.921&19.00&0.009&0.184&32.55&0.921&17.71&0.007&0.121&33.66\\
0.943&19.08&0.009&0.180&31.51&0.943&17.81&0.007&0.128&33.90\\
0.965&19.18&0.009&0.171&32.58&0.965&17.92&0.007&0.133&34.39\\
0.989&19.29&0.009&0.162&31.73&0.989&18.03&0.007&0.148&32.59\\
1.012&19.39&0.009&0.152&31.50&1.012&18.14&0.007&0.158&32.95\\
1.037&19.51&0.009&0.148&32.07&1.037&18.27&0.007&0.165&33.07\\
1.062&19.63&0.010&0.147&31.92&1.062&18.39&0.007&0.174&32.26\\
1.087&19.76&0.010&0.140&31.70&1.087&18.52&0.007&0.174&32.25\\
1.113&19.90&0.010&0.139&31.74&1.113&18.67&0.007&0.168&31.38\\
1.140&20.04&0.010&0.143&30.46&1.140&18.82&0.007&0.166&30.08\\
1.168&20.20&0.010&0.137&31.18&1.168&18.98&0.008&0.158&30.62\\
1.196&20.37&0.010&0.127&31.30&1.196&19.15&0.008&0.146&31.14\\
1.225&20.55&0.010&0.119&31.62&1.225&19.33&0.008&0.135&31.05\\
1.254&20.73&0.011&0.112&31.96&1.254&19.51&0.008&0.127&31.49\\
1.285&20.92&0.011&0.107&31.70&1.285&19.71&0.008&0.110&31.29\\
1.316&21.13&0.011&0.088&31.52&1.316&19.91&0.008&0.099&32.09\\
1.347&21.34&0.012&0.081&31.92&1.347&20.13&0.008&0.085&34.01\\
1.380&21.58&0.012&0.073&32.66&1.380&20.36&0.009&0.073&34.71\\
1.413&21.82&0.013&0.058&33.39&1.413&20.63&0.009&0.048&34.07\\
1.447&22.11&0.019&0.037&35.10&1.447&20.91&0.012&0.031&39.72\\
1.482&22.39&0.021&0.021&50.56&1.482&21.19&0.013&0.015&53.78\\
1.518&22.71&0.024&0.002&-128.20&1.518&21.51&0.015&0.005&36.15\\
1.554&23.00&0.028&0.013&-34.15&1.554&21.80&0.017&0.008&-42.98\\
1.592&23.34&0.032&0.012&-43.40&1.592&22.15&0.020&0.008&-53.55\\
1.630&23.69&0.037&0.022&-41.00&1.630&22.48&0.023&0.023&-52.84\\
1.670&24.10&0.043&0.027&-41.49&1.670&22.85&0.026&0.037&-48.16\\
1.710&24.45&0.049&0.047&-41.49&1.710&23.23&0.031&0.050&-42.17\\
1.751&24.91&0.058&0.051&-49.08&1.751&23.64&0.037&0.070&-45.15\\
1.793&25.30&0.070&0.086&-42.10&1.793&24.11&0.046&0.065&-38.57\\
1.837&25.82&0.086&0.081&-35.41&1.837&24.60&0.059&0.090&-36.89\\
1.881&26.12&0.101&0.144&-33.56&1.881&25.05&0.075&0.096&-24.82\\
1.926&26.94&0.153&0.107&-35.65&1.926&25.58&0.103&0.111&-18.96\\
1.973&27.17&0.166&0.123&-38.56&1.973&25.69&0.112&0.213&-18.96\\
2.020&27.16&0.170&0.241&-33.34&2.020&26.28&0.160&0.221&-21.42\\
2.069&28.10&0.287&0.206&-42.19&2.069&27.43&0.351&0.154&-33.18\\
2.119&28.75&0.420&0.206&-42.19&2.119&27.70&0.426&0.154&-33.18\\
2.170&29.51&0.685&0.206&-42.19&2.170&27.96&0.514&0.154&-33.18\\
\hline
\end{tabular}
\tablefoot {Column1: Semi-major axis in the {\it g} band. Column 2: Azimuthally averaged surface brightness in the {\it g} band. Column 3: Error on the surface brightness in the {\it g} band. Column 4: Ellipticity in the {\it g} band. Column 5: Position angle in the {\it g} band. Column 6: Semi-major axis in the {\it i} band. Column 7: Azimuthally averaged surface brightness in the {\it i} band. Column 8: Error on the surface brightness in the {\it i} band. Column 9: Ellipticity in the {\it i} band. Column 10: Position angle in the {\it i} band.}
\end{table*}

\begin{table*}
\caption{Surface photometry of UGC 7636.}\label{Tab_UGC7636}
\begin{tabular}{lcccccccccc}
\hline\hline
$a^{1/4}_{g}$ & $\mu_{g}$ & $\sigma(\mu_{g})$ & $\epsilon_{g}$ &
$P.A._{g}$ & $a^{1/4}_{i}$ & $\mu_{i}$ & $\sigma(\mu_{i})$ &
$\epsilon_{i}$ & $P.A._{i}$\\

[arcsec] & [mag/arcsec$^{2}$] & [mag/arcsec$^{2}$] & & [deg] & [arcsec] & [mag/arcsec$^{2}$] & [mag/arcsec$^{2}$] & & [deg] \\
\hline
1.012&22.66&0.042&0.390&0.00&1.012&22.26&0.040&0.390&0.00\\
1.037&22.66&0.040&0.390&0.00&1.037&22.26&0.038&0.390&0.00\\
1.062&22.66&0.039&0.390&0.00&1.062&22.26&0.037&0.390&0.00\\
1.087&22.67&0.037&0.390&0.00&1.087&22.25&0.036&0.390&0.00\\
1.113&22.67&0.036&0.390&0.00&1.113&22.24&0.034&0.390&0.00\\
1.140&22.67&0.034&0.390&0.00&1.140&22.24&0.033&0.390&0.00\\
1.168&22.67&0.033&0.390&0.00&1.168&22.24&0.032&0.390&0.00\\
1.196&22.67&0.032&0.390&0.00&1.196&22.23&0.031&0.390&0.00\\
1.225&22.67&0.031&0.390&0.00&1.225&22.23&0.030&0.390&0.00\\
1.254&22.67&0.030&0.390&0.00&1.254&22.22&0.029&0.390&0.00\\
1.285&22.65&0.028&0.390&0.00&1.285&22.22&0.028&0.390&0.00\\
1.316&22.64&0.027&0.390&0.00&1.316&22.24&0.028&0.390&0.00\\
1.347&22.65&0.026&0.390&0.00&1.347&22.24&0.027&0.390&0.00\\
1.380&22.65&0.026&0.390&0.00&1.380&22.27&0.027&0.390&0.00\\
1.413&22.66&0.025&0.390&0.00&1.413&22.27&0.026&0.390&0.00\\
1.447&22.67&0.034&0.390&0.00&1.447&22.29&0.034&0.390&0.00\\
1.482&22.68&0.034&0.390&0.00&1.482&22.28&0.033&0.390&0.00\\
1.518&22.70&0.034&0.390&0.00&1.518&22.29&0.034&0.390&0.00\\
1.554&22.70&0.034&0.390&0.00&1.554&22.28&0.034&0.390&0.00\\
1.592&22.71&0.034&0.390&0.00&1.592&22.32&0.034&0.390&0.00\\
1.630&22.72&0.034&0.390&0.00&1.630&22.35&0.035&0.390&0.00\\
1.670&22.77&0.034&0.390&0.00&1.670&22.37&0.034&0.390&0.00\\
1.710&22.82&0.034&0.390&0.00&1.710&22.40&0.034&0.390&0.00\\
1.751&22.88&0.033&0.390&0.00&1.751&22.45&0.034&0.390&0.00\\
1.793&22.95&0.034&0.390&0.00&1.793&22.50&0.034&0.390&0.00\\
1.837&23.01&0.034&0.390&0.00&1.837&22.55&0.034&0.390&0.00\\
1.881&23.06&0.033&0.390&0.00&1.881&22.59&0.034&0.390&0.00\\
1.926&23.12&0.033&0.390&0.00&1.926&22.65&0.034&0.390&0.00\\
1.973&23.19&0.034&0.390&0.00&1.973&22.70&0.034&0.390&0.00\\
2.020&23.35&0.037&0.390&0.00&2.020&22.83&0.038&0.390&0.00\\
2.069&23.44&0.038&0.390&0.00&2.069&22.92&0.039&0.390&0.00\\
2.119&23.53&0.039&0.390&0.00&2.119&23.00&0.040&0.390&0.00\\
2.170&23.66&0.041&0.390&0.00&2.170&23.12&0.043&0.390&0.00\\
2.222&23.81&0.044&0.390&0.00&2.222&23.22&0.046&0.390&0.00\\
2.276&23.99&0.049&0.390&0.00&2.276&23.37&0.050&0.390&0.00\\
2.331&24.17&0.054&0.390&0.00&2.331&23.50&0.055&0.390&0.00\\
2.387&24.31&0.058&0.390&0.00&2.387&23.64&0.060&0.390&0.00\\
2.444&24.46&0.063&0.390&0.00&2.444&23.75&0.065&0.390&0.00\\
2.503&24.60&0.069&0.390&0.00&2.503&23.88&0.072&0.390&0.00\\
2.564&24.79&0.078&0.390&0.00&2.564&24.02&0.079&0.390&0.00\\
2.626&24.92&0.085&0.390&0.00&2.626&24.13&0.086&0.390&0.00\\
2.689&25.06&0.092&0.390&0.00&2.689&24.21&0.091&0.390&0.00\\
2.754&25.18&0.099&0.390&0.00&2.754&24.30&0.097&0.390&0.00\\
2.820&25.27&0.107&0.390&0.00&2.820&24.42&0.108&0.390&0.00\\
2.888&25.54&0.125&0.390&0.00&2.888&24.66&0.131&0.390&0.00\\
2.958&25.74&0.141&0.390&0.00&2.958&24.83&0.149&0.390&0.00\\
3.029&26.06&0.170&0.390&0.00&3.029&25.08&0.182&0.390&0.00\\
3.102&26.49&0.213&0.390&0.00&3.102&25.37&0.229&0.390&0.00\\
3.177&26.77&0.246&0.390&0.00&3.177&25.51&0.256&0.390&0.00\\
3.254&27.14&0.287&0.390&0.00&3.254&25.69&0.294&0.390&0.00\\
3.332&27.71&0.357&0.390&0.00&3.332&25.91&0.351&0.390&0.00\\
3.412&28.48&0.443&0.390&0.00&3.412&26.27&0.479&0.390&0.00\\
3.495&28.74&0.469&0.390&0.00&3.495&26.31&0.499&0.390&0.00\\
3.579&31.13&0.576&0.390&0.00&3.579&26.87&0.807&0.390&0.00\\
3.665&30.07&0.553&0.390&0.00&3.665&27.08&0.972&0.390&0.00\\
3.844&29.83&0.565&0.390&0.00&3.844&26.83&0.798&0.390&0.00\\
3.937&29.42&0.561&0.390&0.00&3.937&26.93&0.885&0.390&0.00\\
\hline
\end{tabular}
\tablefoot {Column1: Semi-major axis in the {\it g} band. Column 2: Azimuthally averaged surface brightness in the {\it g} band. Column 3: Error on the surface brightness in the {\it g} band. Column 4: Ellipticity in the {\it g} band. Column 5: Position angle in the {\it g} band. Column 6: Semi-major axis in the {\it i} band. Column 7: Azimuthally averaged surface brightness in the {\it i} band. Column 8: Error on the surface brightness in the {\it i} band. Column 9: Ellipticity in the {\it i} band. Column 10: Position angle in the {\it i} band.}
\end{table*}
\end{appendix}

\begin{thebibliography}{}
\bibitem[Aguerri et al.(2005)]{Aguerri05} Aguerri, J.~A.~L., 
Gerhard, O.~E., Arnaboldi, M., et al.\ 2005, \aj, 129, 2585
\bibitem[Andersen et al.(1995)]{Andersen95} Andersen, M.~I.,
  Freyhammer, L., \& Storm, J.\ 1995, Calibrating and Understanding
  HST and ESO Instruments, 53, 87
\bibitem[Arnaboldi et al.(2002)]{Arnaboldi02} Arnaboldi, M., 
Aguerri, J.~A.~L., Napolitano, N.~R., et al.\ 2002, \aj, 123, 760
\bibitem[Arnaboldi et 
al.(2012)]{Arnaboldi12} Arnaboldi, M., Ventimiglia, G., Iodice, E., Gerhard, O., \& Coccato, L.\ 2012, \aap, 545, AA37
\bibitem[Arnold et al.(2011)]{Arnold11} Arnold, J.~A., 
Romanowsky, A.~J., Brodie, J.~P., et al.\ 2011, \apjl, 736, LL26
\bibitem[Arrigoni Battaia et 
al.(2012)]{Battaia12} Arrigoni Battaia, F., Gavazzi, G., Fumagalli, M., et al.\ 2012, \aap, 543, AA112
\bibitem[Ashman 
\& Zepf(1992)]{Ashman92} Ashman, K.~M., \& Zepf, S.~E.\ 1992, \apj, 384, 50
\bibitem[Balcells et al.(2007)]{Balcells07} Balcells, M., Graham, 
A.~W., \& Peletier, R.~F.\ 2007, \apj, 665, 1104
\bibitem[Bender et 
al.(1989)]{Bender89} Bender, R., Surma, P., Doebereiner, S., Moellenhoff, C., \& Madejsky, R.\ 1989, \aap, 217, 35
\bibitem[Bertin 
\& Arnouts(1996)]{Bertin96} Bertin, E., \& Arnouts, S.\ 1996, \aaps, 117, 393
\bibitem[Bertin et al.(2002)]{Bertin02} Bertin, E., Mellier, Y., 
Radovich, M., et al.\ 2002, Astronomical Data Analysis Software and Systems 
XI, 281, 228
\bibitem[Bertin (2006)]{Bertin06} Bertin, E.\ 2006, Astronomical 
Data Analysis Software and Systems XV, 351, 112 
\bibitem[Binney \& Petrou(1985)]{Binney85} Binney, J., \& Petrou,
  M.\ 1985, \mnras, 214, 449
\bibitem[S{\'a}nchez-Bl{\'a}zquez et al.(2007)]{Blazquez07} 
S{\'a}nchez-Bl{\'a}zquez, P., Forbes, D.~A., Strader, J., Brodie, J., 
\& Proctor, R.\ 2007, \mnras, 377, 759
\bibitem[Blom et al.(2012)]{Blom12} Blom, C., Forbes, D.~A., 
Brodie, J.~P., et al.\ 2012, \mnras, 426, 1959
\bibitem[Brodie et al.(2012)]{Brodie12} Brodie, J.~P., Usher, 
C., Conroy, C., et al.\ 2012, \apjl, 759, LL33
\bibitem[Brodie et al.(2014)]{Brodie14} Brodie, J.~P., 
Romanowsky, A.~J., Strader, J., et al.\ 2014, \apj, 796, 52
\bibitem[Bruzual 
\& Charlot(2003)]{BC03} Bruzual, G., \& Charlot, S.\ 2003, \mnras, 344, 1000
\bibitem[Cantiello et al.(2005)]{Cantiello05} Cantiello, M., 
Blakeslee, J.~P., Raimondo, G., et al.\ 2005, \apj, 634, 239
\bibitem[Cantiello 
\& Blakeslee(2007)]{Cantiello07} Cantiello, M., \& Blakeslee, J.~P.\ 2007, \apj, 669, 982
\bibitem[Cantiello et 
al.(2013)]{Cantiello13} Cantiello, M., Grado, A., Blakeslee, J.~P., et al.\ 2013, \aap, 552, AA106
\bibitem[Cantiello et 
al.(2015)]{Cantiello15} Cantiello, M., Capaccioli, M., Napolitano, N., et al.\ 2015, \aap, 576, AA14 
\bibitem[Caon et al.(1993)]{Caon93} Caon, N., Capaccioli, M., 
\& D'Onofrio, M.\ 1993, \mnras, 265, 1013
\bibitem[Caon et 
al.(1994)]{Caon94} Caon, N., Capaccioli, M., \& D'Onofrio, M.\ 1994, \aaps, 106, 199
\bibitem[Capaccioli(1973)]{Capaccioli73} Capaccioli, M.\ 1973, 
\memsai, 44, 417
\bibitem[Capaccioli 
\& de Vaucouleurs(1983)]{Capaccioli83} Capaccioli, M., \& de Vaucouleurs, G.\ 1983, \apjs, 52, 465
\bibitem[Capaccioli et al.(1987)]{Capaccioli87} Capaccioli, M., 
Held, E.~V., \& Nieto, J.-L.\ 1987, \aj, 94, 1519
\bibitem[Capaccioli(1988)]{Capaccioli88} Capaccioli, M.\ 1988, 
Bol.~Acad.~Nac.~Cienc.~Cordoba, Argent., Tomo 58, Nos.~3 - 4, p.~317 - 367, 
58, 317
\bibitem[Capaccioli et al.(1992)]{Capaccioli92} Capaccioli, M., 
Caon, N., \& D'Onofrio, M.\ 1992, \mnras, 259, 323
\bibitem[Capaccioli 
\& Schipani(2011)]{Capaccioli11} Capaccioli, M., \& Schipani, P.\ 2011, The Messenger, 146, 2
\bibitem[Carlberg(1984)]{Carlberg84} Carlberg, R.~G.\ 1984, \apj, 
286, 403
\bibitem[Castro-Rodrigu{\'e}z et 
al.(2009)]{Castro09} Castro-Rodrigu{\'e}z, N., Arnaboldi, M., Aguerri, J.~A.~L., et al.\ 2009, \aap, 507, 621
\bibitem[Chen et al.(2010)]{Chen10} Chen, C.-W., C{\^o}t{\'e}, 
P., West, A.~A., Peng, E.~W., \& Ferrarese, L.\ 2010, \apjs, 191, 1
\bibitem[Coccato et al.(2009)]{Coccato09} Coccato, L., Gerhard, 
O., Arnaboldi, M., et al.\ 2009, \mnras, 394, 1249
\bibitem[Coccato et al.(2010)]{Coccato10} Coccato, L., Gerhard, 
O., \& Arnaboldi, M.\ 2010, \mnras, 407, L26
\bibitem[C{\^o}t{\'e} et al.(1998)]{Cote98} C{\^o}t{\'e}, P., 
Marzke, R.~O., \& West, M.~J.\ 1998, \apj, 501, 554
\bibitem[C{\^o}t{\'e} et al.(2004)]{Cote04} C{\^o}t{\'e}, P., 
Blakeslee, J.~P., Ferrarese, L., et al.\ 2004, \apjs, 153, 223
\bibitem[C{\^o}t{\'e} et al.(2007)]{Cote07} C{\^o}t{\'e}, P., 
Ferrarese, L., Jord{\'a}n, A., et al.\ 2007, \apj, 671, 1456
\bibitem[D'Abrusco et al.(2015)]{Dabrusco15} D'Abrusco, R., 
Fabbiano, G., \& Zezas, A.\ 2015, arXiv:1503.04819
\bibitem[Daddi et al.(2005)]{Daddi05} Daddi, E., Renzini, A., 
Pirzkal, N., et al.\ 2005, \apj, 626, 680
\bibitem[de Jong et al.(2013)]{deJong13} de Jong, J.~T.~A., 
Verdoes Kleijn, G.~A., Kuijken, K.~H., 
\& Valentijn, E.~A.\ 2013, Experimental Astronomy, 35, 25
\bibitem[de Vaucouleurs(1948)]{deV48} de Vaucouleurs, G.\ 
1948, Annales d'Astrophysique, 11, 247
\bibitem[de Vaucouleurs \& Capaccioli(1979)]{deV79} de Vaucouleurs,
  G., \& Capaccioli, M.\ 1979, \apjs, 40, 699
  \bibitem[de Vaucouleurs et al.(1991)]{deV91} de Vaucouleurs, 
G., de Vaucouleurs, A., Corwin, H.~G., Jr., et al.\ 1991, Third Reference 
Catalogue of Bright Galaxies.~Volume I: Explanations and references.~ 
Volume II: Data for galaxies between 0$^{h}$ and 12$^{h}$.~ Volume III: 
Data for galaxies between 12$^{h}$ and 24$^{h}$., by de Vaucouleurs, G.; de 
Vaucouleurs, A.; Corwin, H.~G., Jr.; Buta, R.~J.; Paturel, G.; Fouqu{\'e}, 
P..~Springer, New York, NY (USA), 1991, 2091 p., ISBN 0-387-97552-7, Price 
US\$ 198.00.~ISBN 3-540-97552-7, Price DM 448.00.~ISBN 0-387-97549-7 
(Vol.~I), ISBN 0-387-97550-0 (Vol.~II), ISBN 0-387-97551-9 (Vol.~III).,
\bibitem[Duc et al.(2015)]{Duc15} Duc, P.-A., Cuillandre, 
J.-C., Karabal, E., et al.\ 2015, \mnras, 446, 120
\bibitem[Feldmeier et al.(2004)]{Feldmeier04} Feldmeier, J.~J., 
Ciardullo, R., Jacoby, G.~H., \& Durrell, P.~R.\ 2004, \apj, 615, 196
\bibitem[Ferrarese et al.(2006)]{Ferrarese06} Ferrarese, L., 
C{\^o}t{\'e}, P., Jord{\'a}n, A., et al.\ 2006, \apjs, 164, 334
\bibitem[Ferrarese et al.(2012)]{Ferrarese12} Ferrarese, L., 
C{\^o}t{\'e}, P., Cuillandre, J.-C., et al.\ 2012, \apjs, 200, 4
\bibitem[Forbes et al.(1997)]{Forbes97} Forbes, D.~A., Brodie, 
J.~P., \& Grillmair, C.~J.\ 1997, \aj, 113, 1652
\bibitem[Forbes et al.(2011)]{Forbes11} Forbes, D.~A., Spitler, 
L.~R., Strader, J., et al.\ 2011, \mnras, 413, 2943
\bibitem[Gonzalez et al.(2005)]{Gonzalez05} Gonzalez, A.~H., 
Zabludoff, A.~I., \& Zaritsky, D.\ 2005, \apj, 618, 195
\bibitem[Grado et al.(2012)]{Grado12} Grado, A., Capaccioli, M.,
  Limatola, L., \& Getman, F.\ 2012, Memorie della Societa Astronomica
  Italiana Supplementi, 19, 362
  \bibitem[Halliday et al.(2001)]{Halliday01} Halliday, C., Davies, 
R.~L., Kuntschner, H., et al.\ 2001, \mnras, 326, 473 
\bibitem[Hopkins et al.(2010)]{Hopkins10} Hopkins, P.~F., Bundy, 
K., Hernquist, L., Wuyts, S., \& Cox, T.~J.\ 2010, \mnras, 401, 1099
\bibitem[Huang et al.(2011)]{Huang11} Huang, Z., Radovich, M., Grado,
  A., et al.\ 2011, \aap, 529, A93
\bibitem[Ibata et al.(1994)]{Ibata94} Ibata, R.~A., Gilmore, 
G., \& Irwin, M.~J.\ 1994, \nat, 370, 194
\bibitem[James \& Roos(1975)]{James75} James, F., \& Roos, M.\ 1975,
  Computer Physics Communications, 10, 343
\bibitem[Janowiecki et al.(2010)]{Janowiecki10} Janowiecki, S., 
Mihos, J.~C., Harding, P., et al.\ 2010, \apj, 715, 972
\bibitem[Jedrzejewski(1987)]{iraf} Jedrzejewski, R.~I.\ 
1987, \mnras, 226, 747
\bibitem[Jensen et al.(2003)]{Jensen03} Jensen, J.~B., Tonry, 
J.~L., Barris, B.~J., et al.\ 2003, \apj, 583, 712
\bibitem[Jord{\'a}n et al.(2007)]{Jordan07} Jord{\'a}n, A., 
Blakeslee, J.~P., C{\^o}t{\'e}, P., et al.\ 2007, \apjs, 169, 213
\bibitem[Kim et al.(2000)]{Kim00} Kim, E., Lee, M.~G., 
\& Geisler, D.\ 2000, \mnras, 314, 307
\bibitem[Kormendy(1985)]{Kormendy85} Kormendy, J.\ 1985, \apj, 
295, 73
\bibitem[Kormendy et al.(2009)]{Kormendy09} Kormendy, J., Fisher, 
D.~B., Cornell, M.~E., \& Bender, R.\ 2009, \apjs, 182, 216
\bibitem[Kuijken(2011)]{Kuijken11} Kuijken, K.\ 2011, The 
Messenger, 146, 8
\bibitem[Lee et al.(2000)]{Lee00} Lee, H., Richer, M.~G., 
\& McCall, M.~L.\ 2000, \apjl, 530, L17
\bibitem[Mei et al.(2007)]{Mei07} Mei, S., Blakeslee, J.~P., 
C{\^o}t{\'e}, P., et al.\ 2007, \apj, 655, 144
\bibitem[Mihos et al.(2005)]{Mihos05} Mihos, J.~C., Harding, 
P., Feldmeier, J., \& Morrison, H.\ 2005, \apjl, 631, L41
\bibitem[Mihos et al.(2013)]{Mihos13} Mihos, J.~C., Harding, 
P., Rudick, C.~S., \& Feldmeier, J.~J.\ 2013, \apjl, 764, L20
\bibitem[Napolitano et al.(2009)]{Napolitano09} Napolitano, N.~R., 
Romanowsky, A.~J., Coccato, L., et al.\ 2009, \mnras, 393, 329
\bibitem[Napolitano et al.(2014)]{Napolitano14} Napolitano, N.~R., 
Pota, V., Romanowsky, A.~J., et al.\ 2014, \mnras, 439, 659 
\bibitem[Nilson(1973)]{Nilson73} Nilson, P.\ 1973, Nova Acta 
Regiae Soc.~Sci.~Upsaliensis Ser.~V, 0
\bibitem[Oser et al.(2010)]{Oser10} Oser, L., Ostriker, J.~P., 
Naab, T., Johansson, P.~H., \& Burkert, A.\ 2010, \apj, 725, 2312
\bibitem[Peletier et al.(1990)]{Peletier90} Peletier, R.~F., 
Davies, R.~L., Illingworth, G.~D., Davis, L.~E., 
\& Cawson, M.\ 1990, \aj, 100, 1091
\bibitem[Peng et al.(2006)]{Peng06} Peng, E.~W., Jord{\'a}n, 
A., C{\^o}t{\'e}, P., et al.\ 2006, \apj, 639, 95
\bibitem[Pipino et al.(2008)]{Pipino08} Pipino, A., D'Ercole, A., \& Matteucci, F.\ 2008,
\aap, 484, 679
\bibitem[Pohlen 
\& Trujillo(2006)]{Pohlen06} Pohlen, M., \& Trujillo, I.\ 2006, \aap, 454, 759
\bibitem[Proctor et al.(2009)]{Proctor09} Proctor, R.~N., Forbes, 
D.~A., Romanowsky, A.~J., et al.\ 2009, \mnras, 398, 91
\bibitem[Prugniel 
\& Simien(1996)]{Prugniel96} Prugniel, P., \& Simien, F.\ 1996, \aap, 309, 749
\bibitem[Simien 
\& Prugniel(1997)]{Simien97} Simien, F., \& Prugniel, P.\ 1997, \aaps, 122, 521
\bibitem[Radovich et al. (2004)]{mario}Radovich, M. , Arnaboldi, M., Ripepi, V., et al. 004, A\&A, 417, 51
\bibitem[Romanowsky et al.(2003)]{Romanowsky03} Romanowsky, A.~J., 
Douglas, N.~G., Arnaboldi, M., et al.\ 2003, Science, 301, 1696
\bibitem[Romanowsky et al.(2009)]{Romanowsky09} Romanowsky, A.~J., 
Strader, J., Spitler, L.~R., et al.\ 2009, \aj, 137, 4956
\bibitem[Romanowsky et al.(2012)]{Romanowsky12} Romanowsky, A.~J., 
Strader, J., Brodie, J.~P., et al.\ 2012, \apj, 748, 29
\bibitem[Saglia et al.(2002)]{Saglia02} Saglia, R.~P., Maraston, 
C., Thomas, D., Bender, R., \& Colless, M.\ 2002, \apjl, 579, L13
\bibitem[{{S\'ersic}(1968)}]{1968adga.book.....S}{Sersic}, J.~L. 1968, {Atlas de galaxias australes} (Cordoba, Argentina:  Observatorio Astronomico, 1968)
\bibitem[Shen et al.(2003)]{Shen03} Shen, S., Mo, H.~J., 
White, S.~D.~M., et al.\ 2003, \mnras, 343, 978
\bibitem[Tal et al.(2009)]{Tal09} Tal, T., van Dokkum, P.~G., 
Nelan, J., \& Bezanson, R.\ 2009, \aj, 138, 1417
\bibitem[Tonry 
\& Schneider(1988)]{Tonry88} Tonry, J., \& Schneider, D.~P.\ 1988, \aj, 96, 807
\bibitem[Tonry et al.(2001)]{Tonry01} Tonry, J.~L., Dressler, 
A., Blakeslee, J.~P., et al.\ 2001, \apj, 546, 681
\bibitem[Tortora et al.(2011)]{Tortora11} Tortora, C., Romeo, 
A.~D., Napolitano, N.~R., et al.\ 2011, \mnras, 411, 627
\bibitem[Tortora et al.(2013)]{Tortora13} Tortora, C., Pipino, 
A., D'Ercole, A., Napolitano, N.~R., 
\& Matteucci, F.\ 2013, \mnras, 435, 786
\bibitem[Usher et al.(2012)]{Usher12} Usher, C., Forbes, D.~A., 
Brodie, J.~P., et al.\ 2012, \mnras, 426, 1475
\bibitem[Usher et al.(2015)]{Usher15} Usher, C., Forbes, D.~A., 
Brodie, J.~P., et al.\ 2015, \mnras, 446, 369
\bibitem[van Dokkum et al.(2010)]{vanDokkum10} van Dokkum, P.~G., 
Whitaker, K.~E., Brammer, G., et al.\ 2010, \apj, 709, 1018
\bibitem[Watkins et al.(2014)]{Watkins14} Watkins, A.~E., Mihos, 
J.~C., Harding, P., \& Feldmeier, J.~J.\ 2014, \apj, 791, 38
\bibitem[Whitmore \& Bell(1988)]{Whitmore88} Whitmore, B.~C., \& Bell,
  M.\ 1988, \apj, 324, 741
\bibitem[Wood(2007)]{Wood07} Wood, S.~N.\ 2007, arXiv:0709.3906
\bibitem[(Wood, 2011)]{Wood11} Wood, S.N. (2011) Fast stable restricted maximum likelihood 
and marginal likelihood estimation of semiparametric generalized linear models. Journal of 
the Royal Statistical Society (B) 73(1):3-36
\bibitem[Yoon et al.(2006)]{Yoon06} Yoon, S.-J., Yi, S.~K., 
\& Lee, Y.-W.\ 2006, Science, 311, 1129
\bibitem[Zibetti et al.(2004)]{Zibetti04} Zibetti, S., White, 
S.~D.~M., \& Brinkmann, J.\ 2004, \mnras, 347, 556
\bibitem[Zibetti et al.(2005)]{Zibetti05} Zibetti, S., White, 
S.~D.~M., Schneider, D.~P., \& Brinkmann, J.\ 2005, \mnras, 358, 949

\end{thebibliography}
\end{document}